\documentclass[lettersize,journal]{IEEEtran}
\usepackage{amsmath,amsfonts}
\usepackage{algorithmic}
\usepackage{algorithm}
\usepackage{array}
\usepackage{textcomp}
\usepackage{stfloats}
\usepackage{url}
\usepackage{verbatim}
\usepackage{graphicx}
\usepackage{float} 
\usepackage{subcaption} 
\usepackage{cite}
\usepackage{hyperref}
\usepackage{xcolor} 

\begin{document}

\title{A Comparison of Heuristic, Human-Sensitive, and Self-Play DRL Policies in Human–Agent Teaming}

\author{Ayman bin Kamruddin, Christopher Lam, Gaurav Patil, Mirco Musolesi, Mario di Bernardo, Michael J. Richardson

\thanks{Manuscript received November 18, 2026.}
\thanks{This work has been submitted to the IEEE for possible publication. Copyright may be transferred without notice, after which this version may no longer be accessible.}

\thanks{This work was supported by a PhD fellowship for Dr. bin Kamruddin in Modelling and Engineering Risk and Complexity at the Scuola Superiore Meridionale, Naples, Italy; the Australian Department of Defence—Defence Science and Technology Group's Human Performance Research Network (HPRnet) and the Centre for Advanced Defence Research in Robotics and Autonomous Systems (CADR-RAS); and an Australian Research Council Future Fellowship (FT180100447) awarded to Prof. Richardson.}%

\thanks{Corresponding authors: Christopher Lam (christopher.lam3@hdr.mq.edu.au); Ayman bin Kamruddin (ayman.binkamruddin@gmail.com).}%

\thanks{Ayman bin Kamruddin is with the Scuola Superiore Meridionale, Naples, Italy; the University of Naples Federico II, Naples, Italy; and the School of Psychological Sciences, Macquarie University, Sydney, NSW 2109, Australia.}%

\thanks{Christopher Lam is with the School of Psychological Sciences, Macquarie University, Sydney, NSW 2109, Australia.}%

\thanks{Gaurav Patil is with the School of Psychological Sciences and the Performance and Expertise Research Centre, Macquarie University, Sydney, NSW 2109, Australia.}%

\thanks{Mirco Musolesi is with University College London, London, U.K.}%

\thanks{Mario di Bernardo is with the Scuola Superiore Meridionale, Naples, Italy, and the University of Naples Federico II, Naples, Italy.}%

\thanks{Michael J. Richardson is with the School of Psychological Sciences and the Performance and Expertise Research Centre, Macquarie University, Sydney, NSW 2109, Australia.}%
}

\markboth{Journal of \LaTeX\ Class Files,~Vol.~14, No.~8, August~2021}%
{Shell \MakeLowercase{\textit{et al.}}: A Sample Article Using IEEEtran.cls for IEEE Journals}


\maketitle

\begin{abstract}
Dynamical Perceptual-Motor Primitives (DPMPs) offer a tractable framework for generating human-like movement dynamics in artificial agents (AAs) and have been used to support coordination in human-AI teaming tasks. Traditionally, DPMP controllers rely on heuristic (rule-based) decision policies derived from human behavior to guide goal selection and action sequencing. Although effective in structured tasks, such heuristics can be difficult to generalize to more complex settings, motivating interest in deep reinforcement learning (DRL) approaches for learning decision policies for human-AI interaction. A key challenge, however, is ensuring that DRL policies remain aligned with human strategies, particularly when agents are trained through self-play. To examine this issue, we compared three target-selection policies governing DPMP-based AAs in a first-person herding task: (1) a human-derived heuristic policy, (2) a self-play DRL policy (SP-DRL), and (3) a human-sensitive DRL policy (HS-DRL) trained alongside a heuristic agent. Results showed that heuristic agents most closely matched typical human–human coordination patterns, followed by HS-DRL agents, whereas SP-DRL agents deviated most. These findings highlight the importance of human-informed training when developing DRL-based teammates and suggest that simple heuristic policies may remain an effective approach for designing human-aligned AAs capable of human–AI collaboration.
\end{abstract}

\begin{IEEEkeywords}
human–agent teaming, dynamical perceptual-motor primitives, deep reinforcement learning, human-AI alignment, heuristic decision policies.
\end{IEEEkeywords}

\section{Introduction}
Many real-world tasks require people to coordinate their actions with others in real time under evolving environmental constraints. Concerted success hinges on anticipating others’ actions and intentions \cite{sebanz2009prediction}, organizing movement spatio-temporally \cite{richardson2014complex}, and adapting behavior as interactions unfold \cite{richardson2015symmetry}. These demands are increasingly relevant as artificial agents (AAs) are introduced into roles alongside humans. Accordingly, growing research in human-agent teaming (HAT) examines whether AAs can exhibit coordination patterns similar to those observed in human-human interaction \cite{mcneese2018teaming}. In such settings, effective teaming extends beyond task completion, depending on whether an AA's behavior is predictable, legible, and compatible with human expectations \cite{Curioni2019}.

Modeling human-like behavior in AAs can be framed as a hierarchical control problem, separating action dynamics (i.e., controlling how to move) from higher-order action decisions (i.e., determining which action to perform and when) \cite{patil2023scaffolding}. From a dynamical systems perspective, behavior emerges through continuous coupling between agents and the environment \cite{Warren2006TheAction}. Within this framework, dynamical perceptual-motor primitives (DPMPs)—elementary sets of low-dimensional nonlinear functions—provide a compact architecture for modeling the continuous dynamics of goal-directed movement \cite{saltzman1987skilled, hogan2012dynamic, richardson2015selforganized}. DPMP models have been widely used to generate human-like behaviors in both solo and multiagent tasks, such as navigation \cite{fajen2003behavioral, warren2018collective, bin2024modelling}. When serving as the action dynamics architecture of AA \cite{nalepka2017herd}, DPMPs specify how an agent moves towards a selected goal state but not which goal or subgoal state to pursued when multiple options are available. Modeling such decision policies is substantially more challenging, as action selection must account for the evolving behavior of other agents and changing environmental constraints \cite{Curioni2019, auletta2023predicting}. 

One approach to modeling AA action decisions is to derive heuristic (rule-based) policies from observed human decisions \cite{bin2024modelling, auletta2021humaninspired}. Such heuristics are interpretable and often align well with human decision patterns, supporting predictable coordination in HAT \cite{nalepka2017herd, Curioni2019}. However, they can struggle to generalize to more complex or novel task configurations, limiting their flexibility in dynamic settings \cite{auletta2021humaninspired}. 

As an alternative, deep reinforcement learning (DRL) learns optimized decision polices through simulation-based reward maximization \cite{patil2021dynamical}. A common training paradigm is self-play, where agents learn by interacting with copies of themselves rather than human data \cite{patil2023scaffolding}. While SP-trained agents can achieve high task performance, their strategies often diverge from typical human behavior, reducing suitability for HAT \cite{shek2019takeaways}. Human-sensitive DRL has therefore been proposed to incorporate human behavioral models during training, encouraging policies that better align with human strategies \cite{carroll2020the}.

The present study builds on prior work \cite{bin2024modelling} and related work currently under review \cite{bin2024modellingb}, which introduced a first-person herding task for examining action dynamics and decisions in multi-target environments. There, a DPMP-based locomotion model governed by a human-derived heuristic target-selection policy effectively reproduced human-like herding behavior. Extending this approach, we used the same task to examine how different decision-policy designs shape coordination dynamics in human–AA dyads. Specifically, we compared policies: (1) a heuristic policy \cite{bin2024modelling}, (2) a Self-Play DRL (SP-DRL) policy, and (3) a Human-Sensitive DRL (HS-DRL) policy trained alongside the Heuristic AA. Because all AAs used the same DPMP controller, the study isolated the effect of decision-policy design on human-AI teaming behavior. In particular, we evaluated whether HS-DRL improves human-alignment between DRL agents and human performance.

\subsection{DPMP Model of Human Herding Navigation}
The herding task provides a controlled paradigm for studying goal-directed navigation, decision-making, and multiagent collaboration. Typically implemented as a video game resembling everyday navigation, the task involves one or more participants acting as virtual herder agents (HAs) who corral target agents (TAs; e.g., cattle) into a containment zone (see also Fig. \ref{fig:experiment_setup}). HAs influence TAs by leveraging the TAs’ tendency to move away when approached within a critical distance. Participant HAs in prior work consistently converge onto similar paths, which can be accurately modeled using DPMPs \cite{nalepka2017herd, bin2024modelling}, and subsequently used to control AAs’ action dynamics in human-AA herding tasks \cite{nalepka2019human, rigoli2022comparison}.

\begin{figure}[ht]
     \centering
     \begin{subfigure}[b]{0.24\textwidth}
         \centering
         \includegraphics[width=\textwidth]{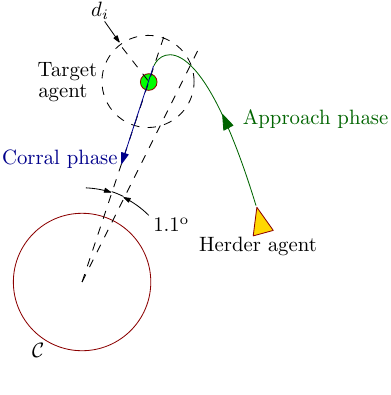}
         \caption{}
         \label{fig:approach_corral}
     \end{subfigure}
     \hfill
     \begin{subfigure}[b]{0.24\textwidth}
         \centering
         \includegraphics[width=\textwidth]{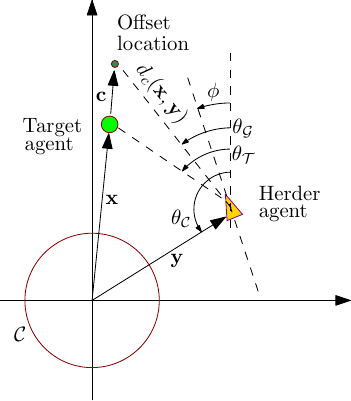}
         \caption{}
         \label{fig:egocentric}
     \end{subfigure}
        \caption{(a) First-person herding phases. The HA first approaches the TA, then transitions to corralling the TA towards the containment zone. The median angle at which human HAs begin influencing a TA is 1.1°. The TA influence radius is $d_i = 10 \text{ m}$. (b) Model variables. $\mathcal{C}$ denotes the containment zone. Positions are given by $\mathbf{x}$ (TA) and $\mathbf{y}$ (HA). Offset goal location is $\mathbf{c}$, and $d_c(\mathbf{x},\mathbf{y})$ is the HA-offset distance. The HA heading angle is $\phi$. $\theta_{\mathcal{G}}, \theta_{\mathcal{T}}, \theta_{\mathcal{C}}$ denote the angles of the offset goal, TA, and containment zone, respectively, measured relative to the vertical axis centered on the HA.}
        \label{fig:my_setup}
\end{figure}

The present model builds on the work of bin Kamruddin et al. \cite{bin2024modelling}, which extended the DPMP-based steering model of human navigation developed by Fajen and Warren \cite{fajen2003behavioral} to a first-person herding task. In this study, participants controlled a HA tasked with corralling a single TA into a centrally-located containment zone. Analysis of HA trajectories revealed two distinct phases: (i) an approach phase and (ii) a corral phase (see Fig. \ref{fig:approach_corral}). During the approach phase, participants steered the HA to an offset position behind the TA, aligned with the direction of the containment zone and located beyond the HA-TA repulsion threshold to avoid prematurely influencing the TA. Once positioned at this offset, the corral phase began, with the HA advancing toward the TA, leveraging the repulsive force to drive it into the containment area.

\cite{bin2024modelling} captured the two phases (Fig.~\ref{fig:egocentric}) by modeling the HA’s heading direction (${\phi}$), turning rate ($\dot{\phi}$), and angular acceleration ($\ddot{\phi}$) as a function of the current positions of the TA ($\mathbf{x}$), HA ($\mathbf{y}$), and environmental obstacles ($\mathbf{x}_{\mathcal{O}_i}$). That is,

\begin{equation}\label{eq:switching_model_egocentric_multi_target}
\begin{split}
\ddot{\phi} = &- b\dot{\phi} + \psi_{\mathcal{G}}(\mathbf{x}, \mathbf{y}, \phi, \theta_{\mathcal{G}}) + \psi_{\mathcal{T}}(\mathbf{x}, \mathbf{y}, \phi, \theta_{\mathcal{T}}) \\
&+  \sum_{i=1}^{N} \psi_{\mathcal{O}_i}(\mathbf{x}_{\mathcal{O}_i}, \mathbf{y}, \phi, \theta_{\mathcal{O}_i}),
\end{split}
\end{equation}

\noindent where $\mathbf{x}(t) \in \mathbb{R}^2$ and $\mathbf{y}(t)\in \mathbb{R}^2$ denote the TA and HA positions relative to the center of the containment zone. The coupling term $\psi_\mathcal{G}$ represent attraction toward the offset goal location, and $\psi_\mathcal{T}$ as repulsion from the selected TA. $\theta_{\mathcal{G}}$, $\theta_{\mathcal{T}}$ and $\theta_{\mathcal{O}_i}$ respectively represent the angular positions of the offset goal location, the TA, and environmental obstacles relative to the HA. $b \in \mathbb{R}^+$ is a positive damping constant that smooths changes in the HA's heading direction. Detailed definitions of these terms are provided in Supplementary Material A. 

The term $\sum_{i=1} ^ N  \psi_{\mathcal{O}_i}$ represents the sum of repulsive obstacle terms arising from the containment zone ($\psi_{\mathcal{C}}$) and each TA not currently selected (non-targeted TAs; $\psi_{\mathcal{NT}}$). The multi-player task warranted a similar obstacle term for the other HA ($\psi_{\mathcal{OH}}$) \cite{bin2024modellingb}. As the containment zone, non-targeted TAs, and the other HA were each modeled by repulsive form,
\begin{equation}\label{eq:N_objects2}
N = \underbrace{1}_{\mathrm{containment\,zone}} + \underbrace{(N_{\mathcal{NT}} - 1)}_{\mathrm{non-targeted\,TAs}} +\underbrace{(N_{\mathcal{OH}} - 1)}_{\mathrm{other\,HA}}.
\end{equation}

This model reproduced prototypical human trajectories in single-player three-target herding (Eq.~\ref{eq:switching_model_egocentric_multi_target} omitting $\psi_{\mathcal{OH}}$) and, when combined with a human-derived heuristic target-selection policy, can simulate multi-target scenarios \cite{bin2024modelling}. However, effective operation of the DPMP controller depends on selecting an appropriate target-selection policy when multiple action possibilities are available.

\subsection{Target-Selection Policies and Action Decisions}
In herding tasks, this action decision corresponds to which TA should be approached and corralled at any given time, while accounting for the behavior of other HAs and positions of TAs in the environment \cite{patil2023scaffolding}. We therefore examined different target-selection policies governing these action decisions. 

\subsubsection{Heuristics for Target-Selection Decisions}
Research employing the herding paradigm has identified heuristic (i.e., rule-based) target-selection policies that capture the decision sequence used by human HAs when selecting and/or switching between TAs in multi-target environments. In a single-herder three-target task, \cite{bin2024modelling} identified the Successive Collinear Angle (SCA); a three-rule heuristic target-selection policy. First, a human HA selects the TA closest in angular distance—defined as the angular deviation between the TA-origin and HA-origin vectors—to their initial position relative to the center of the containment zone (i.e., field origin). Second, after corralling a TA, subsequent TAs are selected based on angular proximity to the HA’s current position. Third, when two TAs are collinear relative to the containment zone (within $20^\circ$), the HA first selects the TA furthest from the zone and, upon approaching the second TA en route, corrals both by switching between them. The SCA policy accurately predicted target-selection order in approximately 79\% of participants’ single-herder, three-target trials and, when integrated with the DPMP model (Eq.~\ref{eq:switching_model_egocentric_multi_target}), reproduced human herding trajectories. 

Subsequent work extended the SCA policy to collaborative multi-herder scenarios, demonstrating that it captures human-like behavior in human–human and human–AA teams \cite{bin2024modellingb}. The interpretability and determinism of heuristic policies make them well suited for integration with DPMP-based models, producing predictable decisions supporting reciprocal coordination in HAT \cite{Curioni2019}. However, heuristics are simplified approximations of human decision-making and lack the flexibility and adaptability required for complex or novel task configurations \cite{auletta2022herding, nalepka2017herd}. These limitations motivate the exploration of learning-based approaches for target-selection policies.

\subsubsection{Deep Reinforcement Learning for Target-Selection Decisions}

An alternative approach to heuristics for action decisions is deep reinforcement learning (DRL), which learns action-selection policies through reward-driven interaction with an environment \cite{patil2021dynamical}. In DRL, a deep neural network approximates a policy that maps observed states to actions, and is iteratively optimized to maximize expected task rewards \cite{sutton1998reinforcement}. This framework has increasingly been used to derive decision policies for DPMP-controlled AAs, offering greater adaptability compared to the rigidity of heuristic-based AAs \cite{nalepka2021interaction}. A common DRL training paradigm is self-play (SP), in which agents learn by interacting with copies of themselves rather than human partners \cite{patil2023scaffolding}. While SP-DRL can produce highly effective, able to exceed expert-level human performance, the resulting strategies often diverge from typical human decision patterns \cite{carroll2020the}. Such divergence can negatively affect human-AA teaming, as human coactors may perceive SP-trained AAs as unpredictable or uncooperative \cite{rigoli2022comparison}. 

To address these limitations, human-sensitive DRL (HS-DRL) has been proposed as a training paradigm that incorporates models of human behavior during learning \cite{carroll2020the}. In this approach, DRL agents are trained alongside simulated human proxies derived from heuristic or behavioral models, encouraging the learned policy to align more closely with human coordination strategies \cite{nalepka2021interaction}. Previous work has shown that such human-sensitive DRL training can improve joint action within human-AI teams compared to traditional SP-DRL agents \cite{carroll2020the}. However, it remains unclear whether HS-DRL produces more effective HAT than SP-DRL or human-derived heuristic policies. The present study addresses this question by comparing heuristic, SP-DRL, and HS-DRL target-selection policies in a first-person herding task.

\subsection{Current Study}
Building on \cite{bin2024modelling} and \cite{bin2024modellingb}, this study aimed to (1) further validate the SCA heuristic policy for controlling DPMP-based HAs in human-AA dyads, and (2) benchmark its efficacy against SP-DRL and HS-DRL policies. While the three types of AAs differed in their target-selection policies, all shared the same DPMP movement model (Eq.~\ref{eq:switching_model_egocentric_multi_target}). This design therefore isolates how different decision-policy approaches influence human-AA coordination. In particular, we examined whether HS-DRL AAs produce more human-compatible behavior than SP-DRL AAs and approach the Heuristic AA's performance.

To evaluate these policies, human-AA dyads completed a multiplayer first-person herding task involving three to five TAs. In a between-subjects design, participants collaborated with either the Heuristic, SP-DRL, or HS-DRL AA. Target-selection decisions and movement trajectories of the human and AAs were analyzed to assess the effects of the policies on HAT. These measures were derived from comparison with human-human dyad data from the same task \cite{bin2024modellingb}, providing a reference for typical human coordination patterns.

Because the SCA was derived from observed human strategies, the Heuristic AA was expected to best capture human herding dynamics, reflected in the highest decision and trajectory overlap with the human-human baseline. Among the DRL policies, the HS-DRL AA was expected to produce more human-aligned behavior than the SP-DRL AA. Consequently, participants collaborating with the Heuristic AA were predicted to exhibit behavior more similar to the human baseline, followed by those paired with the HS-DRL AA, with the least typical behavior for those paired with the SP-DRL AA.

\section{Methodology}
\subsection{Participants}

Seventy-one students from Macquarie University participated in exchange for course credits. Nine participants were excluded due to incomplete sessions or technical issues, leaving a final sample of 62 participants. Participants were randomly assigned to an AA condition: Heuristic ($n$ = 22), HS-DRL ($n$ = 20), or SP-DRL ($n$ = 20). Sample sizes were comparable to prior herding studies \cite{nalepka2017herd,nalepka2019human}. Ages ranged from 17 to 32 years ($M$ = 19.03, $SD$ = 2.58). All participants provided informed consent, and the study was approved by the Macquarie University Human Research Ethics Committee.

\subsection{Apparatus and Task}
\begin{figure*}[ht]
    \centering
    \begin{subfigure}{.5\textwidth}
    \centering
    \includegraphics[width = 1.02\textwidth]{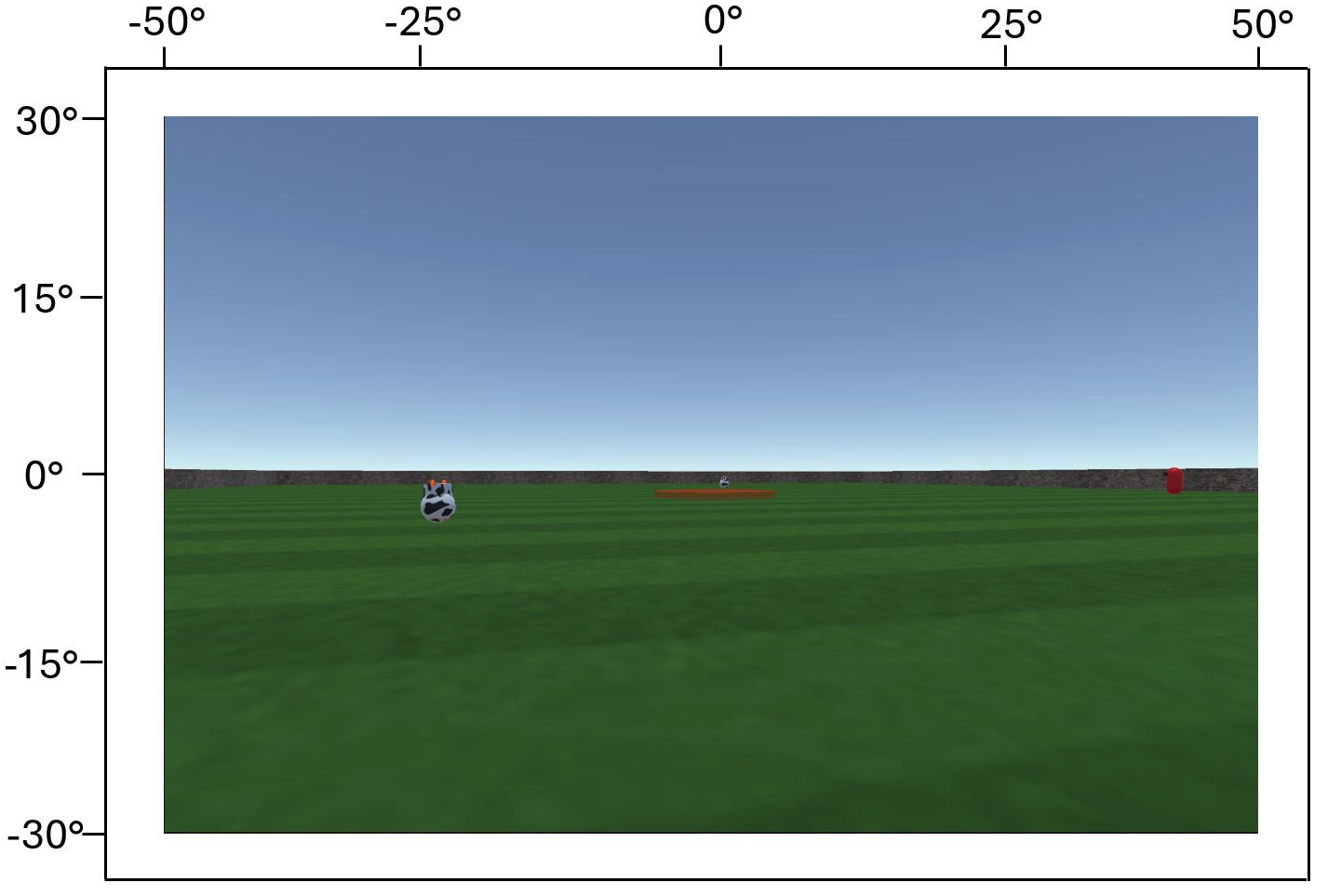}
    \caption{First-person point of view.}
    \label{fig:fp_pov}
    \end{subfigure}
    \begin{subfigure}{.46\textwidth}
    \centering
    \includegraphics[width = 1.05\textwidth]{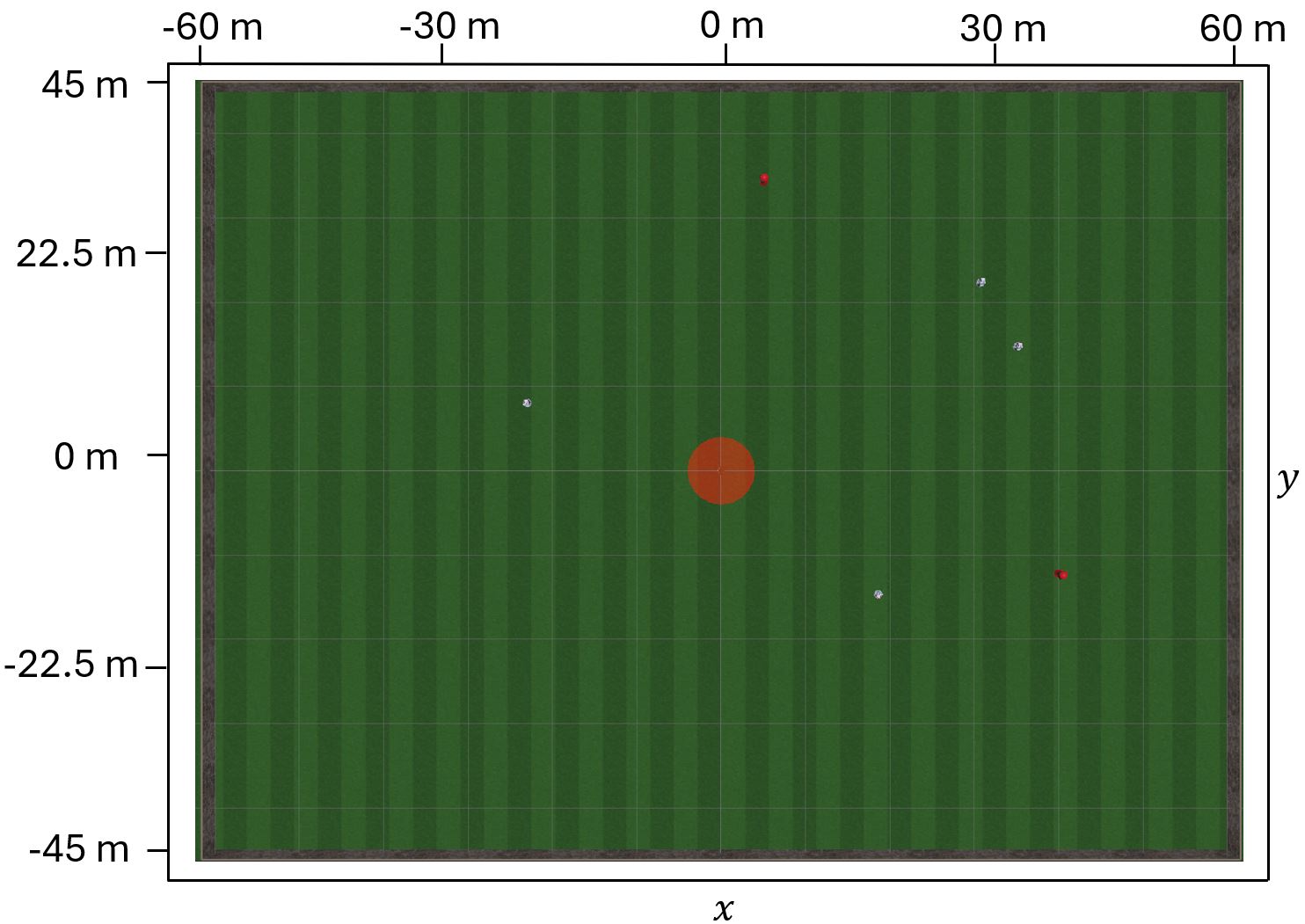}
    \caption{Corresponding bird's eye view.}
    \label{fig:setup}
    \end{subfigure}
    \caption{Task environment. (a) Example first-person view of the game field from a HA. Visible are TAs, the containment zone (red), boundary walls (gray), and partner HA (red capsule) in the distance. (b) Bird's-eye view of the field (not visible to participants), showing HA positions (red dots) and TA positions (white dots).}
    \label{fig:experiment_setup}
\end{figure*}

The first-person herding task was designed using the Unity game engine (version 3.3.0, Unity Technologies, San Francisco, USA), presented to participants on a 27-inch computer monitor. Participants paired with an AA located and corralled TAs into a containment zone (Fig.~\ref{fig:experiment_setup}). The field, enclosed by walls, was 120 m × 90 m. A containment zone was positioned at the field origin, marked by a translucent red circle (radius = 4 m), remaining stationary with agents able to move through freely. HAs were represented as red capsule, while TAs appeared as spherical cows. TAs followed the repulsion-based dynamics described in \cite{bin2024modelling}, remaining stationary until a HA entered an influence radius of 10 m, at which point they moved away from the HA according to the model defined in that work. Game state data (HA and TA positions, heading angles and movement velocities) was recorded at 50 Hz.

\subsection{Herder Agent Control}
Participants controlled the HA using a keyboard and mouse. The keyboard controlled movement, with the W, A, S, and D keys moving the HA forward, left, backward, and right, respectively; diagonal movement was possible by pressing two keys simultaneously. Movement speed was fixed at 5 m/s. The mouse controlled the HA's viewing direction, decoupled from the HA's motion, allowing participants to adjust their angular orientation independently of movement. The first-person view spanned 60° vertically and 97.6° horizontally.

\subsection{Movement Dynamics of the Artificial Agents}
The movement dynamics of all AAs were defined by the same DPMP model (Eq.~\ref{eq:switching_model_egocentric_multi_target}), under the multi-herder condition specified in Eq.~\eqref{eq:N_objects2} \cite{bin2024modellingb}. The assignment of the coupling terms ($\psi_{\mathcal{G}}, \psi_{\mathcal{T}}$, $\psi_{\mathcal{O}_i}$) was determined by an AA's target-selection policy. Consequently, all AAs followed the same approach and corral phases, differing only in their action selections.

\subsection{Target-Selection Policies for Artificial Agents}
Three types of AAs were paired with participants, forming three experimental conditions. While all AAs shared the same DPMP model, they differed in their target-selection policies. Each policy output corresponded to a TA identification (ID) number (1-5; 0 indicating no target). The selected TA determined the offset goal location through the term $\psi_{\mathcal{G}}$ and the targeted TA in $\psi_{\mathcal{T}}$. The containment zone, all remaining TAs, participant HA contributed repulsive obstacle terms in $\psi_{\mathcal{O}_i}$.

\subsubsection{Heuristic (SCA)}
The Heuristic AA used a deterministic rule-based target-selection policy based on the SCA heuristic \cite{bin2024modelling, bin2024modellingb}. The policy selects TAs according to angular distance relative to the containment zone and prioritizes successive targets based on proximity in angular space. When two or more TAs are near-collinear (within $20^\circ$) relative to the containment zone, the furthest TA is selected first, with the near TA influenced en route. The SCA policy was adapted for the multi-herder task, accounting for the other HA by prioritizing TAs closer to the agent and further from the other HA \cite{bin2024modellingb}. A tabulated summary of the generalized SCA policy is provided in Supplementary Material B.


\subsubsection{Self-Play DRL}
SP-DRL AAs were implemented using a fully connected multilayer perceptron (MLP) with two hidden layers (128 nodes each), trained using the Proximal Policy Optimization algorithm (PPO; \cite{schulman2017proximal}). The state space inputs to the network included the positions $(x, y)$ and velocities of all HAs and TAs. Rewards were computed at each environment step and included a small time penalty ($-0.0001$), a positive reward for containing a TA ($0.6/N_{\mathrm{TAs}}$), and a penalty for TA escape ($-1.2/N_{\mathrm{TAs}}$). Training was conducted using 10 SP-DRL AAs interacting in self-play across 20 parallel environment instances running on CPUs at $40\times$ real-time speed, for a total of 150 million environment steps. The two best performing agents were selected based on completion time across the training trial set.

\subsubsection{Human-Sensitive DRL}
HS-DRL AAs used the same MLP architecture, PPO training algorithm, and reward structure as SP-DRL AAs. Training was conducted for 150 million environment steps, with the two best-performing agents selected for the study. HS-DRL AAs were trained using a hybrid approach \cite{patil2021dynamical} (i.e., rather than self-play), where each HS-DRL AA interacted with a Heuristic AA, allowing the policy to adapt to human-like target-selection behavior rather than another DRL agent.

\subsection{Procedure, Trial Sequence, and Initial Conditions}
Upon arrival, participants provided informed consent and completed a brief questionnaire collecting demographic information. The experimenter then explained the task, in which participants collaborated with an AA to locate and corral TAs into a containment zone as quickly as possible. Participants first completed a practice session consisting of five single-HA single-TA trials to familiarize themselves with the controls and task mechanics. After a short break, participants completed the experimental session comprising 24 multi-HA multi-TA trials in which they collaborated with an AA. Trials were evenly distributed across conditions with 3, 4, or 5 TAs per trial.

Following \cite{bin2024modellingb}, the first six human–AA trials also served as practice to familiarize participants with the presence of the AA and were excluded from analysis. The remaining 18 trials (Trials 7–24) were analyzed. Trial order was randomized across participants. Each trial began with agents placed at pre-specified initial positions and ended once all TAs stopped within the containment zone.

Participants were assigned to either Player 1 or Player 2 initial positions, counterbalanced across conditions to control for starting-location effects. The sequence and initial conditions of the analyzed trials matched those used in the human–human baseline experiment of \cite{bin2024modellingb}, enabling direct comparison with human dyad performance. Participants were not informed which policy controlled their AA partner.


\subsection{Measurement of Target-Selection Decisions}
Target-selection decisions was quantified using a dependent variable termed \textit{decision overlap}, which measured the similarity between a HA’s target-selection sequence (human or AA) and those of typical human HAs from the human-human dataset reported in \cite{bin2024modellingb}. For each trial, a 50 Hz time series was extracted, indicating when the HA was actively corralling a TA. At each timestep a binary vector of dimension $N_{\mathrm{TA}}$ was constructed, where each element indicated whether the HA was interacting with a particular TA (1 = engagement, 0 = no engagement). This produced a binary time series representing the sequence of target selections during a trial.

The resulting decision sequences were compared with those of the 21 human participants occupying the player role (Player 1 or Player 2) in the human-human baseline dataset \cite{bin2024modellingb}. Similarity between sequences was computed using Dynamic Time Warping (DTW). The resulting DTW distance was normalized by the product of the lengths of the two time series, converted to a similarity score by subtracting the normalized distance from 1, yielding values between 0 (no overlap) and 1 (perfect overlap). For each trial, the decision overlap for a HA was obtained by averaging the similarity values computed against the 21 human reference sequences. This produced 18 decision overlap scores per HA across the analyzed trials. 

\begin{algorithm}[H]
\caption{Calculating DTW Target-Selection Overlap Scores for two HAs' timeseries $i$ and $j$. $t$ is time, and $k$ ranges from $1$ to $N_{TA}$.}\label{alg:DTW-TS}
\begin{algorithmic}
\REQUIRE $HA_{i}$ and $HA_{j}$: player timeseries data
\STATE Initialize $\text{binary-encoding}_{i \text{ or } j}(t) \gets \underbrace{(0,0,\dots,0)}_{N_{TA}}$ for all $t$
\FOR{$t = 1$ \TO $\text{len}(HA_{i \text{ or }j})$}
    \FOR{$k = 1$ \TO $N_{TA}$}
        \IF{$HA_{i \text{ or } j}(t)$ interacted with $TA_k(t)$}
            \STATE $\text{binary-encoding}_{i \text{ or } j}(t)_k \gets 1$
        \ENDIF
    \ENDFOR
\ENDFOR
\STATE $\text{DTW} \gets \text{DTW}(\text{binary-encoding}_i, \text{binary-encoding}_j)$
\STATE $\text{normalized DTW} \gets \frac{\text{DTW}}{\text{len}(HA_{i}) \times \text{len}(HA_{j})}$
\RETURN $1 - \text{normalized DTW}$
\end{algorithmic}
\end{algorithm}

\subsection{Measurement of Navigation Trajectories}
\begin{figure*}
    \centering
    \includegraphics[width=\linewidth]{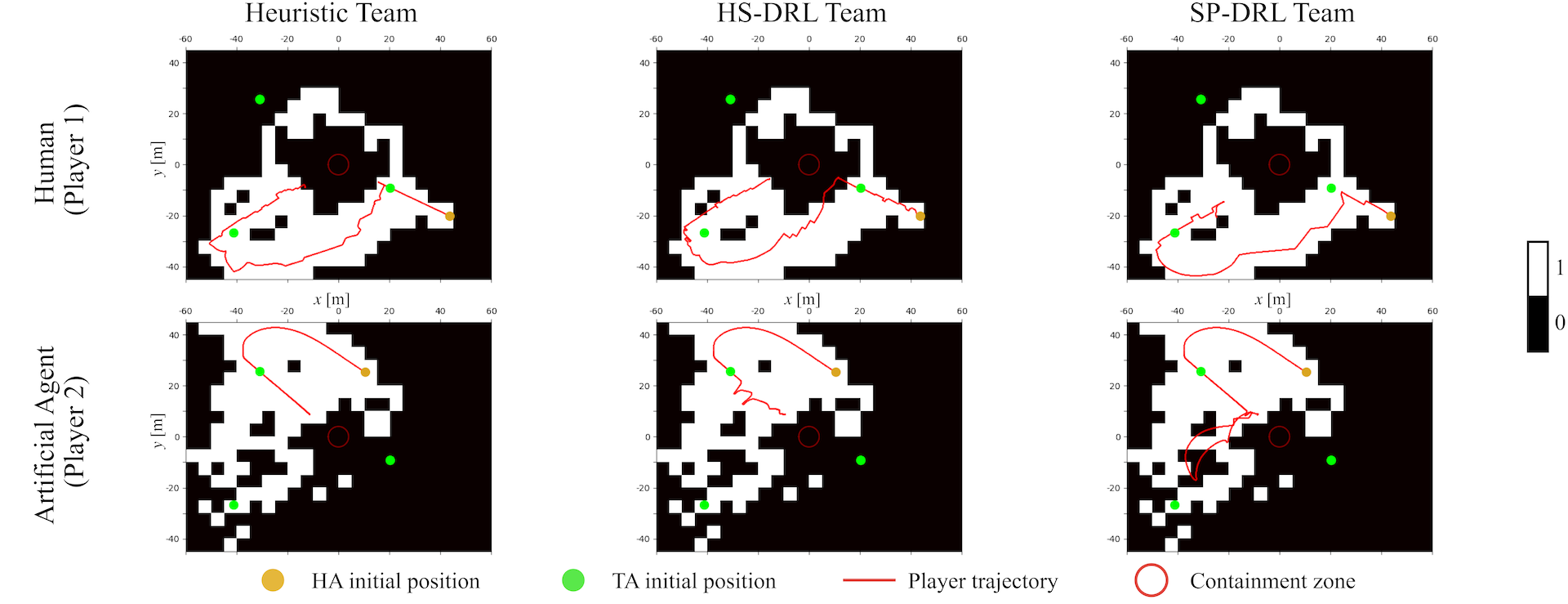}
    \caption{Example binary trace maps (Trial 7) with participant and AA trajectories. Binary trace maps were derived from the human-human dataset \cite{bin2024modellingb}. Red lines indicate the HA paths used to compute trajectory overlap with typical human navigation.}
    \label{fig:traj_plots_on_binary}
\end{figure*}

Navigation behavior was quantified using \textit{binary trace overlap}, which measured the similarity between a HA's trajectory (human or AA) and typical human trajectories from the human-human dataset \cite{bin2024modellingb}. Binary trace maps were generated from the human-huma dataset by dividing the game field into $5 \times 5$ spatial bins and recording the frequency of trajectory points passing through each bin (see Fig. \ref{fig:traj_plots_on_binary} and Supplementary Material B). Bins containing more than 10 trajectory points were marked white, while bins with fewer were marked black. One binary trace map was created for each of the 18 analyzed trials using the corresponding human-human trajectory data. 

For each trial, a HA's trajectory overlap score was computed as the proportion of trajectory points falling within the white bins, normalized by the total number of bins visited by the trajectory. Scores ranged from 0 to 1, where higher values indicated greater similarity to typical human navigation patterns.

\section{Results}
Separate mixed-design ANOVAs were conducted to examine differences in target-selection (decision) overlap and binary trace (trajectory) overlap scores for both AAs and participants. For the AA analyses, the between-subjects factor was \textit{AA Type} (Heuristic, HS-DRL, SP-DRL). For participant analyses, the corresponding factor was \textit{AA Coherder}. The number of TAs in each trial (\textit{TA Num}: 3, 4, or 5) was included as a repeated-measures factor. Sphericity violations were corrected using the Greenhouse-Geisser adjustment where applicable. Post-hoc pairwise comparisons were Bonferroni corrected with $\alpha = .05$.

\subsection{Artificial Agents on Target-Selection Decision Overlap}
The mixed-design ANOVA revealed significant main effects for \textit{AA Type}, \(F(2, 59) = 28.73, p < .001, \eta_p^2 = .493\), and \textit{TA Num}, \(F(1.76, 104.09) = 702.23, p < .001, \eta_p^2 = .922\), as well as a significant \textit{AA Type} \(\times\) \textit{TA Num} interaction, \(F(3.53, 104.09) = 6.88, p < .001, \eta_p^2 = .189\). 

Simple effects analyses showed differences for 3-TA trials, \(F(2, 59) = 11.67, p < .001\), with the Heuristic AA exhibiting higher decision overlap than the HS-DRL (\(\text{contrast} = .014\), \(p = .010\)) and SP-DRL AAs (\(\text{contrast} = .022\), \(p < .001\)), with no difference between the HS- and SP-DRL AAs (\(p = .307\)). 

Again, AA Type was significant for 4-TA trials, \(F(2, 59) = 4.90, p = .011\), with the Heuristic AA showing greater overlap than the SP-DRL AA (\(\text{contrast} = .023\), \(p = .008\)). The HS-DRL AA did not differ from either agent ($p\mathrm{s} > .346$).

For 5-TA trials, \(F(2, 59) = 31.81, p < .001\), the Heuristic AA produced higher overlap than HS-DRL (\(\text{contrast} = .029\), \(p < .001\)) and SP-DRL AAs (\(\text{contrast} = .060\), \(p < .001\)); HS-DRL exceeded the SP-DRL AA (\(\text{contrast} = .031\), \(p < .001\)).

\subsection{Artificial Agents on Binary Trace Overlap}
The mixed-design ANOVA on trajectory overlap revealed significant main effects of \textit{AA Type}, \(F(2, 59) = 36.33, p < .001, \eta_p^2 = .552\), and \textit{TA Num}, \(F(1.85, 109.35) = 4.59, p = .014, \eta_p^2 = .072\). The \textit{AA Type} \(\times\) \textit{TA Num} interaction was not significant, \(F(3.71, 109.35) = 2.47, p = .053, \eta_p^2 = .077\).

Post-hoc comparisons for \textit{AA Type} indicated that the Heuristic AA produced significantly higher trajectory overlap scores than both the HS-DRL (\(\text{contrast} = .069, p < .001\)) and SP-DRL AAs (\(\text{contrast} = .118, p < .001\)). The HS-DRL AA also exceeded the SP-DRL AA (\(\text{contrast} = .049, p = .003\)).

For \textit{TA Num}, trajectory overlap was lower in 3-TA trials than with 4 TAs (\(\text{contrast} = -.026, p = .029\)). 5-TA trials did not differ from either 3-TA (\(p = .284\)) or 4-TA trials (\(p = .366\)).

\subsection{Participants on Target-Selection Decision Overlap}
The mixed-design ANOVA revealed no main effect of \textit{AA Coherder}, \(F(2, 59) = 1.76, p = .181, \eta_p^2 = .056\), but significant effects of \textit{TA Num}, \(F(1.83, 108.12) = 363.89, p < .001, \eta_p^2 = .860\), and the \textit{AA Coherder}~\(\times\)~\textit{TA Num} interaction, \(F(3.66, 108.12) = 3.29, p = .016, \eta_p^2 = .100\).

Simple effects showed no differences in decision overlap among participants for 3-TA, \(F(2, 59) = .39, p = .677\), or 4-TA trials, \(F(2, 29) = .426, p = .655\), but a significant effect of \textit{AA Coherder} for 5-TA trials, \(F(2, 59) = 3.69, p = .031\). Post-hoc tests showed higher scores for participants paired with the Heuristic AA than the HS-DRL AA (\(\text{contrast} = .052, p < .042\)), with no differences between SP-DRL and either Heuristic (\(p = 1.000\)) or HS-DRL (\(p = .112\)) conditions.

\subsection{Participants on Binary Trace Overlap}
The mixed-design ANOVA revealed no main effect of \textit{AA Coherder}, \(F(2, 59) = 1.00, p = .373, \eta_p^2 = .033\), nor an \textit{AA Coherder}~\(\times\)~\textit{TA Num} interaction, \(F(3.84, 113.16) = 1.94, p = .112, \eta_p^2 = .062\). However, a significant main effect of \textit{TA Num} was observed, \(F(1.92, 113.16) = 4.51, p = .014, \eta_p^2 = .071\).

Post-hoc comparisons showed lower trajectory overlap in 3- than 5-TA trials (\(\text{contrast} = -.034, p = .018\)). 4-TA trials did not differ from 3-TA (\(p = .530)\) or 5-TA trials (\(p = .251)\).

\begin{figure*}
    \centering
    \includegraphics[width=\linewidth]{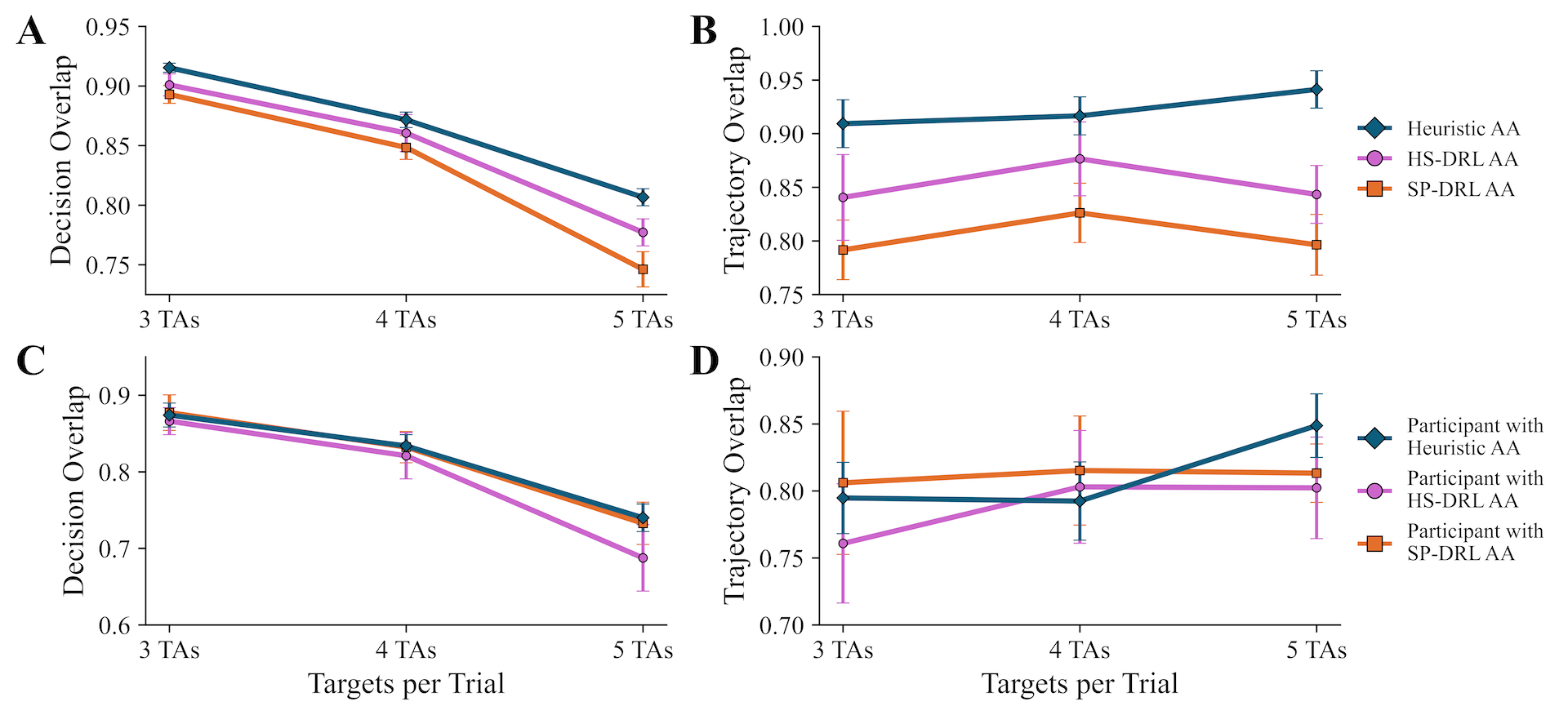}
    \caption{Average decision and trajectory overlap scores for AAs (Panels A--B) and participants (Panels C--D) across target-number conditions; error bars represent $95\%$ confidence intervals.}
    \label{fig:fig_chris}
\end{figure*}

\section{Discussion}
This study evaluated whether the SCA heuristic target-selection policy can effectively control DPMP-based AAs in human-AI dyads performing a first-person herding task. To benchmark its efficacy, the SCA heuristic was compared with SP-DRL and HS-DRL policies. Evaluation focused on how closely AA and participant behavior aligned with typical human dynamics observed in human-human dyads \cite{bin2024modellingb}.

Overall, the Heuristic AA produced the most human-aligned behavior, exhibiting higher decision and trajectory overlap scores than both DRL-based AAs. The HS-DRL AA showed greater human-alignment than the SP-DRL AA, though both remained less human-like than the heuristic agent. For participants, decision overlap differed across AA coherder conditions, whereas trajectory overlap remained consistent. These findings suggest that human-derived heuristic policies can promote human-compatible behavior in AAs, while human partners adjust their target-selection decisions across AI policies without substantial changes to their movement trajectories.

\subsection{Interpreting Behavior in Human-AI Dyads}
Since all AAs used the same DPMP model, differences in behavioral alignment reflect target-selection policies. The advantage of the Heuristic AA was most pronounced in trials with five targets (Fig.~\ref{fig:fig_chris}, Panels A--B), where demands were highest. These findings support human-derived heuristics for capturing goal-directed behavior in multiagent tasks \cite{nalepka2017herd}. 

This outcome is consistent with prior work showing that the SCA policy captures the dominant target-selection strategy used by human herders in the first-person task \cite{bin2024modelling}. Because the SCA policy was derived from observed human behavior, it naturally aligns with the perceptual and decision constraints underlying human action. When combined with the DPMP model, the heuristic policy enables agents to adapt continuously to the positions of targets and partners, reflecting the behavioral dynamics framework proposed by Fajen and Warren \cite{fajen2003behavioral}. This coupling between human-derived decision rules and dynamical movement control likely contributed to the strong human-behavioral alignment observed for the Heuristic AA.

In contrast, both DRL agents exhibited lower alignment with typical human behavior. Reinforcement learning policies optimize task rewards rather than human similarity, often producing strategies that diverge from typical human decisions \cite{shek2019takeaways}. The HS-DRL AA showed greater alignment with human behavior than the SP-DRL AA, consistent with prior findings that incorporating human behavioral models during training can guide policies towards more human-compatible strategies\cite{carroll2020the, nalepka2021interaction}. However, the HS-DRL AA did not outperform the Heuristic AA, suggesting that while human-sensitive training can improve DRL alignment with human behavior, simple human-derived heuristics may remain a highly effective approach for human-AI teaming in structured coordination tasks.

Participant decision overlap varied across AA conditions, whereas trajectory overlap remained stable, suggesting that participants adapted their target-selection strategies while maintaining consistent movement patterns. Such flexibility is consistent with theories of joint action, which emphasize that individuals adjust their decisions in response to their coactor's behavior and intentions to maintain coordination \cite{sebanz2009prediction}. The similar decision overlap observed in the Heuristic and SP-DRL participant conditions may reflect compensatory behavior, where participants adjusted their choices to accommodate the SP-DRL AA's variable yet less human-like decisions \cite{carroll2020the}. Lower decision overlap in the HS-DRL participant condition, however, may reflect reduced engagement or strategic diference, potentially arising when participants perceive the agent as highly capable and defer TA selection. This interpretation remains speculative and warrants direct investigation in future work  \cite{harbarth2024overtrusting}. 

\subsection{Implications for Human--AI Teaming}
The present findings support combining rule-based heuristic decision policies with DPMP-based movement models to design AA teammates that exhibit human-compatible behavior in dynamic human-AI tasks. Such architectures may be particularly useful for synthetic teammates in first-person training environments and collaborative navigation tasks where predictable and interpretable behavior is important \cite{nalepka2019human, nalepka2022assessing}. 

Comparable participant outcomes with DRL-based agents to the Heuristic AA condition suggest that reinforcement learning remains viable for human--AI teaming. Deriving heuristic policies often requires extensive analysis of human strategies, whereas DRL methods can learn policies autonomously through simulation \cite{rigoli2022comparison}. In complex environments where human decision rules are difficult to specify, DRL offers a scalable alternative.

\section{Conclusion and Future Work}
The present study examined heuristic, HS-DRL, and SP-DRL target-selection policies for AAs collaborating with humans in a first-person herding task. Results showed that the Heuristic AA most closely reproduced human decisions and trajectories, supporting the use of human-derived heuristics for DPMP control. Participants also demonstrated robust adaptability across AA conditions, suggesting that DRL policies can still support effective human--AI teaming. Overall, the findings underscore the value of combining human-inspired heuristics with dynamical systems models to develop AA teammates capable of human-compatible behavior. 

One limitation of this study is the relatively low task complexity, with trials containing a maximum of five targets, which may have constrained variability in target-selection decisions. Future work is underway examining larger first-person herding environments with up to 12 targets and three herders to test the generality of the SCA heuristic under greater decision demands. In addition, all AAs used the same DPMP movement model, which may have reduced observable differences in participant coordination across policies; future work could compare the DPMP model with DRL-parameterized movement controllers. Finally, supervised machine learning may provide a promising direction for modeling human decision policies, potentially leveraging synthetic pre-training using human-like simulation data generated by Heuristic or HS-DRL AAs when large human datasets are unavailable. 

As human--AI collaboration become increasingly common, understanding how agent decision policies and movement dynamics shape coordination will be critical for designing effective AAs teammates within human-agent systems.

\section*{Acknowledgments}
The authors thank Hannah Sandison, Sala Ghanem and Drs Elliot Saltzman, Benoît G. Bardy and Simon Hosking for assistance and helpful comments on this and related work. This work was completed as part of Dr bin Kamruddin's PhD and Christopher Lam's honours theses. Christopher Lam led the data collection and analysis and was the primary author of the manuscript.

\section*{Data Availability}
Data can be accessed through \href{https://data.mendeley.com/datasets/y3d4b3z59j/1}{Mendeley Data} (DOI: 10.17632/y3d4b3z59j.1). Accompanying code and game programs can be accessed through \href{https://github.com/MInD-Laboratory/Comparison-of-Heuristic-and-DRL-Policies-in-Human-Agent-Teaming.git}{GitHub}.

%
\bibliographystyle{IEEEtran}
\bibliography{references.bib}

\newpage

\end{document}


\title{Supplementary Material A}

\author{Ayman bin Kamruddin, Christopher Lam, Gaurav Patil, Mirco Musolesi, Mario di Bernardo, Michael J. Richardson}

\maketitle

\section*{First-Person Herding Model}

\subsection*{Definition of Terms in Navigational Model}
To accurately capture the movement of human participants, the repulsive and attractive coupling terms in the model were defined to reflect requisite behaviors. 

The term $\psi_\mathcal{G}$, representing the attractive coupling between the HA heading direction and the angular position of the offset location $\mathbf{x}_c = \mathbf{x} + \mathbf{c}$, where $\mathbf{c} := c\frac{\mathbf{x}}{\lVert\mathbf{x}\rVert}$, is defined as:
\begin{equation}
    \psi_\mathcal{G}(\mathbf{x}, \mathbf{y}, \phi, \theta_\mathcal{G}) := - k_g(\phi - \theta_\mathcal{G})(e^{-c_1 d_c(\mathbf{x},\mathbf{y})} + c_2),
\end{equation}
where $\theta_\mathcal{G}$ is the angle of the offset location with respect to the vertical axis centered on the HA, $k_g$ reflects the strength of attraction, $c_1$ and $c_2$ are positive constants, and $d_c(\mathbf{x}, \mathbf{y}) = \lVert \mathbf{x} + \mathbf{c} - \mathbf{y} \rVert$ is the distance between the offset location and the HA.

The term $\psi_\mathcal{T}$, which repels the HA from the TA's current location, is defined as:
\begin{equation}
    \psi_\mathcal{T}(\mathbf{x}, \mathbf{y}, \phi, \theta_\mathcal{T}) := \zeta (\mathbf{x}, \mathbf{y}) k_o (\phi - \theta_\mathcal{T}) e^{-c_3\lvert \phi - \theta_\mathcal{T} \rvert}e^{-c_4\lVert\mathbf{x} - \mathbf{y}\rVert},
\end{equation}
where $\theta_\mathcal{T}$ is the angle between the TA and vertical axis centered on the HA, $k_o$ represents the strength of repulsion, $c_3$ and $c_4$ are positive constants, and $\zeta (\mathbf{x}, \mathbf{y})$ is a sigmoidal function:
\begin{equation}
    \zeta (\mathbf{x}, \mathbf{y}) = 1 - \sigma\Bigg(\epsilon - \cos^{-1}\Big(\frac{\mathbf{x}\cdot\mathbf{y}}{\lVert\mathbf{x}\rVert\lVert\mathbf{y}\rVert}\Big)\Bigg),
\end{equation}
where $\sigma(z) = \frac{1}{1+e^{-z}}$. The parameter $\epsilon$ was set to $5^\circ$ based on trajectory data, indicating the angle when the HA first influences the TA.

The term $\psi_\mathcal{C}$, which repels the HA away from the containment zone $\mathcal{C}$, is defined as:
\begin{equation}
    \psi_\mathcal{C}(\mathbf{x}_\mathcal{C}, \mathbf{y}, \phi, \theta_{\mathcal{C}}) := k_o (\phi - \theta_\mathcal{C}) e^{-c_5\lvert \phi - \theta_\mathcal{C} \rvert}e^{-c_6\lVert\mathbf{x}_\mathcal{C}-\mathbf{y}\rVert},
\end{equation}
where $\theta_{\mathcal{C}}$ is the angle between the center of the containment zone and the vertical axis centered at the HA, and $c_5$ and $c_6$ are positive constants.

\begin{figure}[ht]
    \centering
    \includegraphics[width=0.5\linewidth]{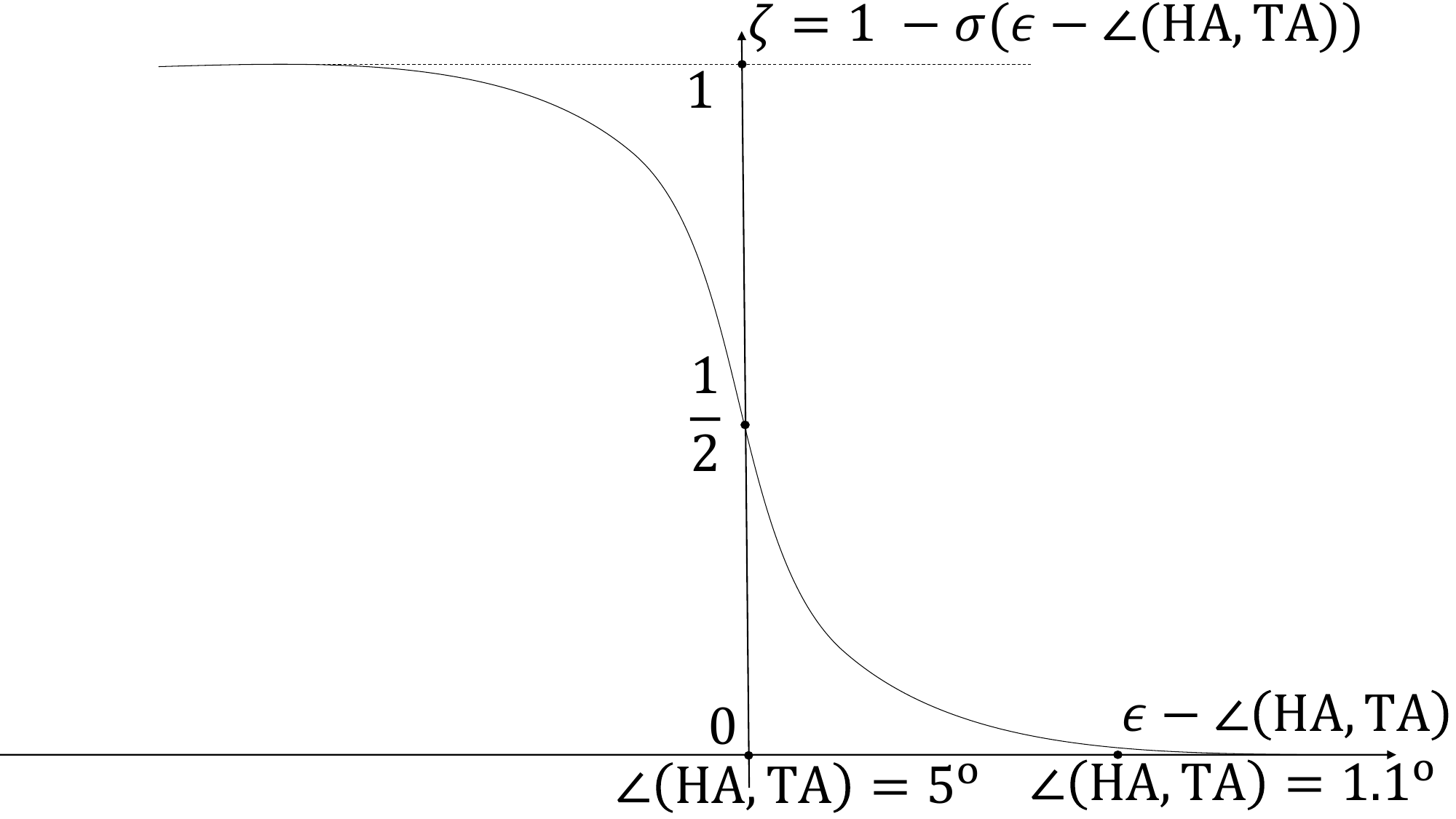}
    \caption{Justification of choice of $\epsilon = 5^\circ$. $\zeta = 0$ when $\angle(\mathrm{HA,TA}) = 5^\circ$, ensuring the repulsive term decays as the HA approaches the TA.}
    \label{fig:zeta_explanation}
\end{figure}


\title{Supplementary Material B}

\author{Ayman bin Kamruddin, Christopher Lam, Gaurav Patil, Mirco Musolesi, Mario di Bernardo, Michael J. Richardson}

\maketitle


\begin{table*}[ht]
\flushleft
\caption{Rule logic of the successive collinear angle (SCA) target-selection policy adapted for multiple herders}
\label{tab:sca-heuristic-policy}
\centering
\begin{tabular}{p{1cm} p{3cm} p{11cm}}
\hline
\textbf{\footnotesize Step} & 
\textbf{\footnotesize Condition} & 
\textbf{\footnotesize Action} \\
\hline
\footnotesize 1 & \footnotesize HA current target = null (otherwise skip to Step 2) &
\footnotesize Select the closest TA to the HA based on the angular distance of TAs from the HA.
\begin{list}{--}{\setlength{\leftmargin}{1em}}
    \item \footnotesize If the selected TA is closest to other-HA, skip this TA and select the next closest TA. 
    \item \footnotesize If all TAs are closer to other-HA, select TA furthest from the other-HA.
    \item \footnotesize Assign (return) the selected TA as HA's current target.
\end{list} 
\\
\footnotesize 2 & \footnotesize Check for near-collinear TAs & 
\footnotesize Identify if other TAs are within 20° of the currently selected TA (with respect to CZ). 
\begin{list}{--}{\setlength{\leftmargin}{1em}}
    \item \footnotesize If no collinear TAs, skip to Step 3, otherwise...
    \item \footnotesize Order the collinear TAs based on their Euclidean distance to CZ. 
    \item \footnotesize Assign (return) furthest TA from CZ as HA's current target (resulting in target switching when corralling collinear TAs).
\end{list} 
\\
\footnotesize 3 & \footnotesize Check for TA containment & 
\footnotesize If HA's currently selected TA is in the containment zone and stationary, current target = null, wait 1.0 seconds.

\\
\footnotesize 4 & \footnotesize Pause and Return & 
\footnotesize Wait 0.25 seconds then return to Step 1.
\\
\hline
\end{tabular}
\flushleft
\footnotesize{Note: HA = Herding Agent; TA = Target Agent; CZ = Containment Zone.}
\end{table*}


\title{Supplementary Material C - Check}

\author{Ayman bin Kamruddin, Christopher Lam, Gaurav Patil, Mirco Musolesi, Mario di Bernardo, Michael J. Richardson}

\maketitle

\section{Figures}

\noindent This supplementary material contains figures of the individual player trajectories from the human-AA teams (in red), against the background of the binary heatmap generated from the human-human dyad.

\section*{Heuristic Team}

    \subsection*{Trial \#7}
        \begin{figure}[H]
            \begin{subfigure}{0.49\textwidth}
                \includegraphics[width=\linewidth]{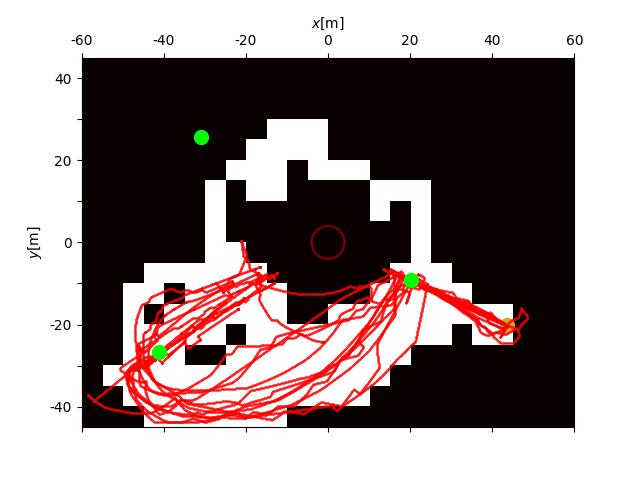}
                \caption{Player 1 - Human}
            \end{subfigure}
            \hfill
            \begin{subfigure}{0.49\textwidth}
                \includegraphics[width=\linewidth]{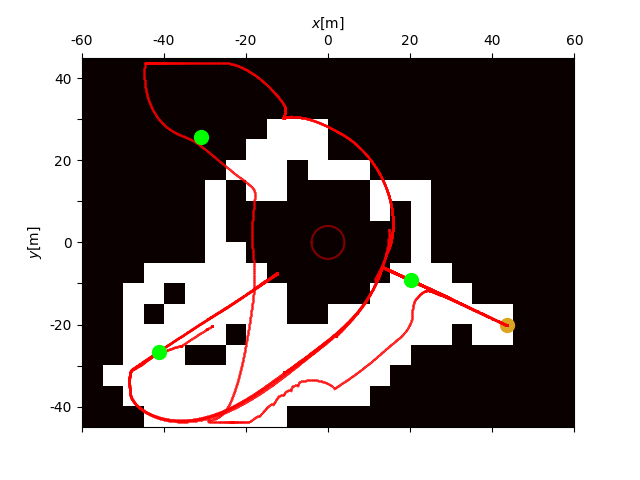}
                \caption{Player 1 - AA}
            \end{subfigure}

            \medskip
            \begin{subfigure}{0.49\textwidth}
                \includegraphics[width=\linewidth]{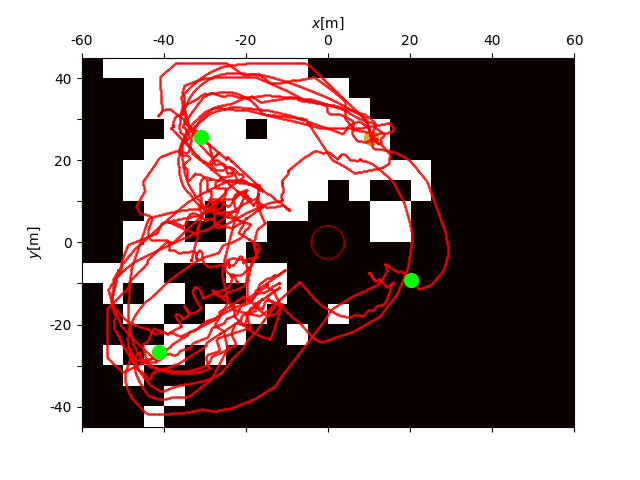}
                \caption{Player 2 - Human}
            \end{subfigure}
            \hfill
            \begin{subfigure}{0.49\textwidth}
                \includegraphics[width=\linewidth]{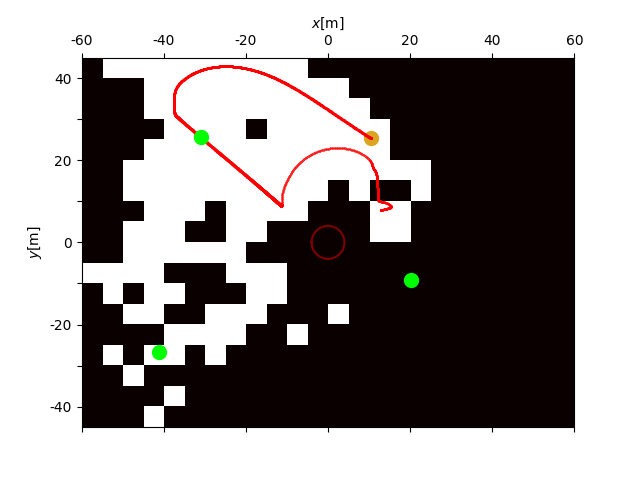}
                \caption{Player 2 - AA}
            \end{subfigure}
        \end{figure}
        \newpage

    \subsection*{Trial \#8}
        \begin{figure}[H]
            \begin{subfigure}{0.49\textwidth}
                \includegraphics[width=\linewidth]{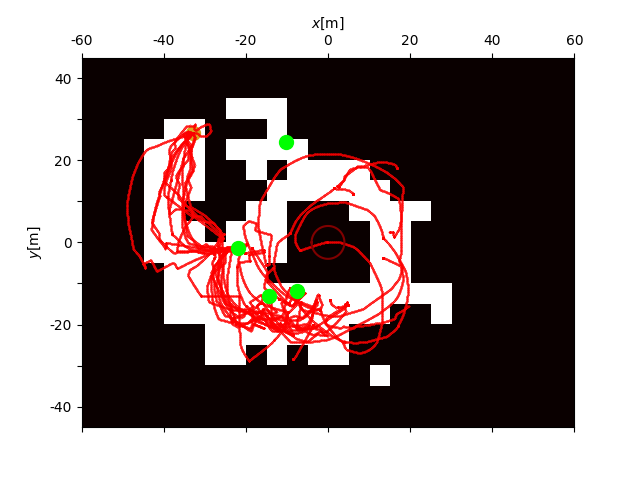}
                \caption{Player 1 - Human}
            \end{subfigure}
            \hfill
            \begin{subfigure}{0.49\textwidth}
                \includegraphics[width=\linewidth]{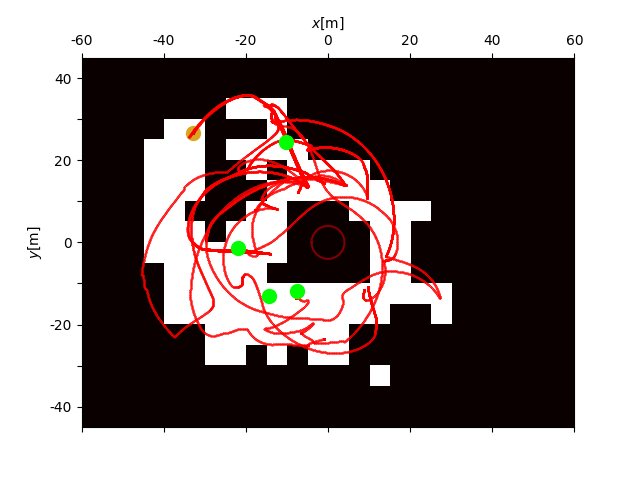}
                \caption{Player 1 - AA}
            \end{subfigure}

            \medskip
            \begin{subfigure}{0.49\textwidth}
                \includegraphics[width=\linewidth]{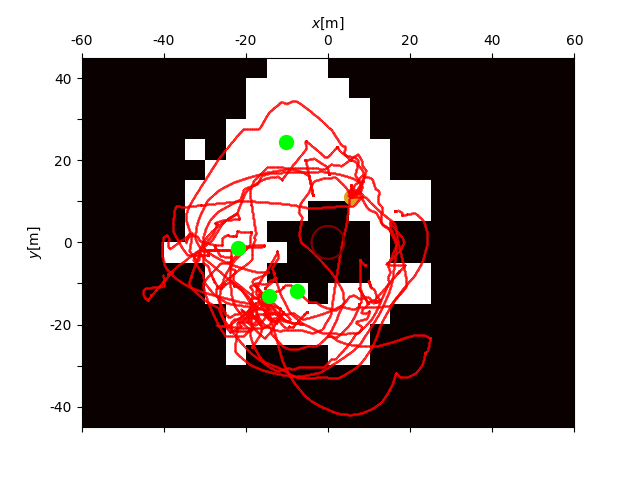}
                \caption{Player 2 - Human}
            \end{subfigure}
            \hfill
            \begin{subfigure}{0.49\textwidth}
                \includegraphics[width=\linewidth]{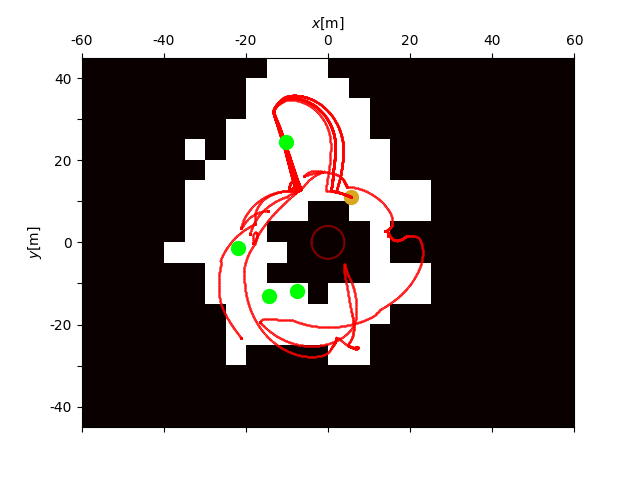}
                \caption{Player 2 - AA}
            \end{subfigure}
        \end{figure}
        \newpage

    \subsection*{Trial \#9}
        \begin{figure}[H]
            \begin{subfigure}{0.49\textwidth}
                \includegraphics[width=\linewidth]{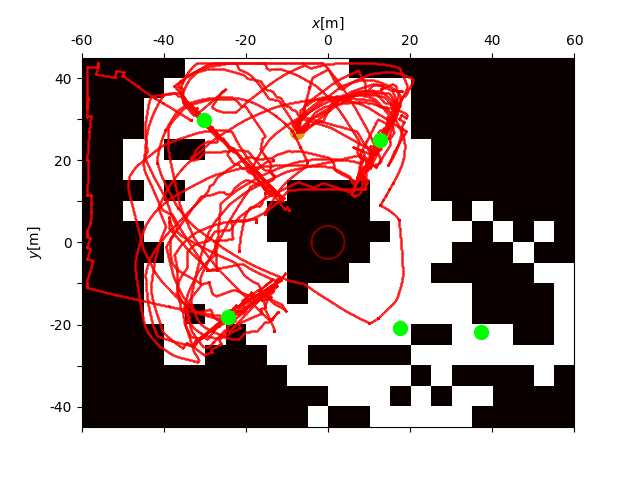}
                \caption{Player 1 - Human}
            \end{subfigure}
            \hfill
            \begin{subfigure}{0.49\textwidth}
                \includegraphics[width=\linewidth]{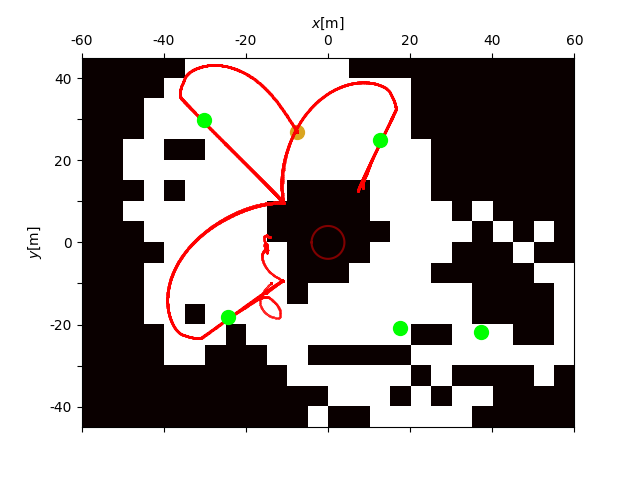}
                \caption{Player 1 - AA}
            \end{subfigure}

            \medskip
            \begin{subfigure}{0.49\textwidth}
                \includegraphics[width=\linewidth]{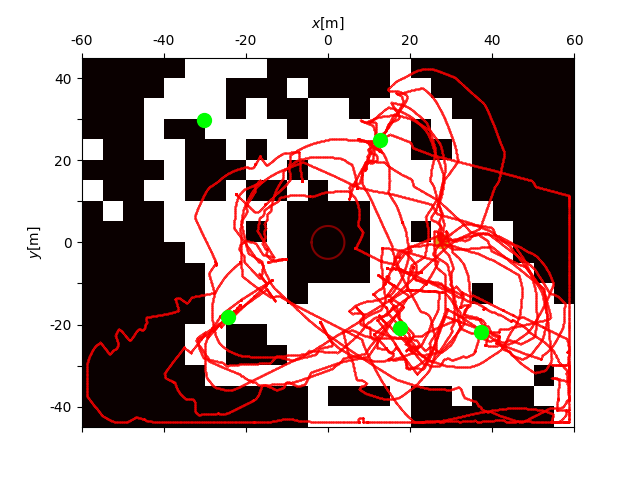}
                \caption{Player 2 - Human}
            \end{subfigure}
            \hfill
            \begin{subfigure}{0.49\textwidth}
                \includegraphics[width=\linewidth]{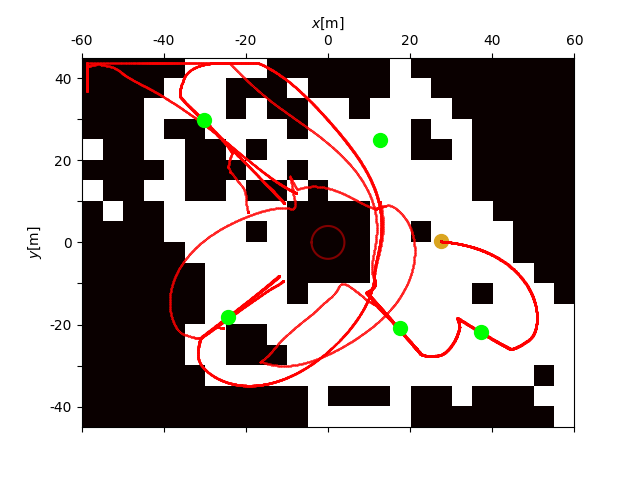}
                \caption{Player 2 - AA}
            \end{subfigure}
        \end{figure}
        \newpage

    \subsection*{Trial \#10}
        \begin{figure}[H]
            \begin{subfigure}{0.49\textwidth}
                \includegraphics[width=\linewidth]{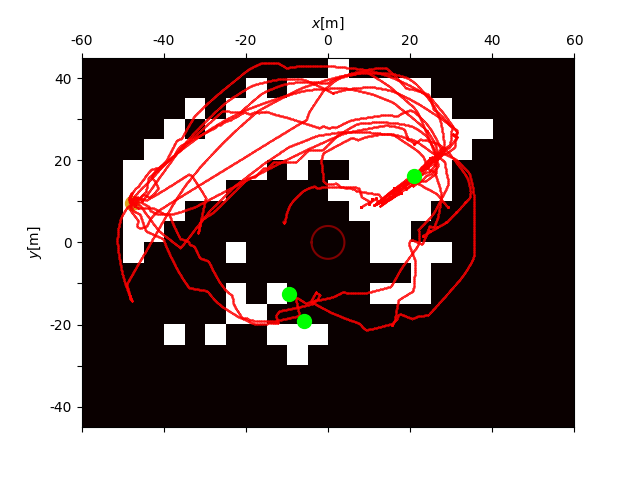}
                \caption{Player 1 - Human}
            \end{subfigure}
            \hfill
            \begin{subfigure}{0.49\textwidth}
                \includegraphics[width=\linewidth]{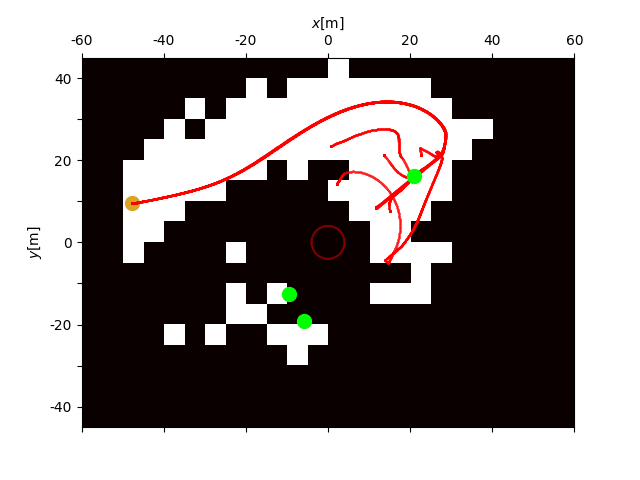}
                \caption{Player 1 - AA}
            \end{subfigure}

            \medskip
            \begin{subfigure}{0.49\textwidth}
                \includegraphics[width=\linewidth]{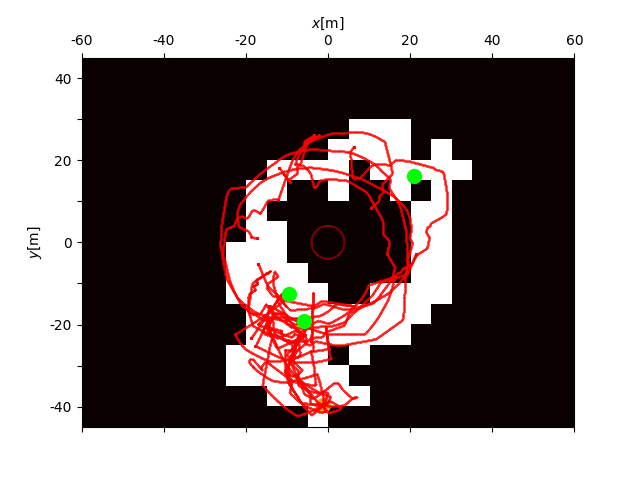}
                \caption{Player 2 - Human}
            \end{subfigure}
            \hfill
            \begin{subfigure}{0.49\textwidth}
                \includegraphics[width=\linewidth]{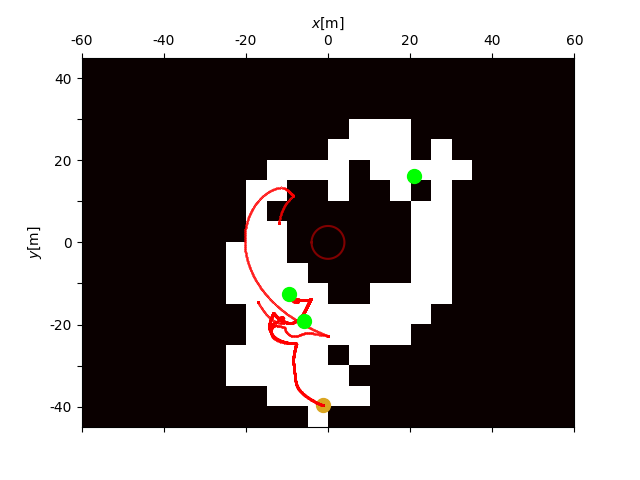}
                \caption{Player 2 - AA}
            \end{subfigure}
        \end{figure}
        \newpage

    \subsection*{Trial \#11}
        \begin{figure}[H]
            \begin{subfigure}{0.49\textwidth}
                \includegraphics[width=\linewidth]{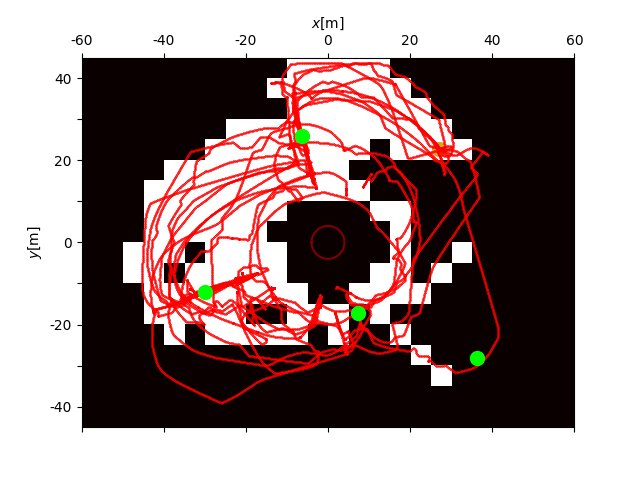}
                \caption{Player 1 - Human}
            \end{subfigure}
            \hfill
            \begin{subfigure}{0.49\textwidth}
                \includegraphics[width=\linewidth]{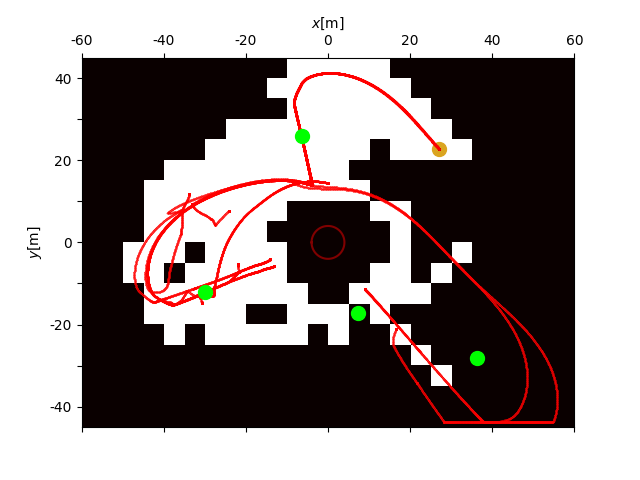}
                \caption{Player 1 - AA}
            \end{subfigure}

            \medskip
            \begin{subfigure}{0.49\textwidth}
                \includegraphics[width=\linewidth]{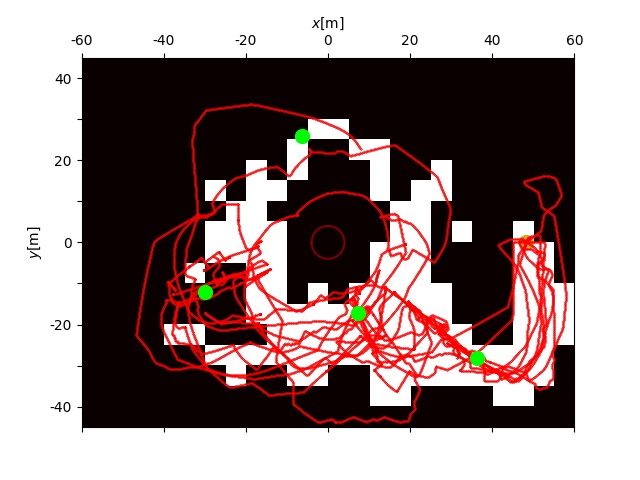}
                \caption{Player 2 - Human}
            \end{subfigure}
            \hfill
            \begin{subfigure}{0.49\textwidth}
                \includegraphics[width=\linewidth]{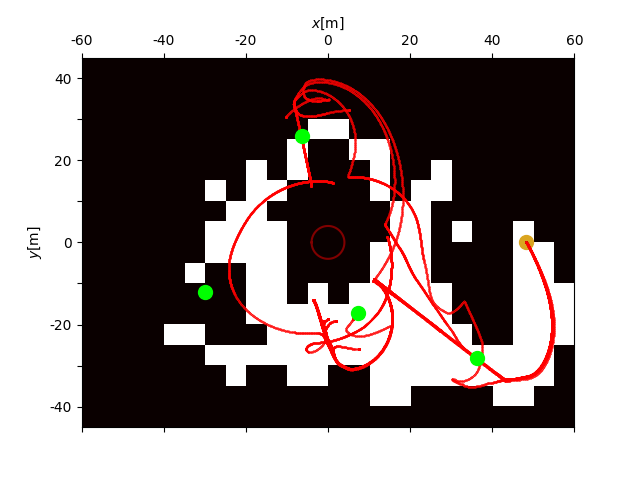}
                \caption{Player 2 - AA}
            \end{subfigure}
        \end{figure}
        \newpage

    \subsection*{Trial \#12}
        \begin{figure}[H]
            \begin{subfigure}{0.49\textwidth}
                \includegraphics[width=\linewidth]{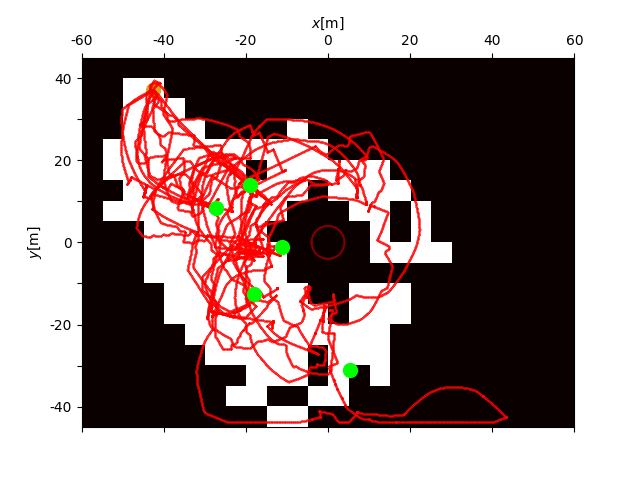}
                \caption{Player 1 - Human}
            \end{subfigure}
            \hfill
            \begin{subfigure}{0.49\textwidth}
                \includegraphics[width=\linewidth]{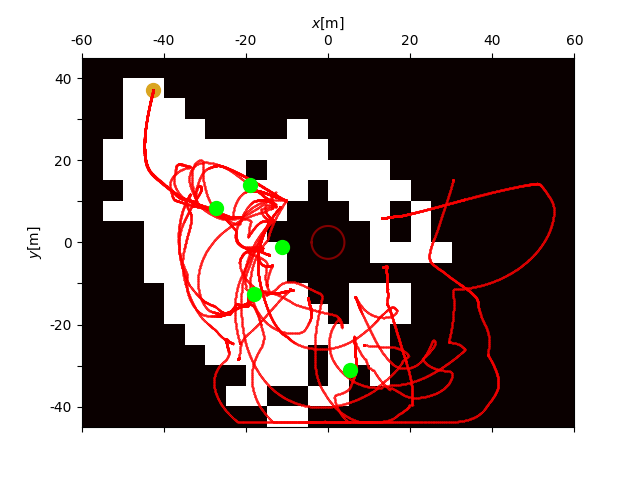}
                \caption{Player 1 - AA}
            \end{subfigure}

            \medskip
            \begin{subfigure}{0.49\textwidth}
                \includegraphics[width=\linewidth]{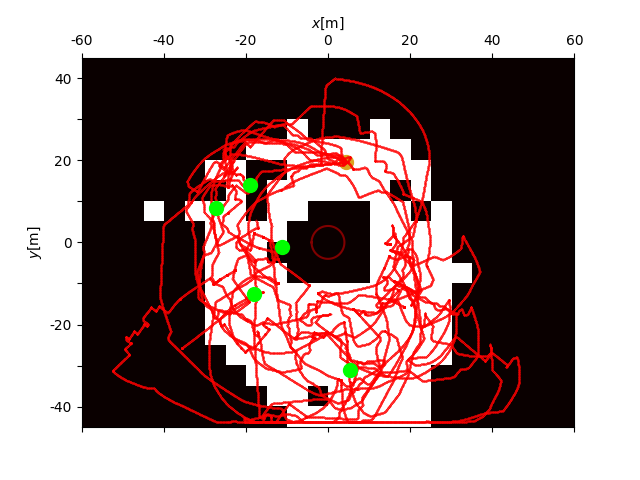}
                \caption{Player 2 - Human}
            \end{subfigure}
            \hfill
            \begin{subfigure}{0.49\textwidth}
                \includegraphics[width=\linewidth]{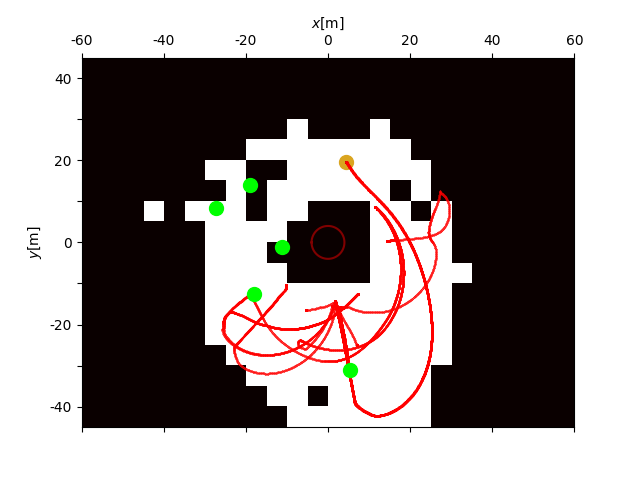}
                \caption{Player 2 - AA}
            \end{subfigure}
        \end{figure}
        \newpage

    \subsection*{Trial \#13}
        \begin{figure}[H]
            \begin{subfigure}{0.49\textwidth}
                \includegraphics[width=\linewidth]{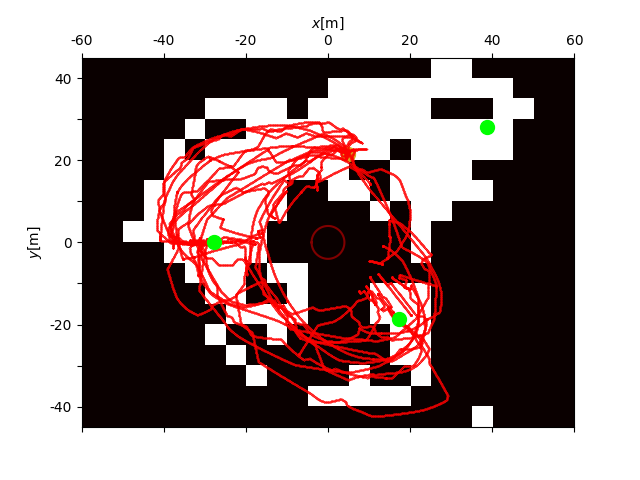}
                \caption{Player 1 - Human}
            \end{subfigure}
            \hfill
            \begin{subfigure}{0.49\textwidth}
                \includegraphics[width=\linewidth]{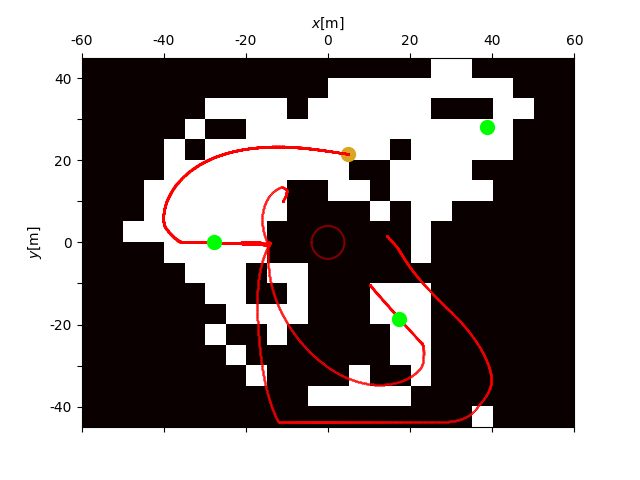}
                \caption{Player 1 - AA}
            \end{subfigure}

            \medskip
            \begin{subfigure}{0.49\textwidth}
                \includegraphics[width=\linewidth]{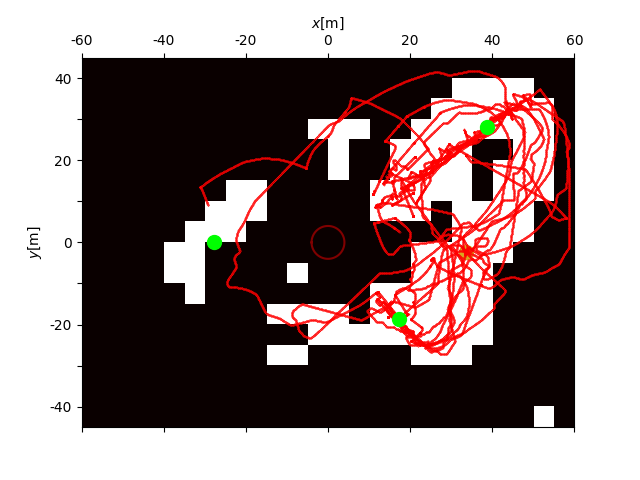}
                \caption{Player 2 - Human}
            \end{subfigure}
            \hfill
            \begin{subfigure}{0.49\textwidth}
                \includegraphics[width=\linewidth]{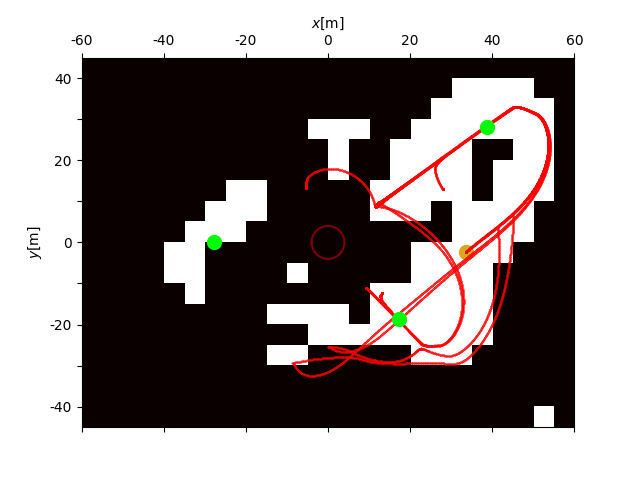}
                \caption{Player 2 - AA}
            \end{subfigure}
        \end{figure}
        \newpage

    \subsection*{Trial \#14}
        \begin{figure}[H]
            \begin{subfigure}{0.49\textwidth}
                \includegraphics[width=\linewidth]{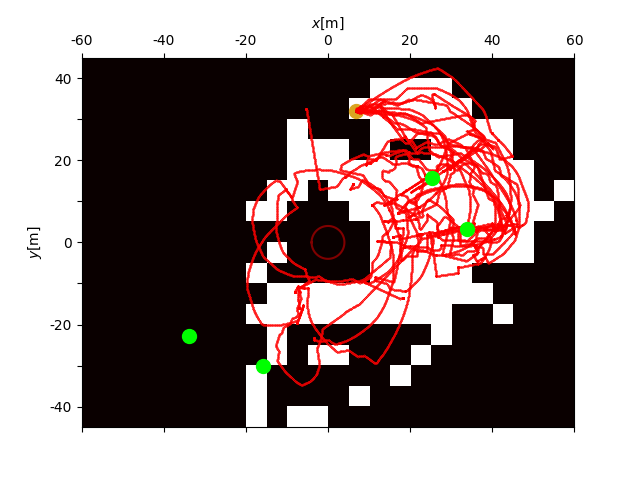}
                \caption{Player 1 - Human}
            \end{subfigure}
            \hfill
            \begin{subfigure}{0.49\textwidth}
                \includegraphics[width=\linewidth]{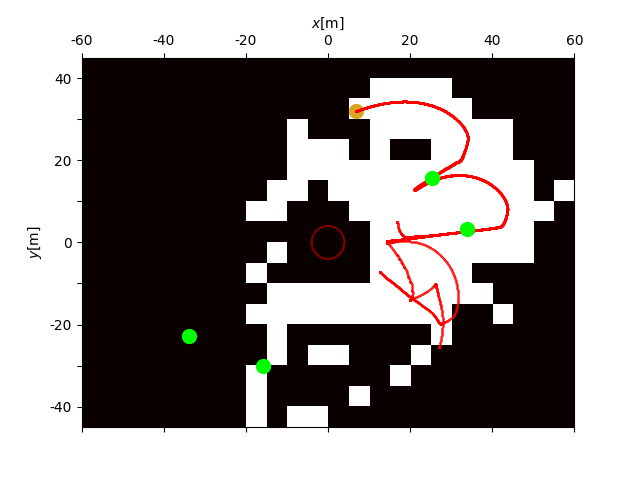}
                \caption{Player 1 - AA}
            \end{subfigure}

            \medskip
            \begin{subfigure}{0.49\textwidth}
                \includegraphics[width=\linewidth]{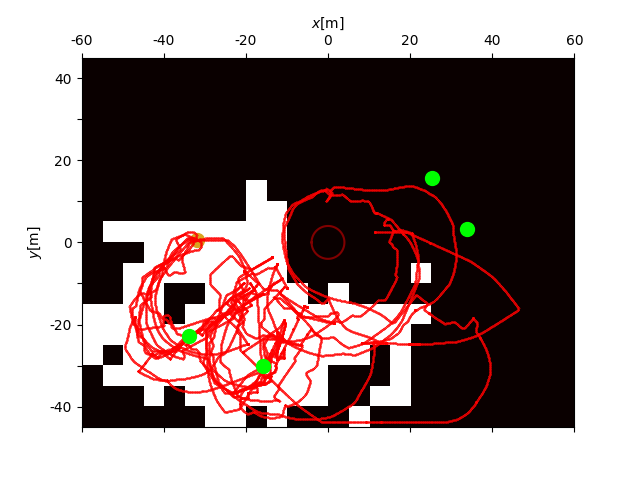}
                \caption{Player 2 - Human}
            \end{subfigure}
            \hfill
            \begin{subfigure}{0.49\textwidth}
                \includegraphics[width=\linewidth]{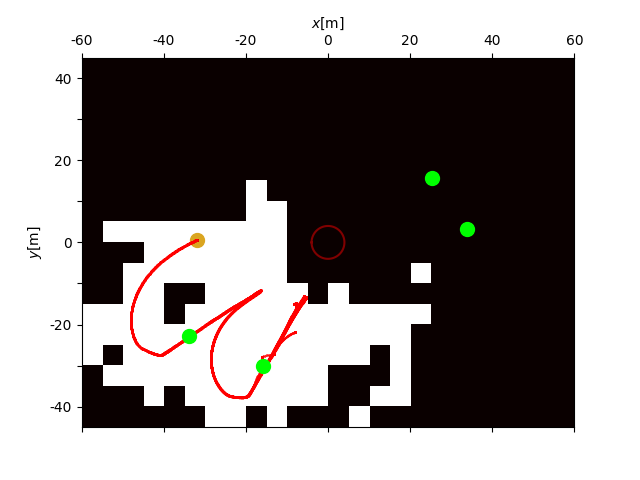}
                \caption{Player 2 - AA}
            \end{subfigure}
        \end{figure}
        \newpage

    \subsection*{Trial \#15}
        \begin{figure}[H]
            \begin{subfigure}{0.49\textwidth}
                \includegraphics[width=\linewidth]{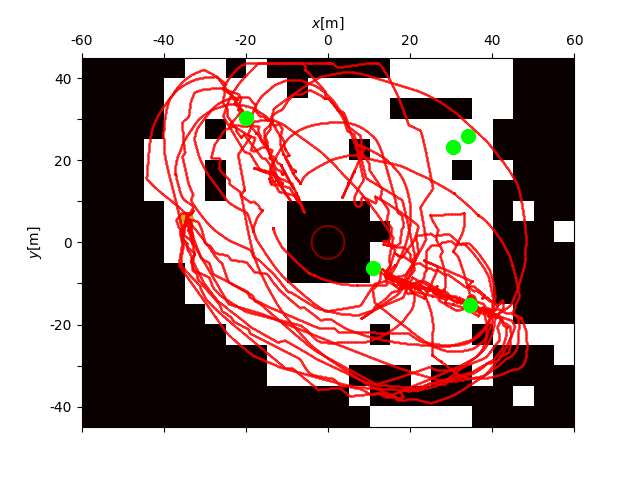}
                \caption{Player 1 - Human}
            \end{subfigure}
            \hfill
            \begin{subfigure}{0.49\textwidth}
                \includegraphics[width=\linewidth]{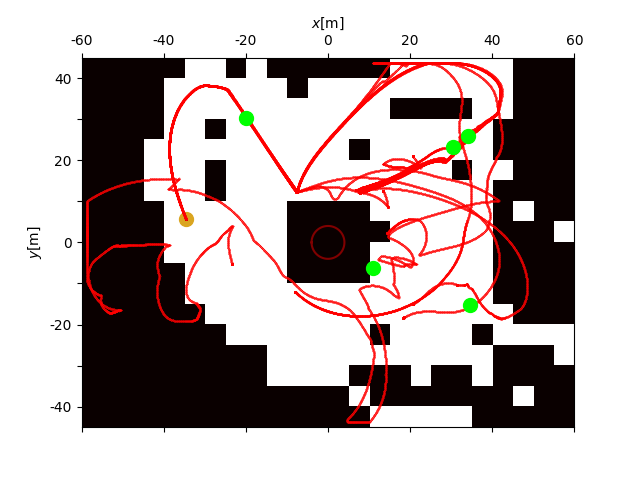}
                \caption{Player 1 - AA}
            \end{subfigure}

            \medskip
            \begin{subfigure}{0.49\textwidth}
                \includegraphics[width=\linewidth]{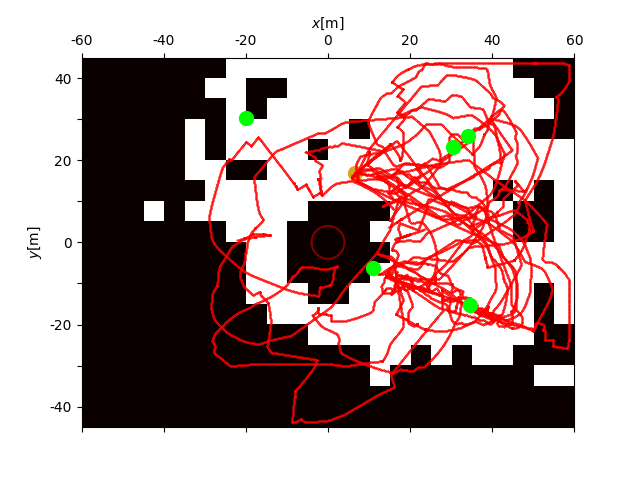}
                \caption{Player 2 - Human}
            \end{subfigure}
            \hfill
            \begin{subfigure}{0.49\textwidth}
                \includegraphics[width=\linewidth]{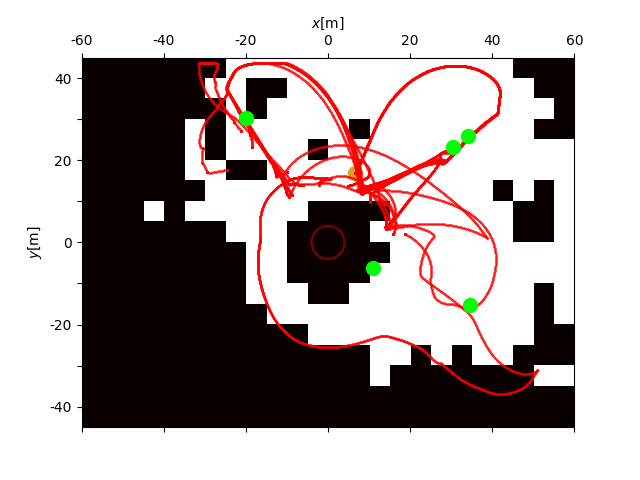}
                \caption{Player 2 - AA}
            \end{subfigure}
        \end{figure}
        \newpage

    \subsection*{Trial \#16}
        \begin{figure}[H]
            \begin{subfigure}{0.49\textwidth}
                \includegraphics[width=\linewidth]{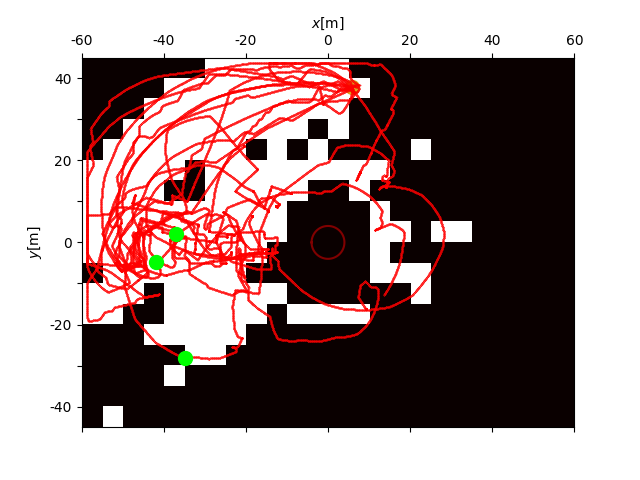}
                \caption{Player 1 - Human}
            \end{subfigure}
            \hfill
            \begin{subfigure}{0.49\textwidth}
                \includegraphics[width=\linewidth]{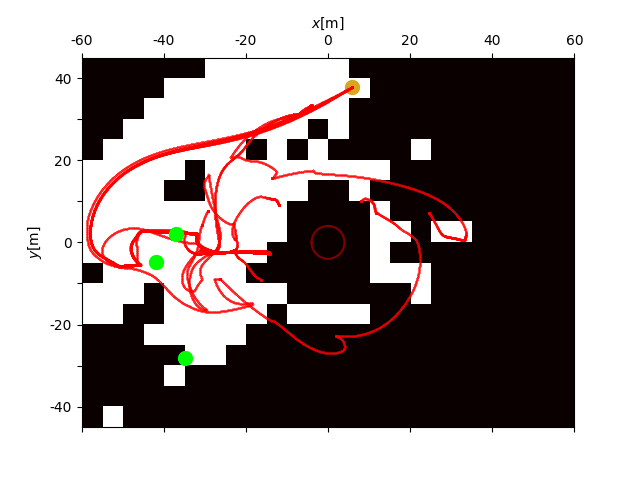}
                \caption{Player 1 - AA}
            \end{subfigure}

            \medskip
            \begin{subfigure}{0.49\textwidth}
                \includegraphics[width=\linewidth]{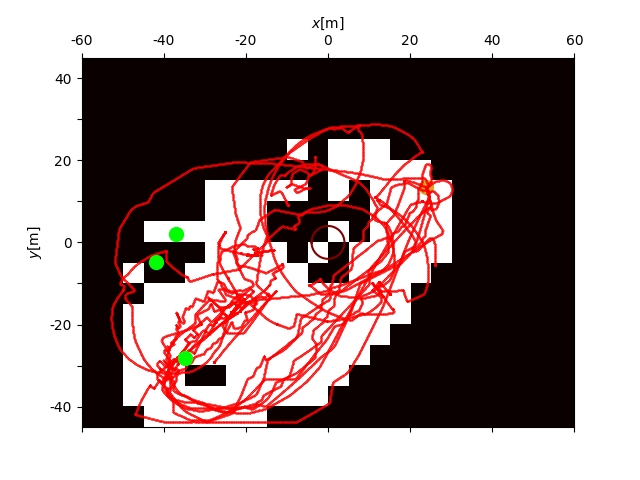}
                \caption{Player 2 - Human}
            \end{subfigure}
            \hfill
            \begin{subfigure}{0.49\textwidth}
                \includegraphics[width=\linewidth]{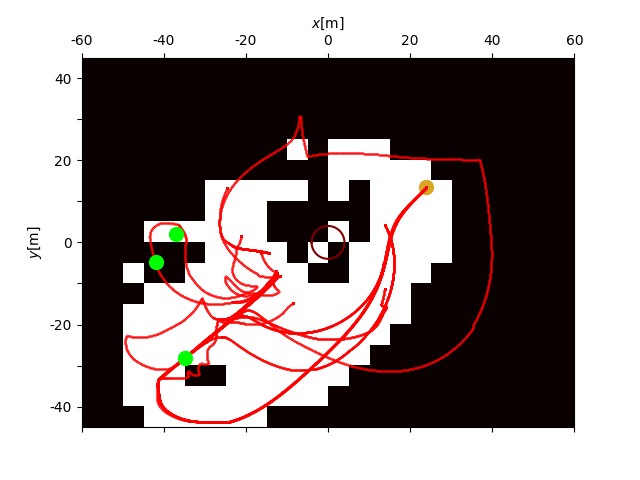}
                \caption{Player 2 - AA}
            \end{subfigure}
        \end{figure}
        \newpage

    \subsection*{Trial \#17}
        \begin{figure}[H]
            \begin{subfigure}{0.49\textwidth}
                \includegraphics[width=\linewidth]{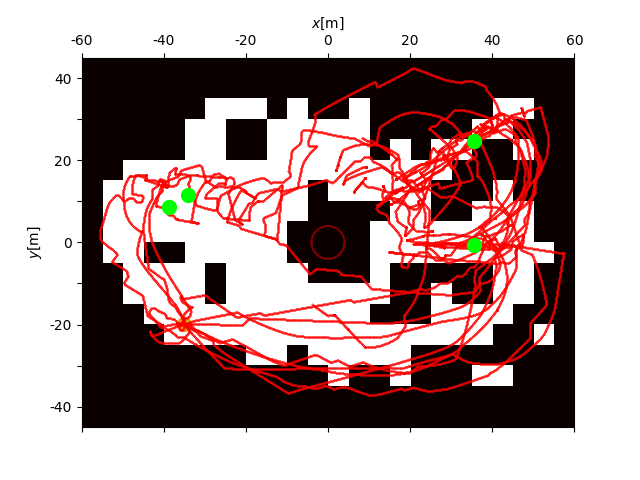}
                \caption{Player 1 - Human}
            \end{subfigure}
            \hfill
            \begin{subfigure}{0.49\textwidth}
                \includegraphics[width=\linewidth]{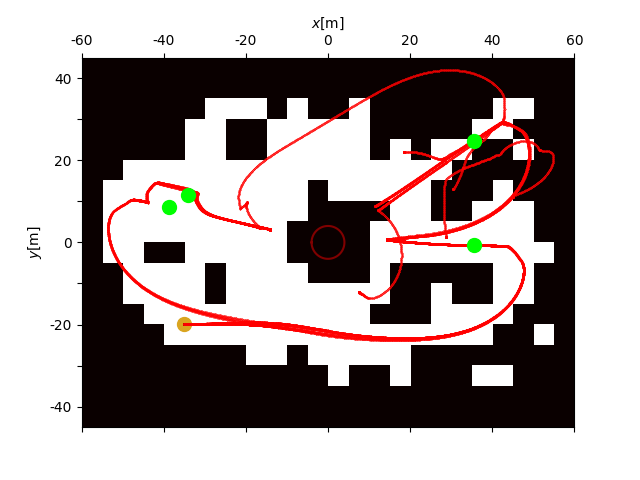}
                \caption{Player 1 - AA}
            \end{subfigure}

            \medskip
            \begin{subfigure}{0.49\textwidth}
                \includegraphics[width=\linewidth]{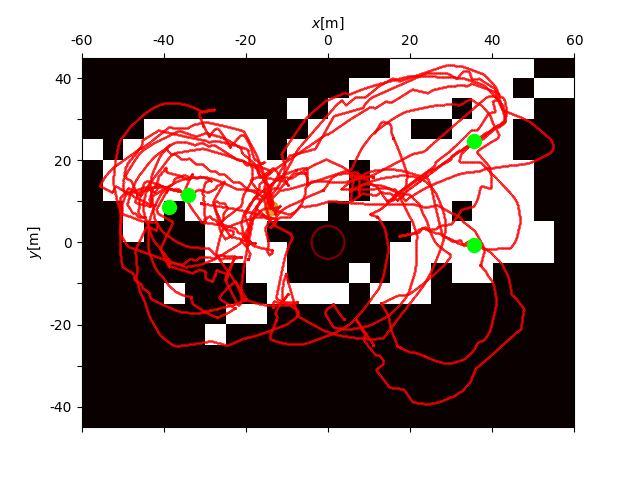}
                \caption{Player 2 - Human}
            \end{subfigure}
            \hfill
            \begin{subfigure}{0.49\textwidth}
                \includegraphics[width=\linewidth]{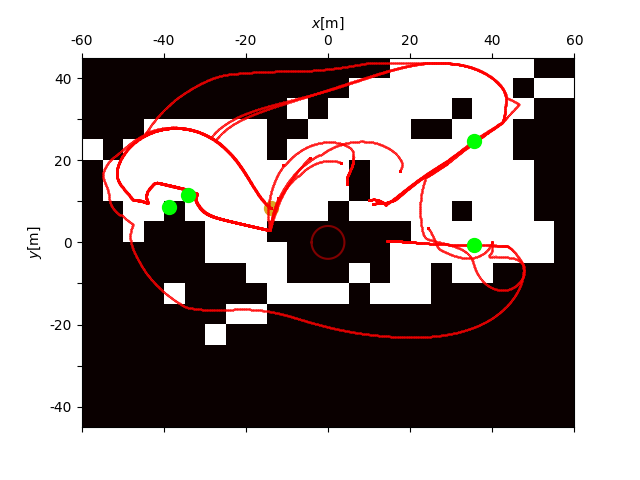}
                \caption{Player 2 - AA}
            \end{subfigure}
        \end{figure}
        \newpage

    \subsection*{Trial \#18}
        \begin{figure}[H]
            \begin{subfigure}{0.49\textwidth}
                \includegraphics[width=\linewidth]{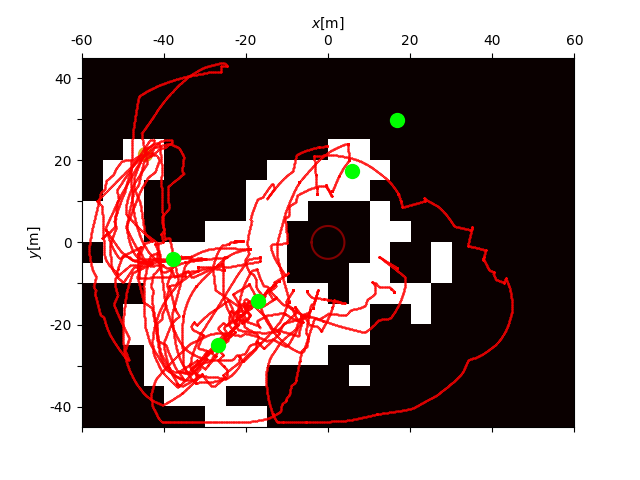}
                \caption{Player 1 - Human}
            \end{subfigure}
            \hfill
            \begin{subfigure}{0.49\textwidth}
                \includegraphics[width=\linewidth]{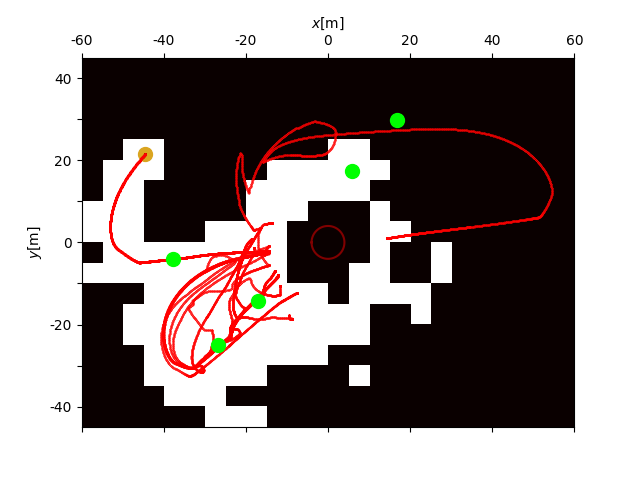}
                \caption{Player 1 - AA}
            \end{subfigure}

            \medskip
            \begin{subfigure}{0.49\textwidth}
                \includegraphics[width=\linewidth]{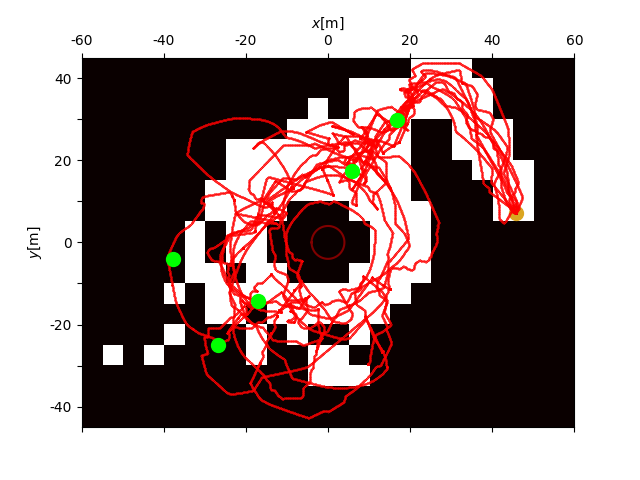}
                \caption{Player 2 - Human}
            \end{subfigure}
            \hfill
            \begin{subfigure}{0.49\textwidth}
                \includegraphics[width=\linewidth]{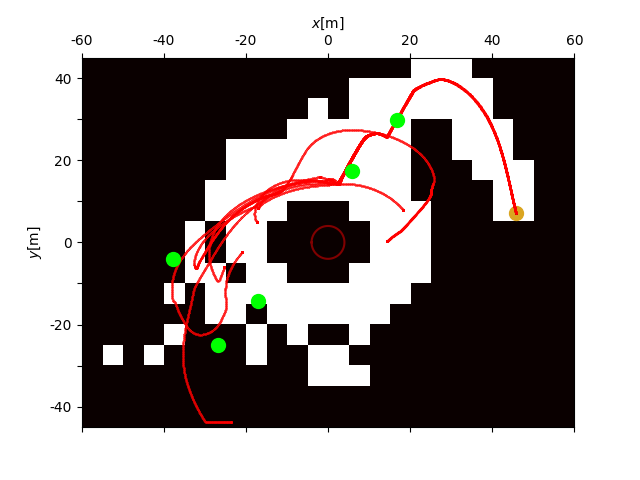}
                \caption{Player 2 - AA}
            \end{subfigure}
        \end{figure}
        \newpage

    \subsection*{Trial \#19}
        \begin{figure}[H]
            \begin{subfigure}{0.49\textwidth}
                \includegraphics[width=\linewidth]{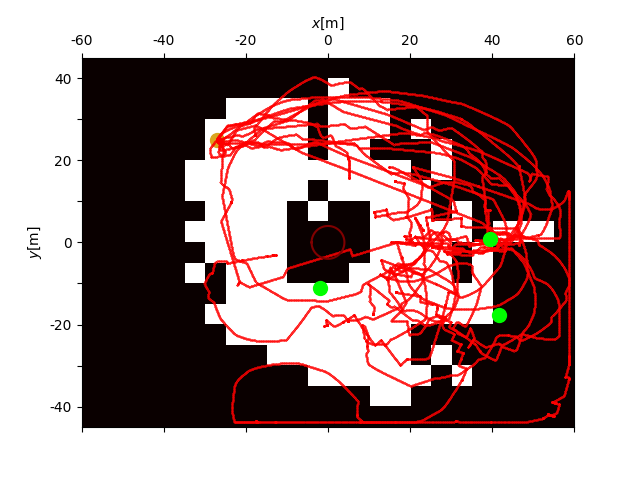}
                \caption{Player 1 - Human}
            \end{subfigure}
            \hfill
            \begin{subfigure}{0.49\textwidth}
                \includegraphics[width=\linewidth]{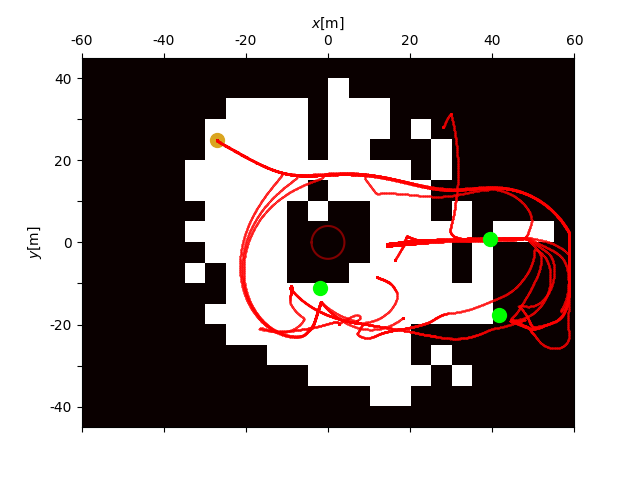}
                \caption{Player 1 - AA}
            \end{subfigure}

            \medskip
            \begin{subfigure}{0.49\textwidth}
                \includegraphics[width=\linewidth]{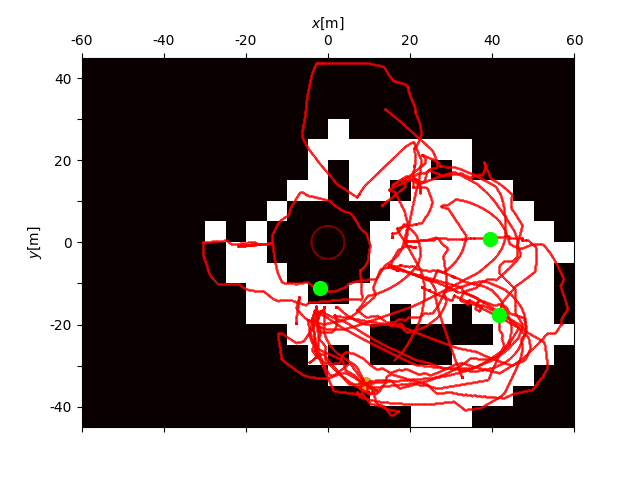}
                \caption{Player 2 - Human}
            \end{subfigure}
            \hfill
            \begin{subfigure}{0.49\textwidth}
                \includegraphics[width=\linewidth]{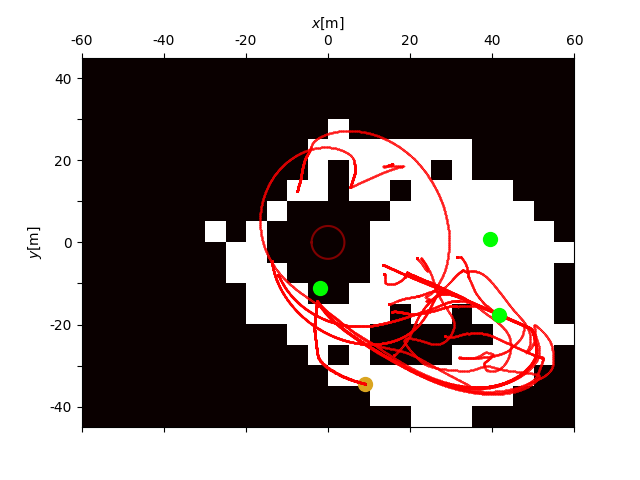}
                \caption{Player 2 - AA}
            \end{subfigure}
        \end{figure}
        \newpage

    \subsection*{Trial \#20}
        \begin{figure}[H]
            \begin{subfigure}{0.49\textwidth}
                \includegraphics[width=\linewidth]{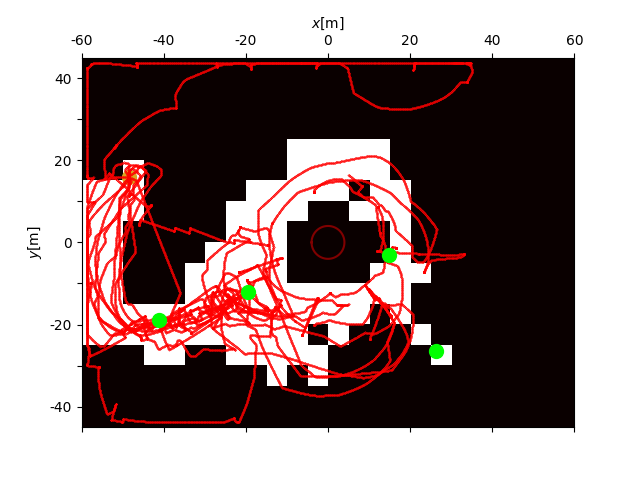}
                \caption{Player 1 - Human}
            \end{subfigure}
            \hfill
            \begin{subfigure}{0.49\textwidth}
                \includegraphics[width=\linewidth]{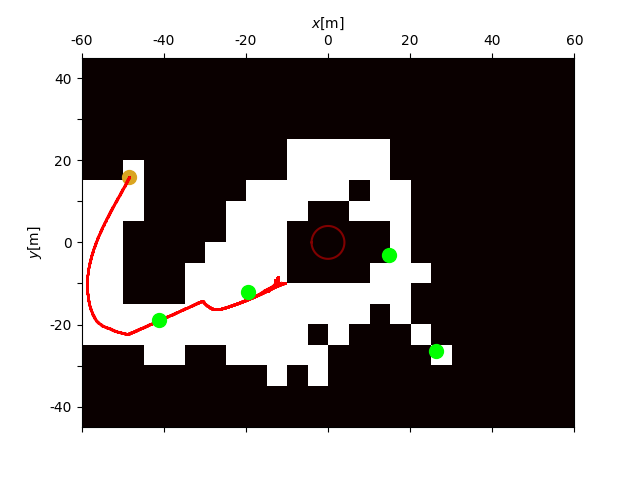}
                \caption{Player 1 - AA}
            \end{subfigure}

            \medskip
            \begin{subfigure}{0.49\textwidth}
                \includegraphics[width=\linewidth]{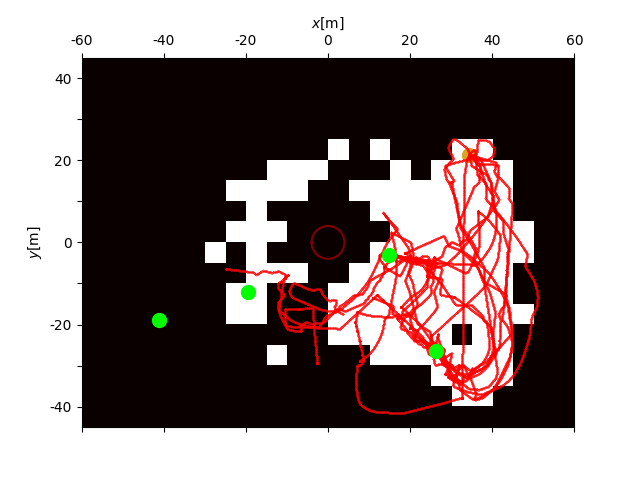}
                \caption{Player 2 - Human}
            \end{subfigure}
            \hfill
            \begin{subfigure}{0.49\textwidth}
                \includegraphics[width=\linewidth]{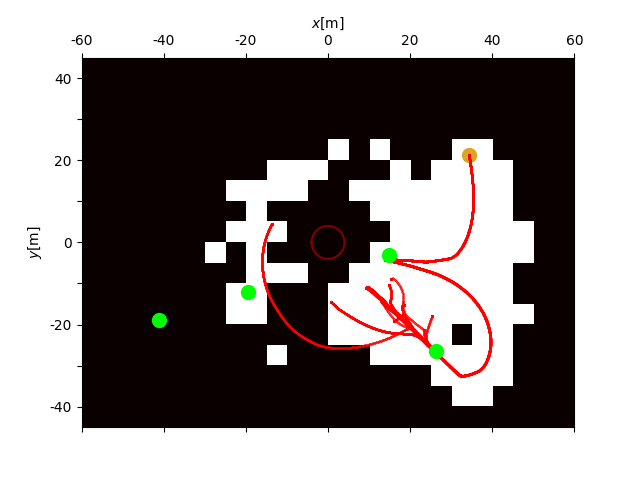}
                \caption{Player 2 - AA}
            \end{subfigure}
        \end{figure}
        \newpage

    \subsection*{Trial \#21}
        \begin{figure}[H]
            \begin{subfigure}{0.49\textwidth}
                \includegraphics[width=\linewidth]{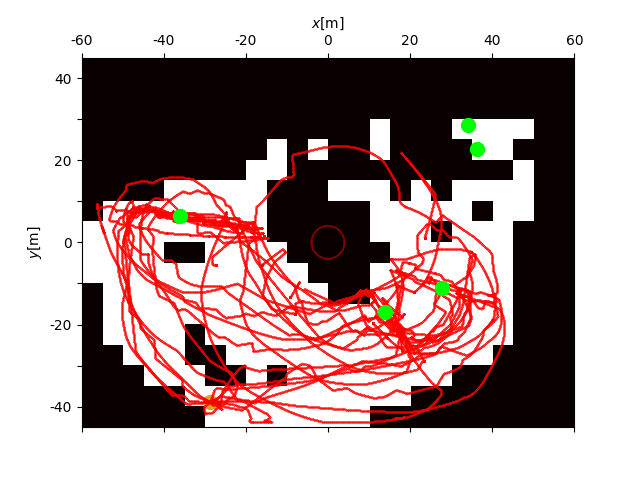}
                \caption{Player 1 - Human}
            \end{subfigure}
            \hfill
            \begin{subfigure}{0.49\textwidth}
                \includegraphics[width=\linewidth]{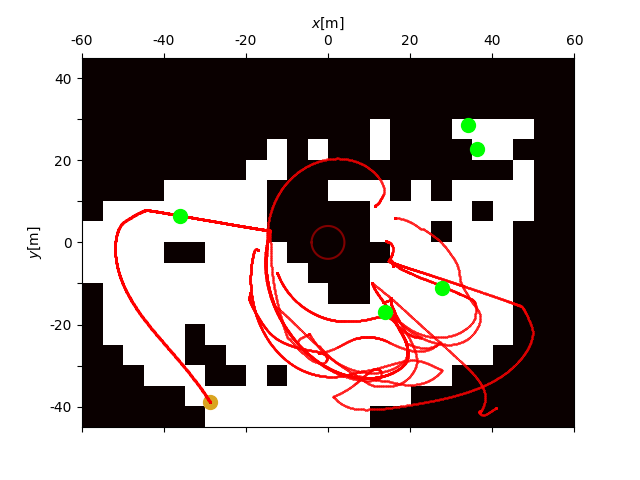}
                \caption{Player 1 - AA}
            \end{subfigure}

            \medskip
            \begin{subfigure}{0.49\textwidth}
                \includegraphics[width=\linewidth]{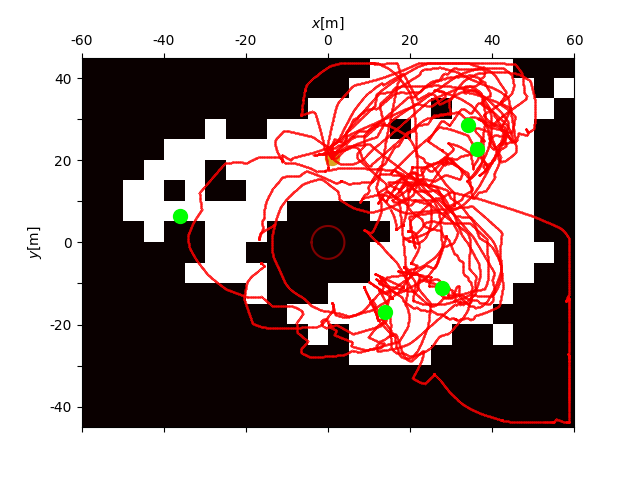}
                \caption{Player 2 - Human}
            \end{subfigure}
            \hfill
            \begin{subfigure}{0.49\textwidth}
                \includegraphics[width=\linewidth]{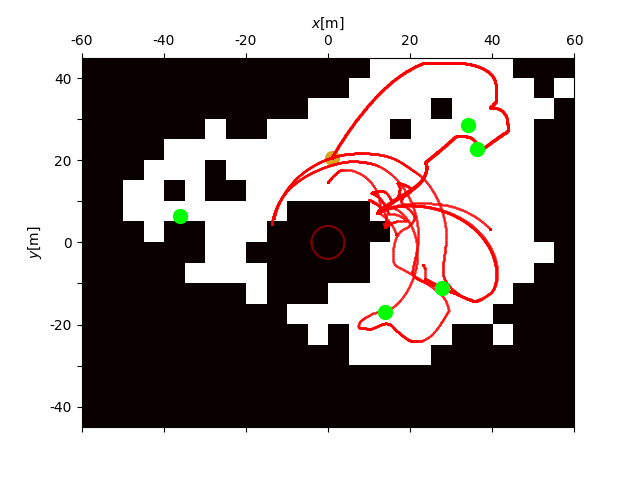}
                \caption{Player 2 - AA}
            \end{subfigure}
        \end{figure}
        \newpage

    \subsection*{Trial \#22}
        \begin{figure}[H]
            \begin{subfigure}{0.49\textwidth}
                \includegraphics[width=\linewidth]{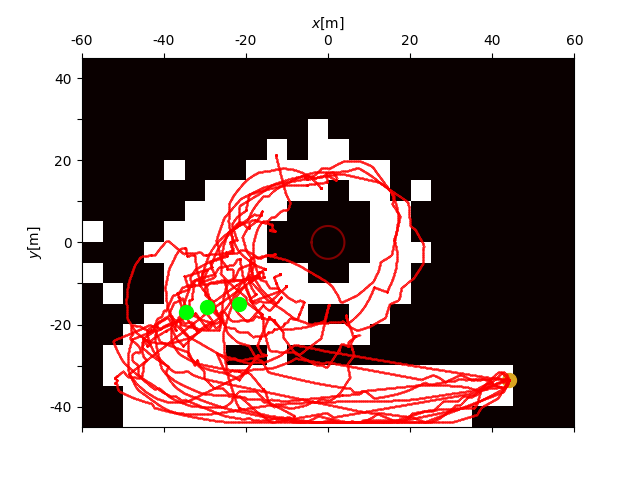}
                \caption{Player 1 - Human}
            \end{subfigure}
            \hfill
            \begin{subfigure}{0.49\textwidth}
                \includegraphics[width=\linewidth]{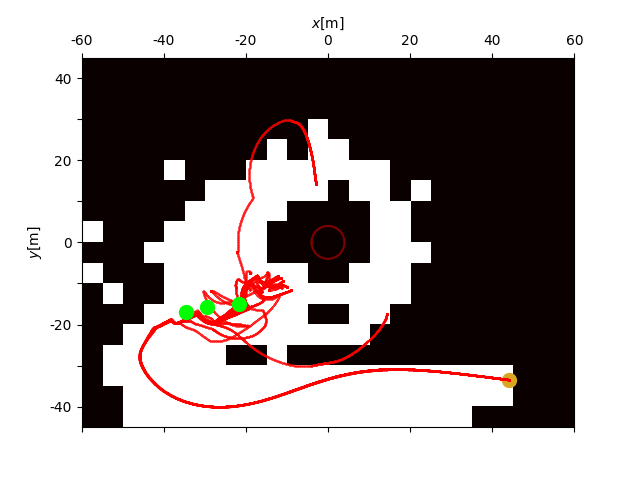}
                \caption{Player 1 - AA}
            \end{subfigure}

            \medskip
            \begin{subfigure}{0.49\textwidth}
                \includegraphics[width=\linewidth]{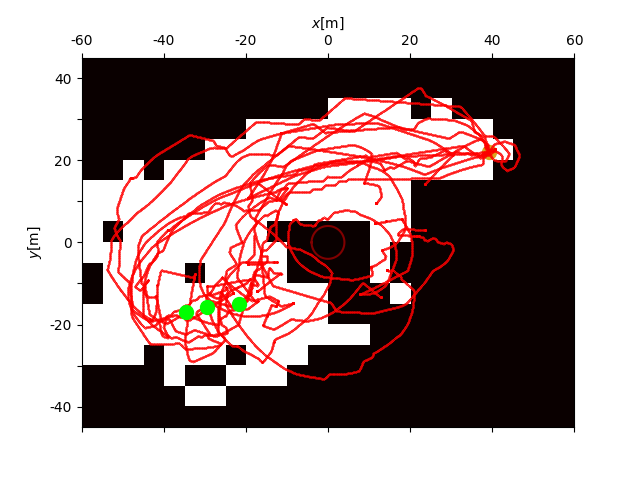}
                \caption{Player 2 - Human}
            \end{subfigure}
            \hfill
            \begin{subfigure}{0.49\textwidth}
                \includegraphics[width=\linewidth]{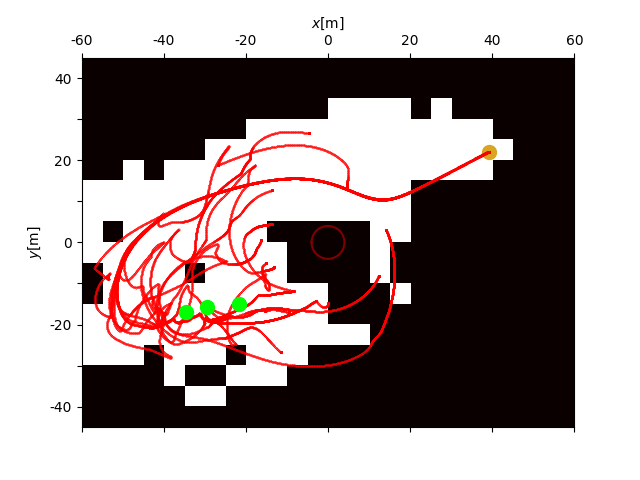}
                \caption{Player 2 - AA}
            \end{subfigure}
        \end{figure}
        \newpage

    \subsection*{Trial \#23}
        \begin{figure}[H]
            \begin{subfigure}{0.49\textwidth}
                \includegraphics[width=\linewidth]{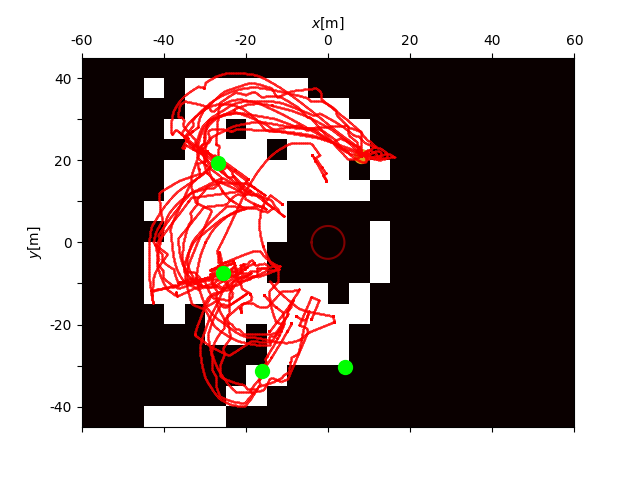}
                \caption{Player 1 - Human}
            \end{subfigure}
            \hfill
            \begin{subfigure}{0.49\textwidth}
                \includegraphics[width=\linewidth]{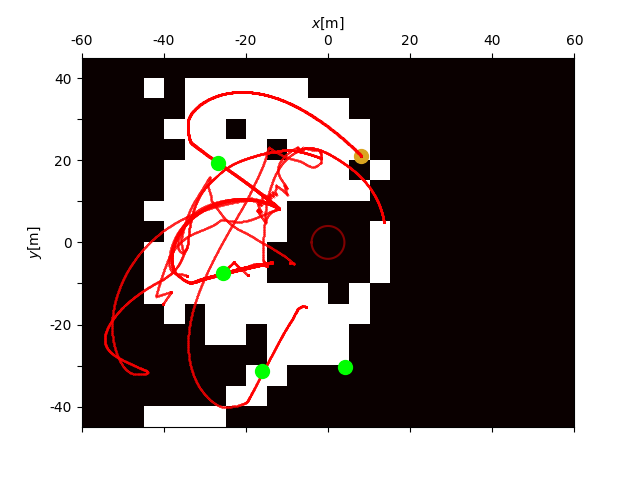}
                \caption{Player 1 - AA}
            \end{subfigure}

            \medskip
            \begin{subfigure}{0.49\textwidth}
                \includegraphics[width=\linewidth]{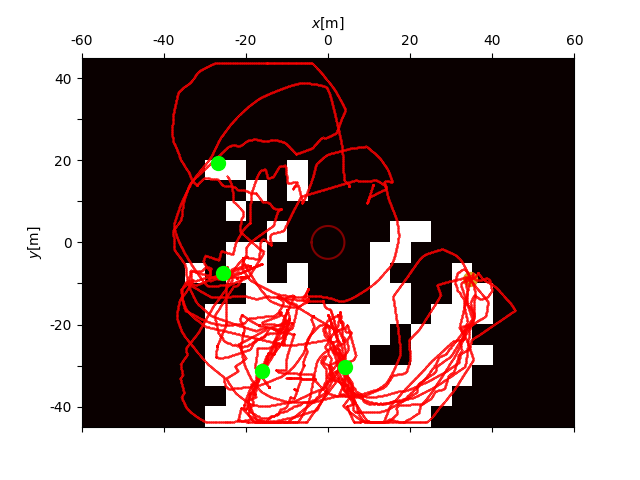}
                \caption{Player 2 - Human}
            \end{subfigure}
            \hfill
            \begin{subfigure}{0.49\textwidth}
                \includegraphics[width=\linewidth]{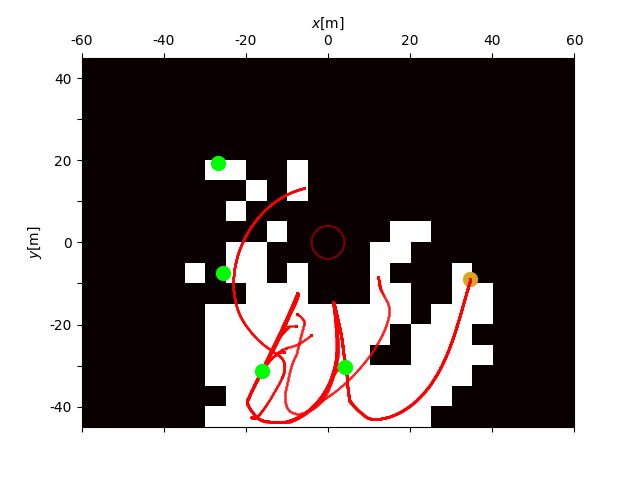}
                \caption{Player 2 - AA}
            \end{subfigure}
        \end{figure}
        \newpage

    \subsection*{Trial \#24}
        \begin{figure}[H]
            \begin{subfigure}{0.49\textwidth}
                \includegraphics[width=\linewidth]{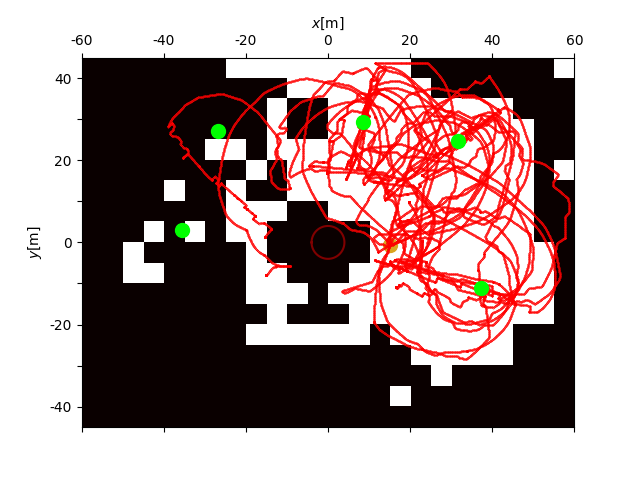}
                \caption{Player 1 - Human}
            \end{subfigure}
            \hfill
            \begin{subfigure}{0.49\textwidth}
                \includegraphics[width=\linewidth]{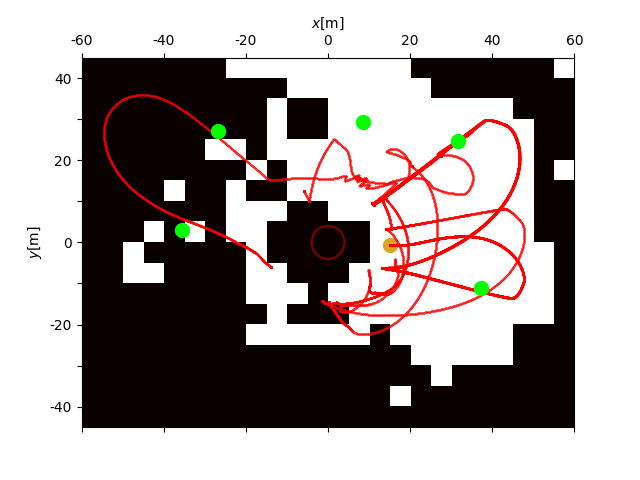}
                \caption{Player 1 - AA}
            \end{subfigure}

            \medskip
            \begin{subfigure}{0.49\textwidth}
                \includegraphics[width=\linewidth]{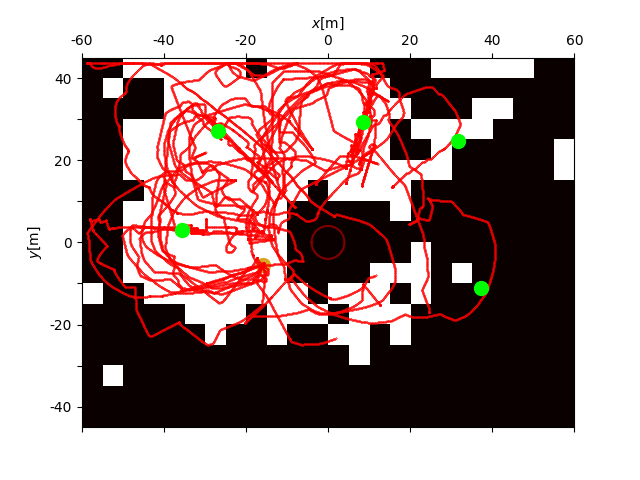}
                \caption{Player 2 - Human}
            \end{subfigure}
            \hfill
            \begin{subfigure}{0.49\textwidth}
                \includegraphics[width=\linewidth]{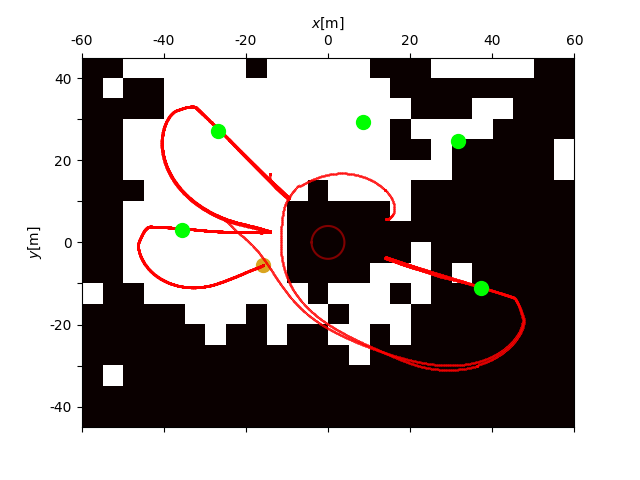}
                \caption{Player 2 - AA}
            \end{subfigure}
        \end{figure}
        \newpage

\section*{Human-Sensitive Team}

    \subsection*{Trial \#7}
        \begin{figure}[H]
            \begin{subfigure}{0.49\textwidth}
                \includegraphics[width=\linewidth]{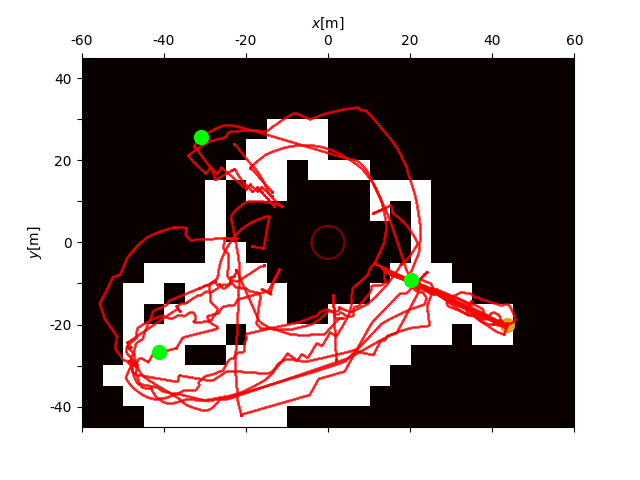}
                \caption{Player 1 - Human}
            \end{subfigure}
            \hfill
            \begin{subfigure}{0.49\textwidth}
                \includegraphics[width=\linewidth]{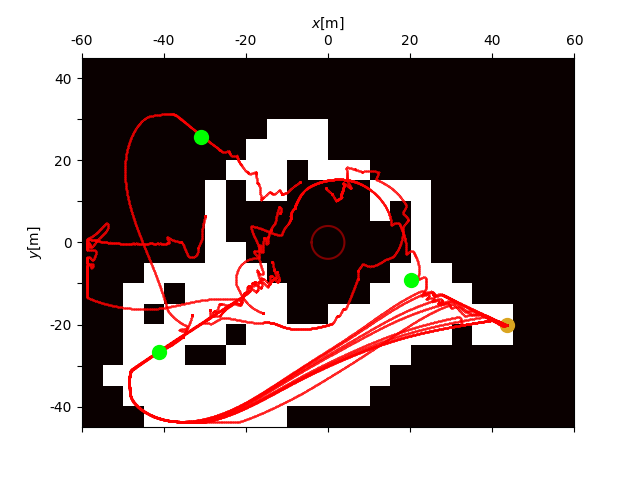}
                \caption{Player 1 - AA}
            \end{subfigure}

            \medskip
            \begin{subfigure}{0.49\textwidth}
                \includegraphics[width=\linewidth]{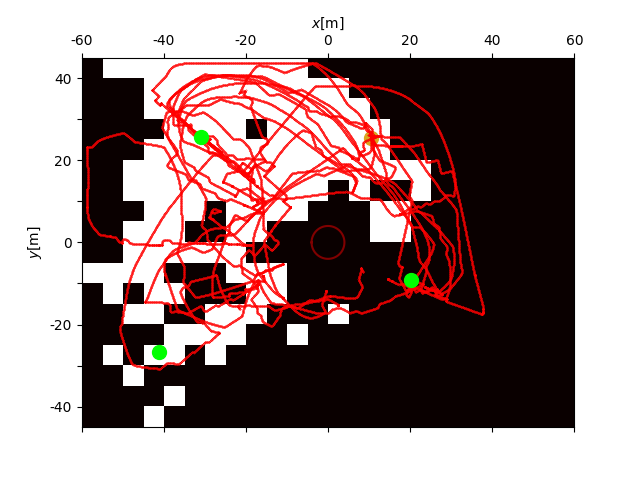}
                \caption{Player 2 - Human}
            \end{subfigure}
            \hfill
            \begin{subfigure}{0.49\textwidth}
                \includegraphics[width=\linewidth]{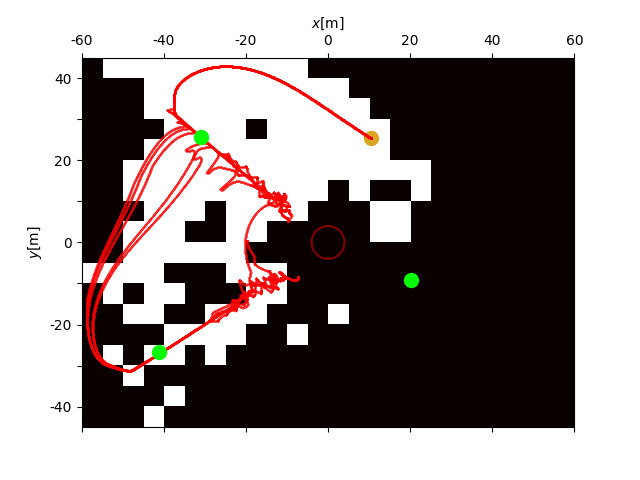}
                \caption{Player 2 - AA}
            \end{subfigure}
        \end{figure}
        \newpage

    \subsection*{Trial \#8}
        \begin{figure}[H]
            \begin{subfigure}{0.49\textwidth}
                \includegraphics[width=\linewidth]{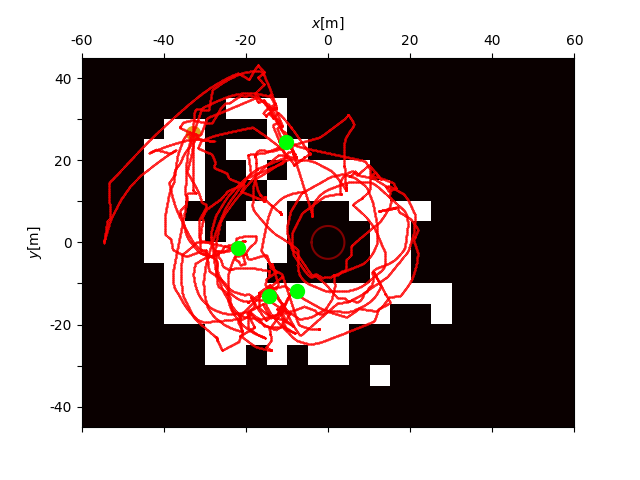}
                \caption{Player 1 - Human}
            \end{subfigure}
            \hfill
            \begin{subfigure}{0.49\textwidth}
                \includegraphics[width=\linewidth]{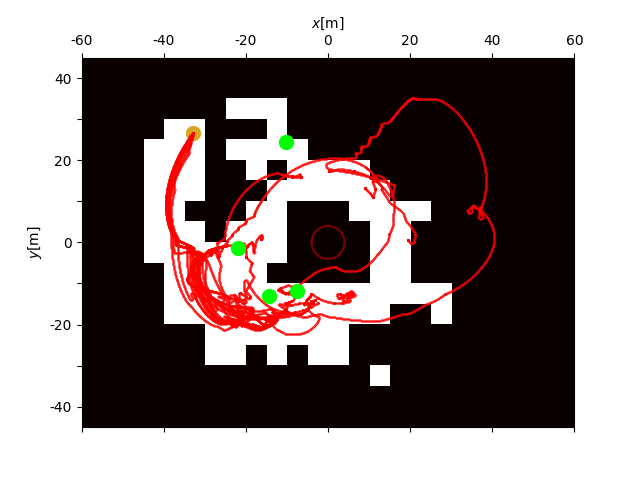}
                \caption{Player 1 - AA}
            \end{subfigure}

            \medskip
            \begin{subfigure}{0.49\textwidth}
                \includegraphics[width=\linewidth]{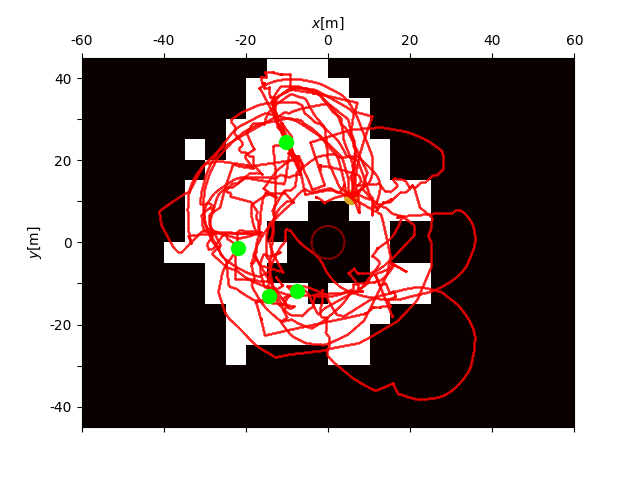}
                \caption{Player 2 - Human}
            \end{subfigure}
            \hfill
            \begin{subfigure}{0.49\textwidth}
                \includegraphics[width=\linewidth]{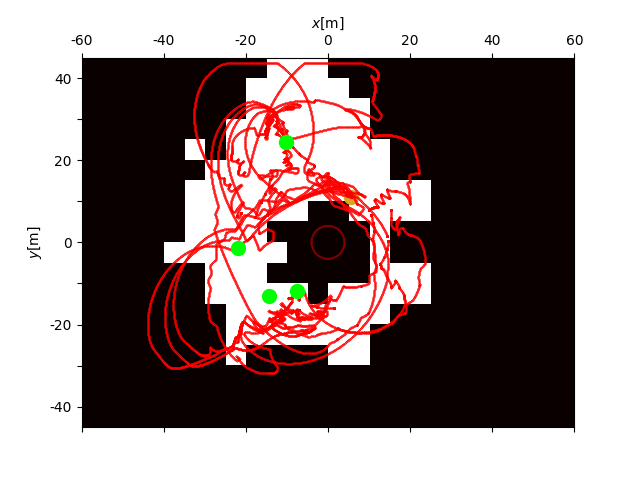}
                \caption{Player 2 - AA}
            \end{subfigure}
        \end{figure}
        \newpage

    \subsection*{Trial \#9}
        \begin{figure}[H]
            \begin{subfigure}{0.49\textwidth}
                \includegraphics[width=\linewidth]{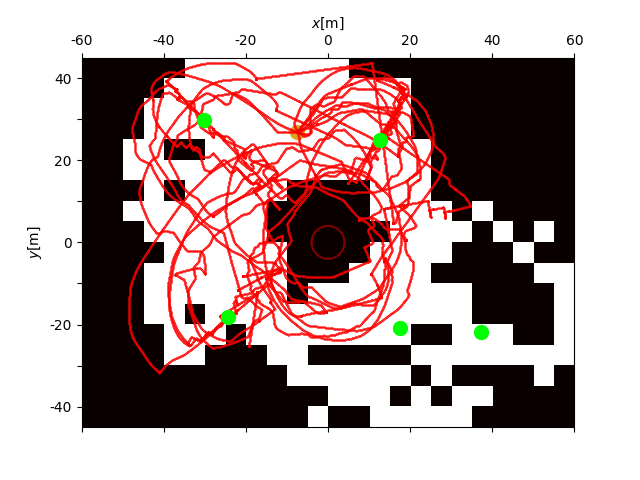}
                \caption{Player 1 - Human}
            \end{subfigure}
            \hfill
            \begin{subfigure}{0.49\textwidth}
                \includegraphics[width=\linewidth]{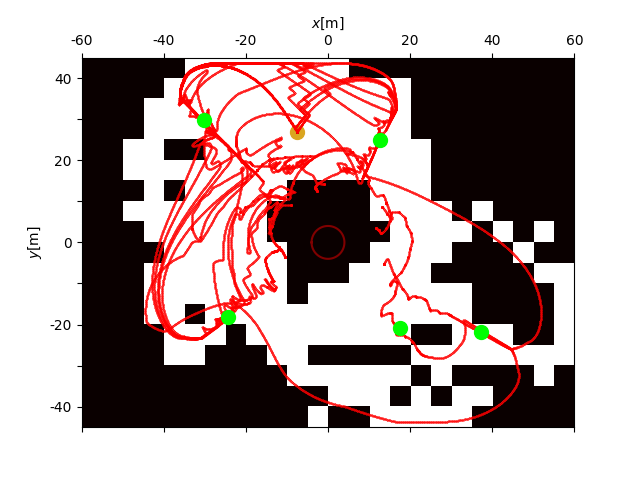}
                \caption{Player 1 - AA}
            \end{subfigure}

            \medskip
            \begin{subfigure}{0.49\textwidth}
                \includegraphics[width=\linewidth]{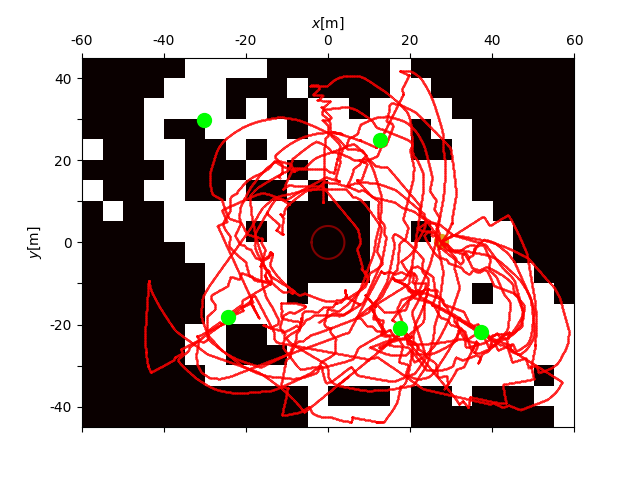}
                \caption{Player 2 - Human}
            \end{subfigure}
            \hfill
            \begin{subfigure}{0.49\textwidth}
                \includegraphics[width=\linewidth]{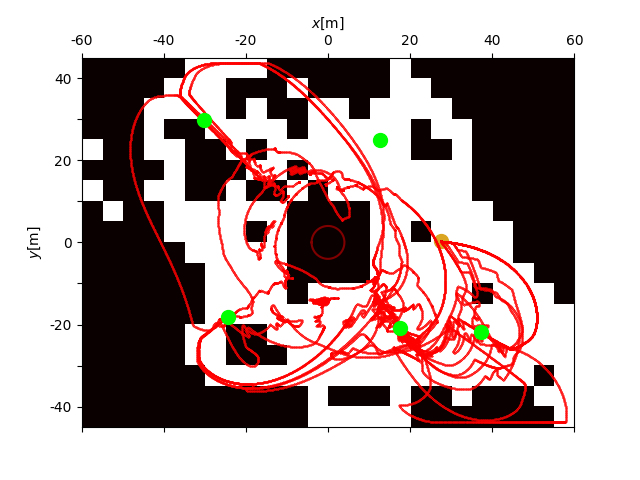}
                \caption{Player 2 - AA}
            \end{subfigure}
        \end{figure}
        \newpage

    \subsection*{Trial \#10}
        \begin{figure}[H]
            \begin{subfigure}{0.49\textwidth}
                \includegraphics[width=\linewidth]{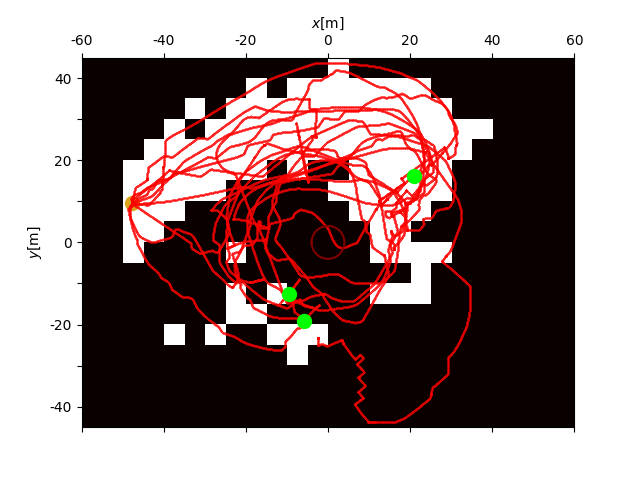}
                \caption{Player 1 - Human}
            \end{subfigure}
            \hfill
            \begin{subfigure}{0.49\textwidth}
                \includegraphics[width=\linewidth]{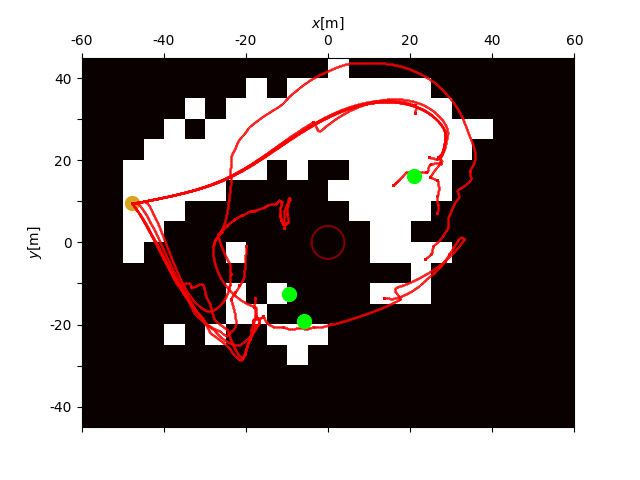}
                \caption{Player 1 - AA}
            \end{subfigure}

            \medskip
            \begin{subfigure}{0.49\textwidth}
                \includegraphics[width=\linewidth]{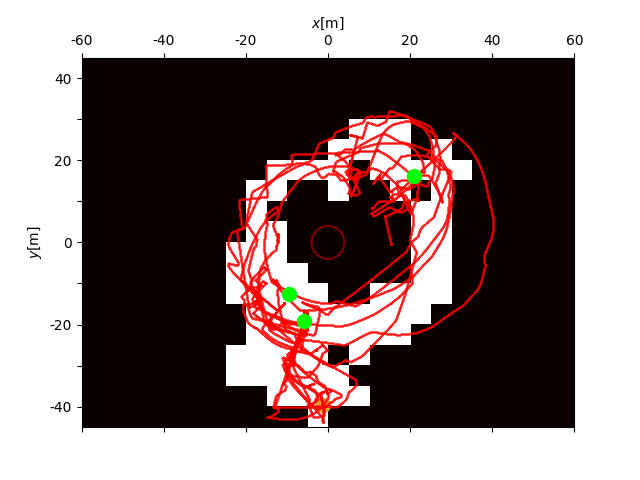}
                \caption{Player 2 - Human}
            \end{subfigure}
            \hfill
            \begin{subfigure}{0.49\textwidth}
                \includegraphics[width=\linewidth]{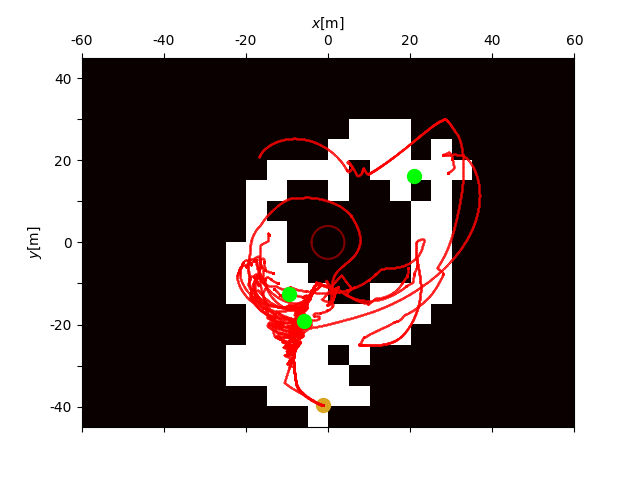}
                \caption{Player 2 - AA}
            \end{subfigure}
        \end{figure}
        \newpage

    \subsection*{Trial \#11}
        \begin{figure}[H]
            \begin{subfigure}{0.49\textwidth}
                \includegraphics[width=\linewidth]{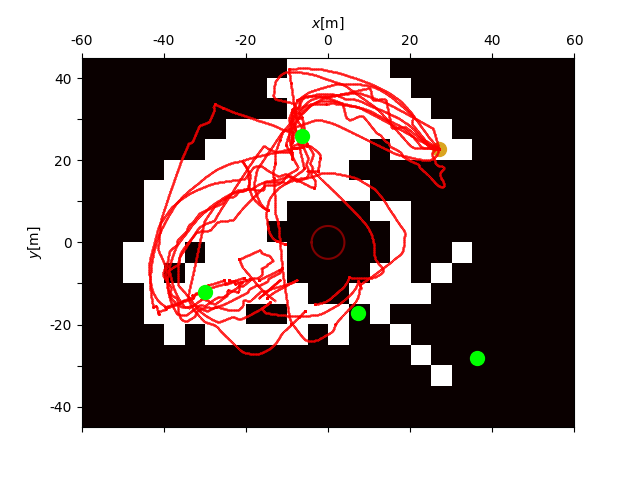}
                \caption{Player 1 - Human}
            \end{subfigure}
            \hfill
            \begin{subfigure}{0.49\textwidth}
                \includegraphics[width=\linewidth]{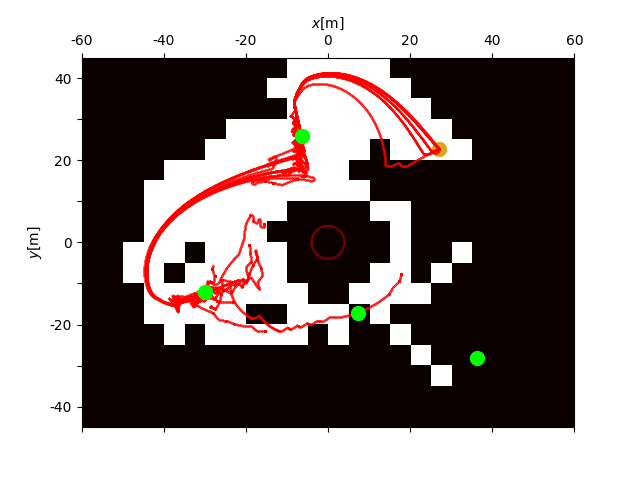}
                \caption{Player 1 - AA}
            \end{subfigure}

            \medskip
            \begin{subfigure}{0.49\textwidth}
                \includegraphics[width=\linewidth]{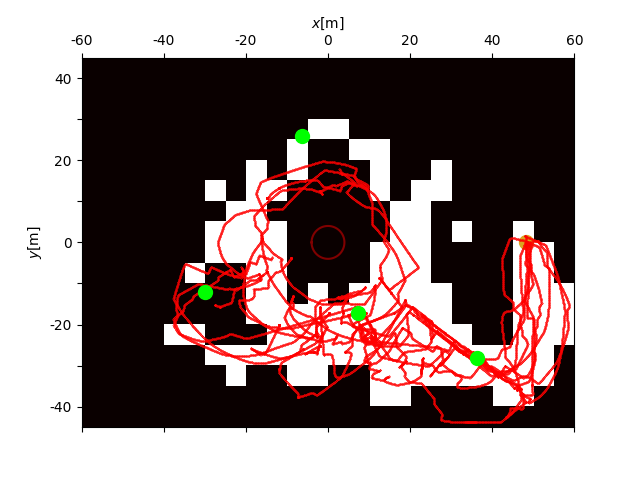}
                \caption{Player 2 - Human}
            \end{subfigure}
            \hfill
            \begin{subfigure}{0.49\textwidth}
                \includegraphics[width=\linewidth]{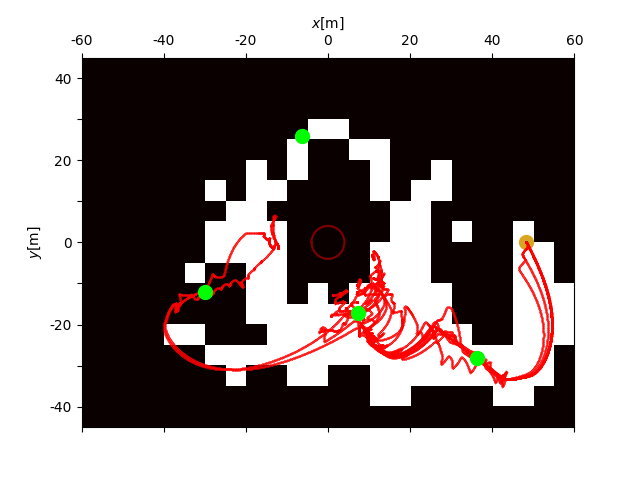}
                \caption{Player 2 - AA}
            \end{subfigure}
        \end{figure}
        \newpage

    \subsection*{Trial \#12}
        \begin{figure}[H]
            \begin{subfigure}{0.49\textwidth}
                \includegraphics[width=\linewidth]{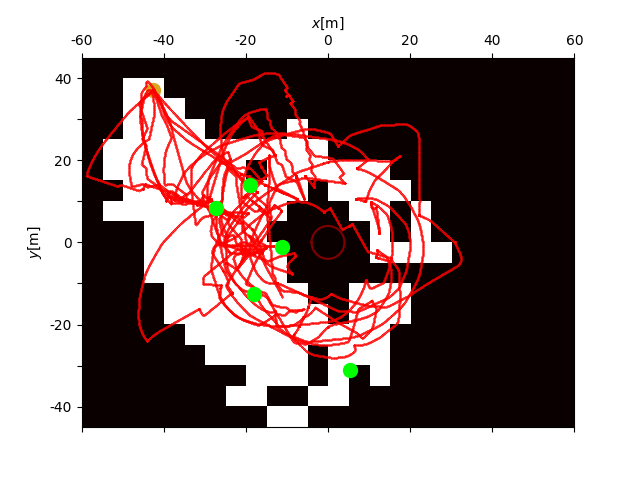}
                \caption{Player 1 - Human}
            \end{subfigure}
            \hfill
            \begin{subfigure}{0.49\textwidth}
                \includegraphics[width=\linewidth]{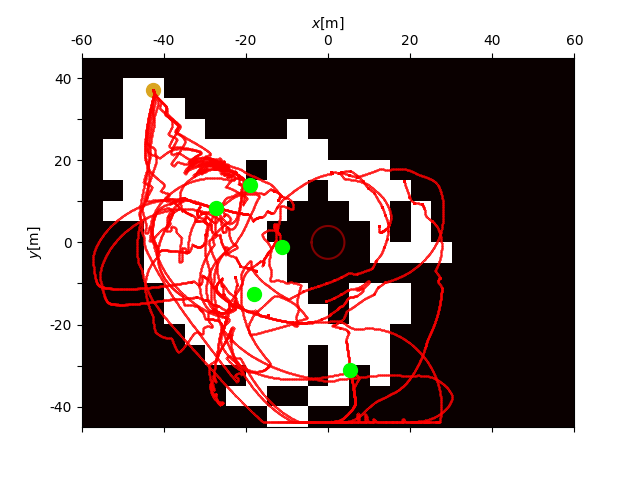}
                \caption{Player 1 - AA}
            \end{subfigure}

            \medskip
            \begin{subfigure}{0.49\textwidth}
                \includegraphics[width=\linewidth]{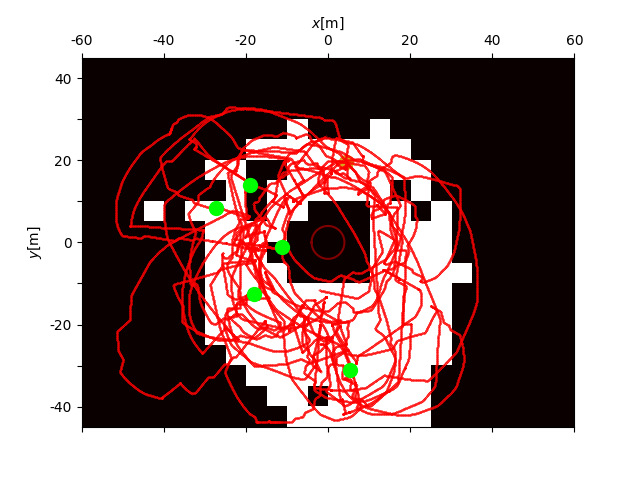}
                \caption{Player 2 - Human}
            \end{subfigure}
            \hfill
            \begin{subfigure}{0.49\textwidth}
                \includegraphics[width=\linewidth]{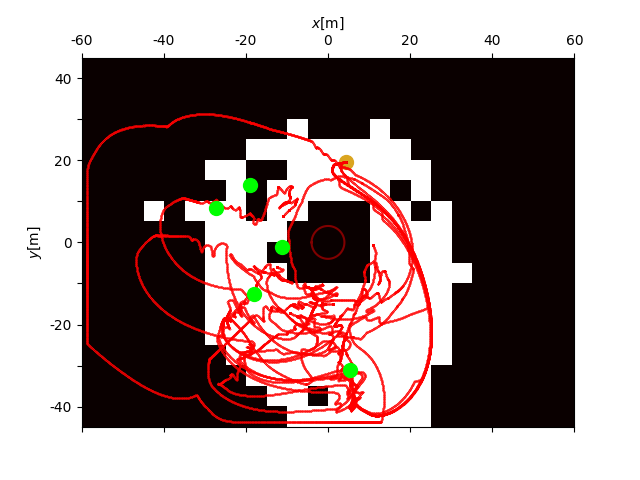}
                \caption{Player 2 - AA}
            \end{subfigure}
        \end{figure}
        \newpage

    \subsection*{Trial \#13}
        \begin{figure}[H]
            \begin{subfigure}{0.49\textwidth}
                \includegraphics[width=\linewidth]{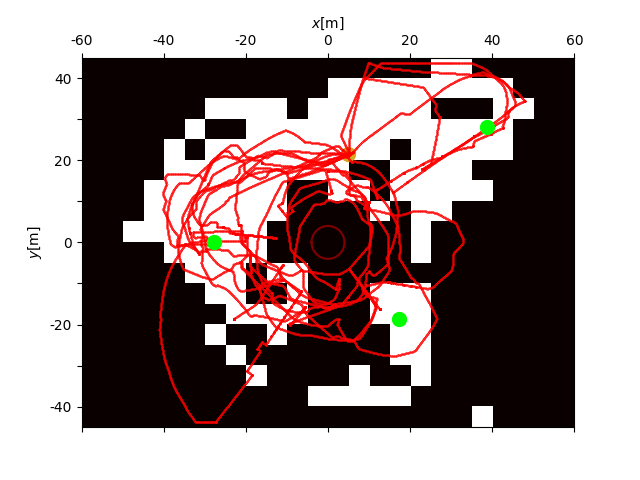}
                \caption{Player 1 - Human}
            \end{subfigure}
            \hfill
            \begin{subfigure}{0.49\textwidth}
                \includegraphics[width=\linewidth]{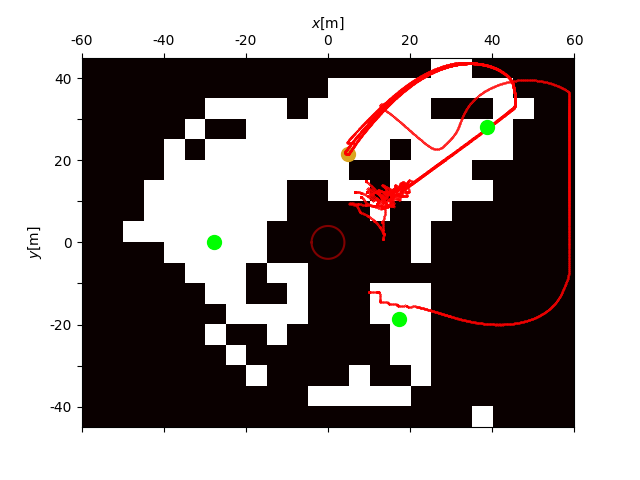}
                \caption{Player 1 - AA}
            \end{subfigure}

            \medskip
            \begin{subfigure}{0.49\textwidth}
                \includegraphics[width=\linewidth]{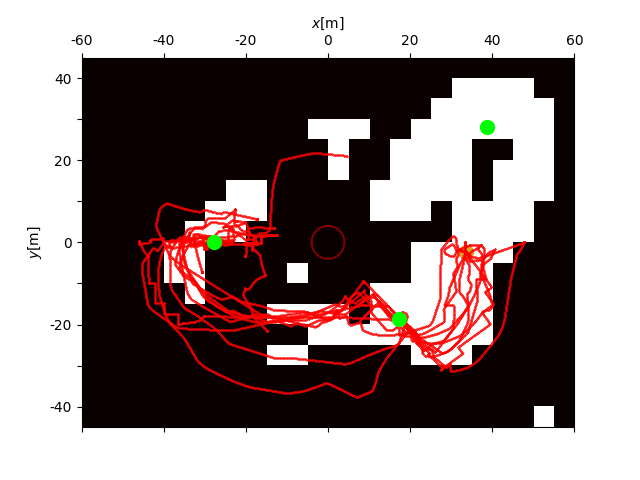}
                \caption{Player 2 - Human}
            \end{subfigure}
            \hfill
            \begin{subfigure}{0.49\textwidth}
                \includegraphics[width=\linewidth]{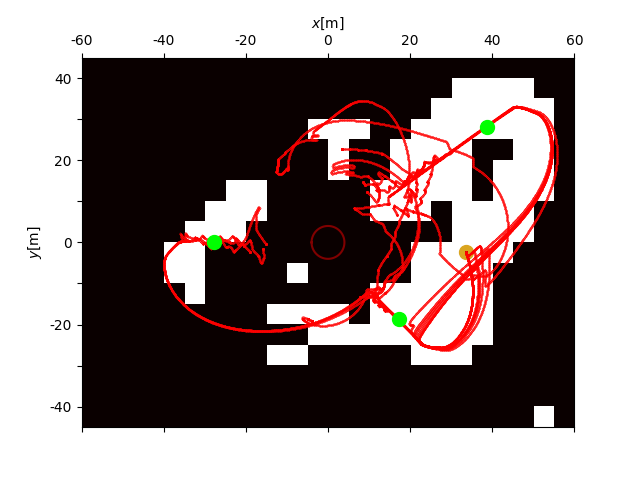}
                \caption{Player 2 - AA}
            \end{subfigure}
        \end{figure}
        \newpage

    \subsection*{Trial \#14}
        \begin{figure}[H]
            \begin{subfigure}{0.49\textwidth}
                \includegraphics[width=\linewidth]{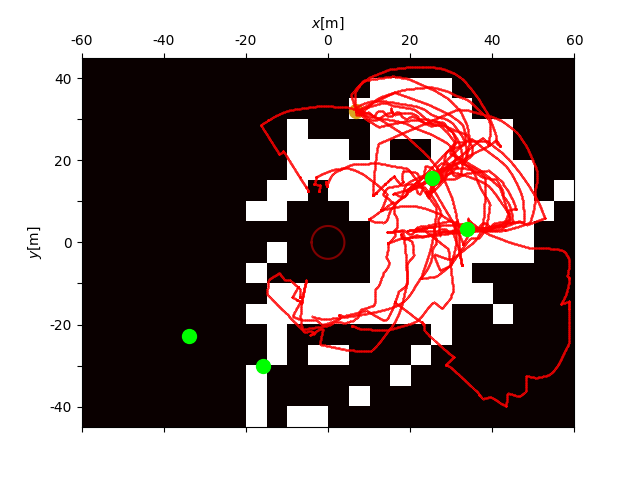}
                \caption{Player 1 - Human}
            \end{subfigure}
            \hfill
            \begin{subfigure}{0.49\textwidth}
                \includegraphics[width=\linewidth]{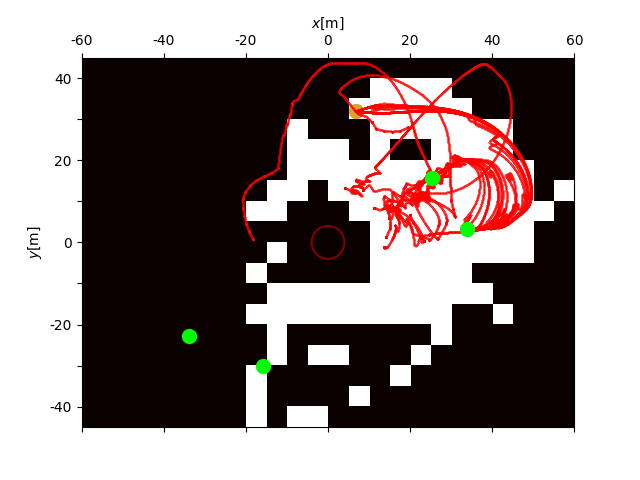}
                \caption{Player 1 - AA}
            \end{subfigure}

            \medskip
            \begin{subfigure}{0.49\textwidth}
                \includegraphics[width=\linewidth]{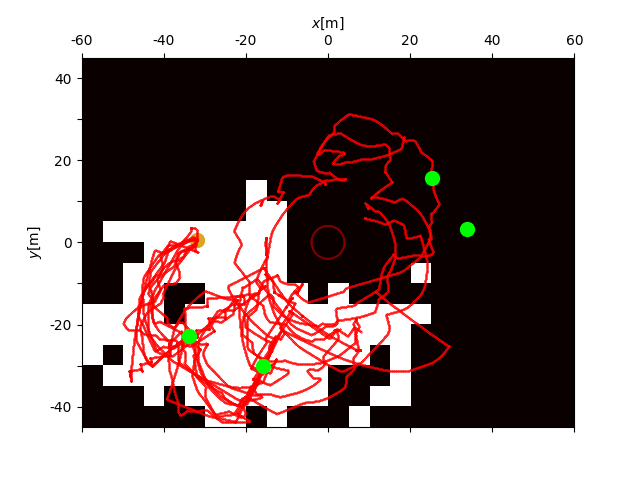}
                \caption{Player 2 - Human}
            \end{subfigure}
            \hfill
            \begin{subfigure}{0.49\textwidth}
                \includegraphics[width=\linewidth]{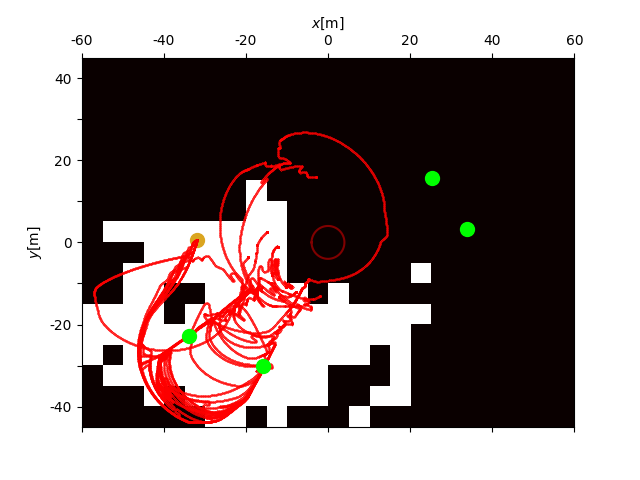}
                \caption{Player 2 - AA}
            \end{subfigure}
        \end{figure}
        \newpage

    \subsection*{Trial \#15}
        \begin{figure}[H]
            \begin{subfigure}{0.49\textwidth}
                \includegraphics[width=\linewidth]{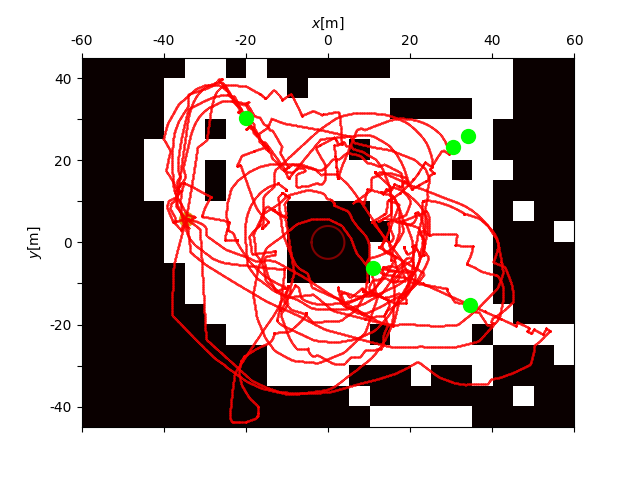}
                \caption{Player 1 - Human}
            \end{subfigure}
            \hfill
            \begin{subfigure}{0.49\textwidth}
                \includegraphics[width=\linewidth]{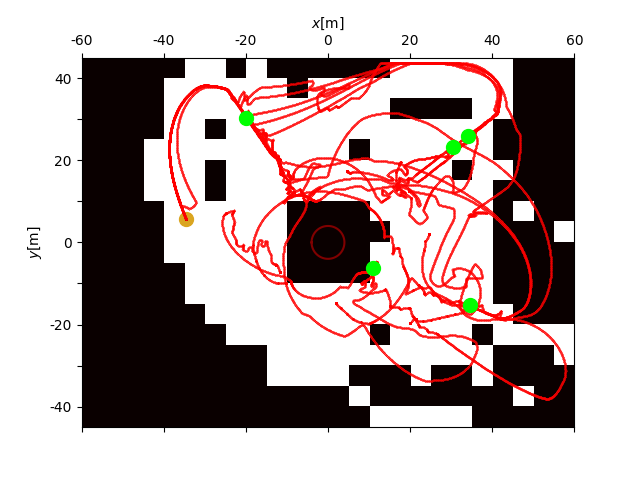}
                \caption{Player 1 - AA}
            \end{subfigure}

            \medskip
            \begin{subfigure}{0.49\textwidth}
                \includegraphics[width=\linewidth]{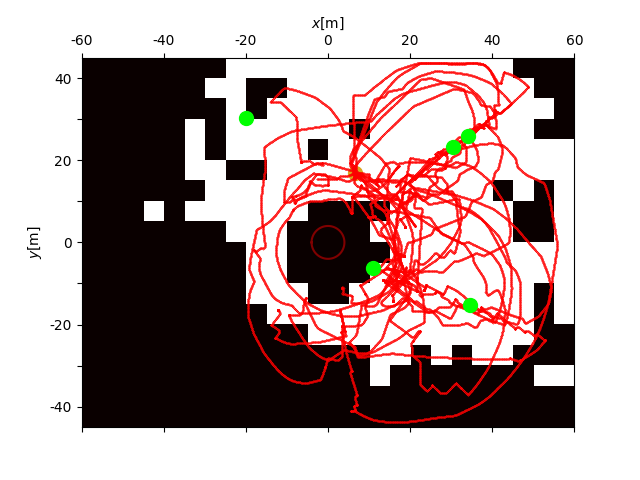}
                \caption{Player 2 - Human}
            \end{subfigure}
            \hfill
            \begin{subfigure}{0.49\textwidth}
                \includegraphics[width=\linewidth]{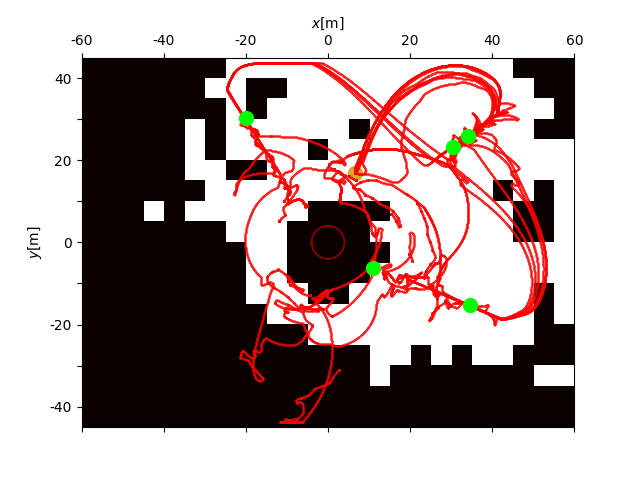}
                \caption{Player 2 - AA}
            \end{subfigure}
        \end{figure}
        \newpage

    \subsection*{Trial \#16}
        \begin{figure}[H]
            \begin{subfigure}{0.49\textwidth}
                \includegraphics[width=\linewidth]{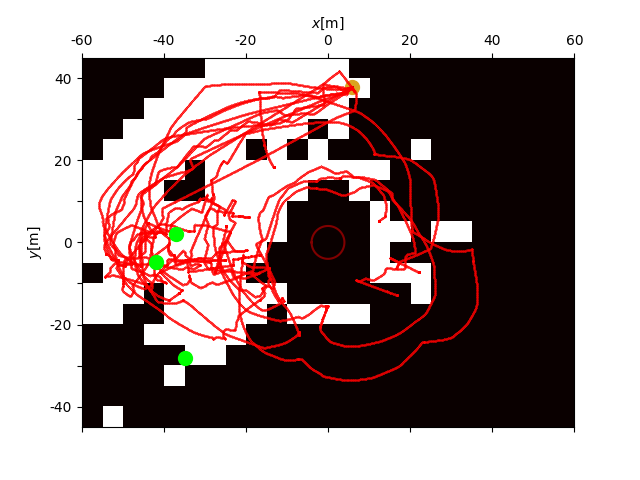}
                \caption{Player 1 - Human}
            \end{subfigure}
            \hfill
            \begin{subfigure}{0.49\textwidth}
                \includegraphics[width=\linewidth]{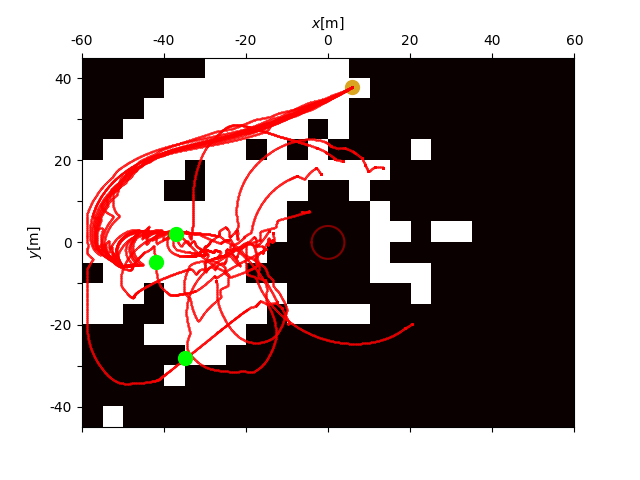}
                \caption{Player 1 - AA}
            \end{subfigure}

            \medskip
            \begin{subfigure}{0.49\textwidth}
                \includegraphics[width=\linewidth]{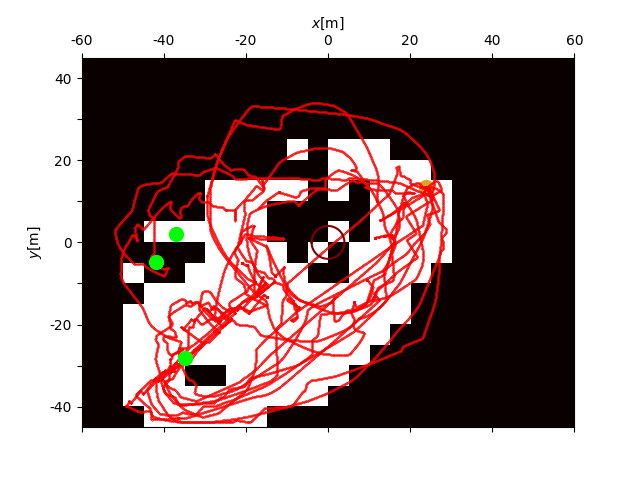}
                \caption{Player 2 - Human}
            \end{subfigure}
            \hfill
            \begin{subfigure}{0.49\textwidth}
                \includegraphics[width=\linewidth]{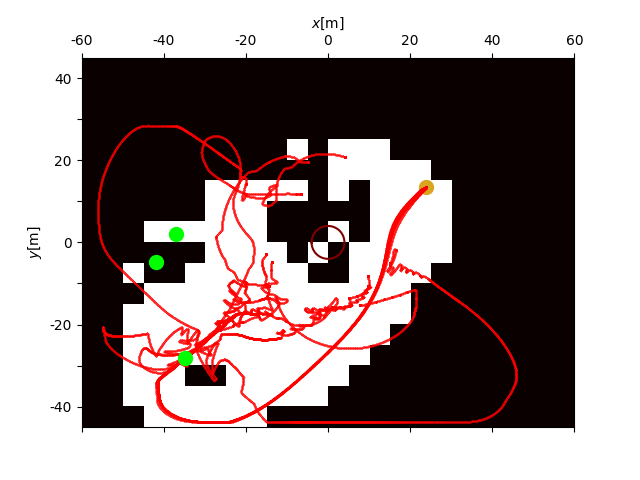}
                \caption{Player 2 - AA}
            \end{subfigure}
        \end{figure}
        \newpage

    \subsection*{Trial \#17}
        \begin{figure}[H]
            \begin{subfigure}{0.49\textwidth}
                \includegraphics[width=\linewidth]{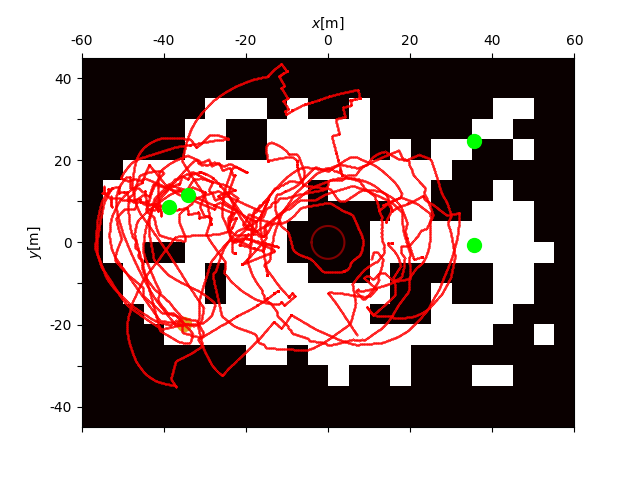}
                \caption{Player 1 - Human}
            \end{subfigure}
            \hfill
            \begin{subfigure}{0.49\textwidth}
                \includegraphics[width=\linewidth]{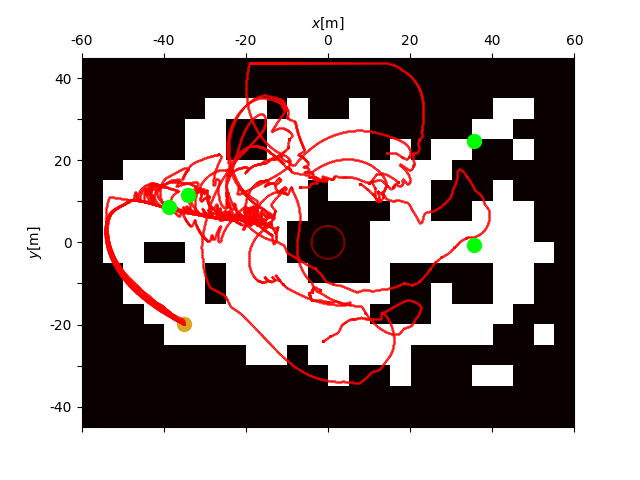}
                \caption{Player 1 - AA}
            \end{subfigure}

            \medskip
            \begin{subfigure}{0.49\textwidth}
                \includegraphics[width=\linewidth]{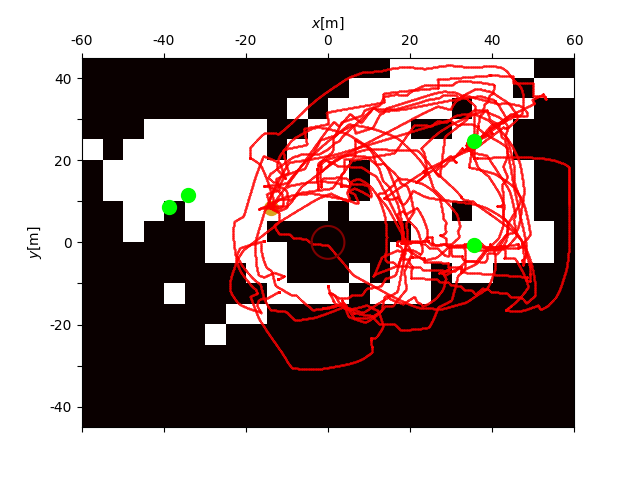}
                \caption{Player 2 - Human}
            \end{subfigure}
            \hfill
            \begin{subfigure}{0.49\textwidth}
                \includegraphics[width=\linewidth]{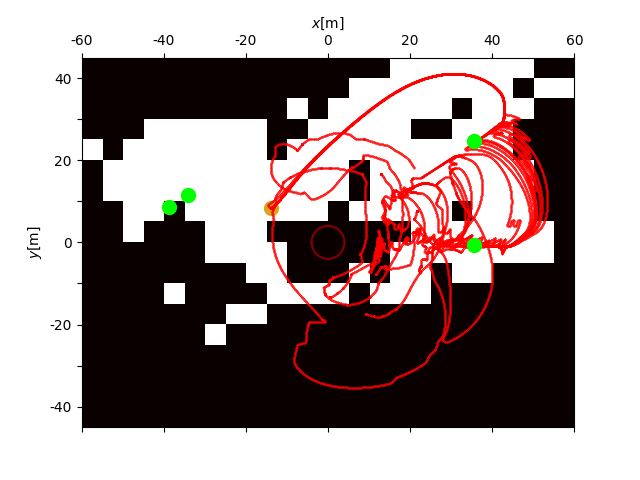}
                \caption{Player 2 - AA}
            \end{subfigure}
        \end{figure}
        \newpage

    \subsection*{Trial \#18}
        \begin{figure}[H]
            \begin{subfigure}{0.49\textwidth}
                \includegraphics[width=\linewidth]{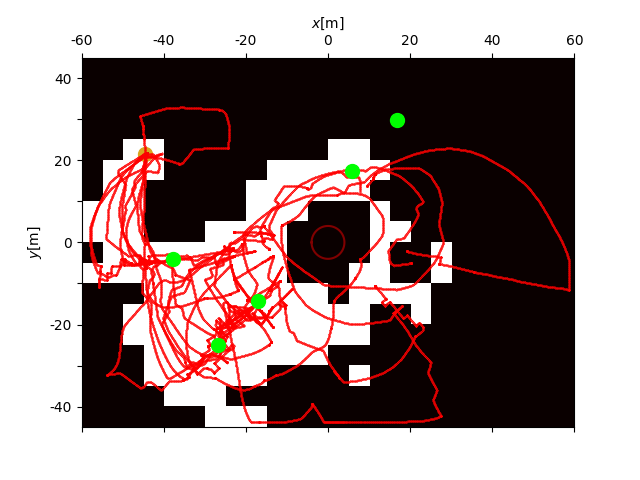}
                \caption{Player 1 - Human}
            \end{subfigure}
            \hfill
            \begin{subfigure}{0.49\textwidth}
                \includegraphics[width=\linewidth]{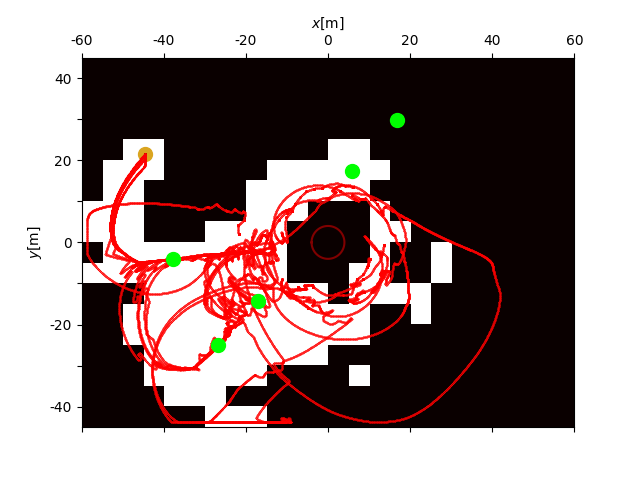}
                \caption{Player 1 - AA}
            \end{subfigure}

            \medskip
            \begin{subfigure}{0.49\textwidth}
                \includegraphics[width=\linewidth]{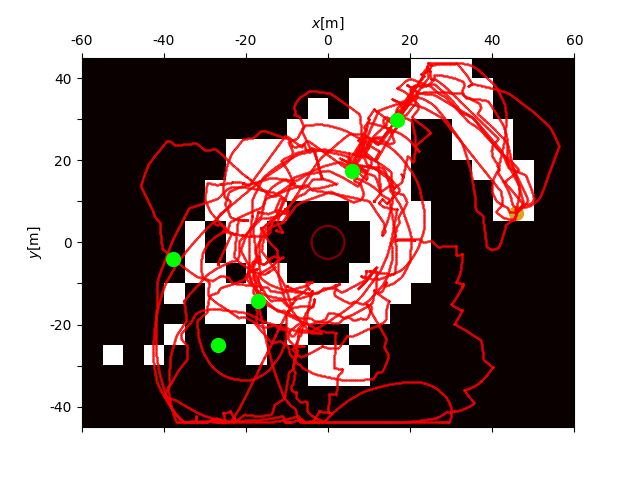}
                \caption{Player 2 - Human}
            \end{subfigure}
            \hfill
            \begin{subfigure}{0.49\textwidth}
                \includegraphics[width=\linewidth]{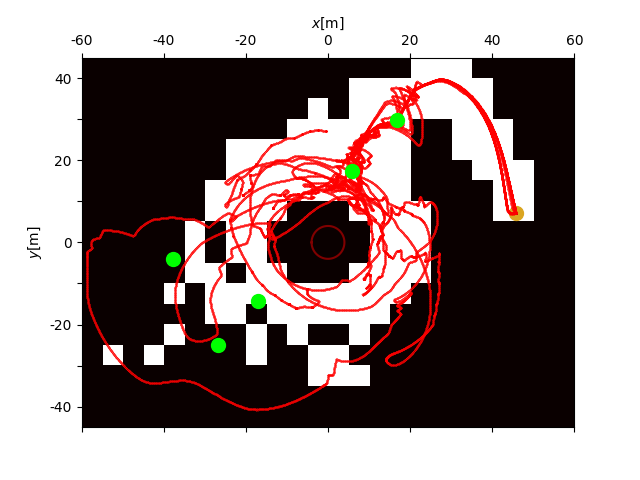}
                \caption{Player 2 - AA}
            \end{subfigure}
        \end{figure}
        \newpage

    \subsection*{Trial \#19}
        \begin{figure}[H]
            \begin{subfigure}{0.49\textwidth}
                \includegraphics[width=\linewidth]{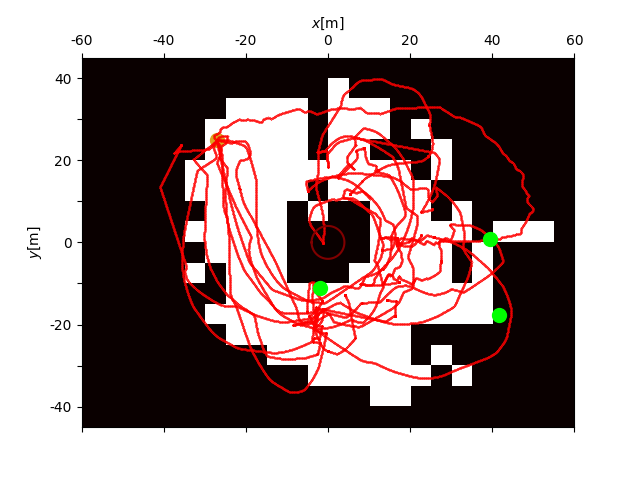}
                \caption{Player 1 - Human}
            \end{subfigure}
            \hfill
            \begin{subfigure}{0.49\textwidth}
                \includegraphics[width=\linewidth]{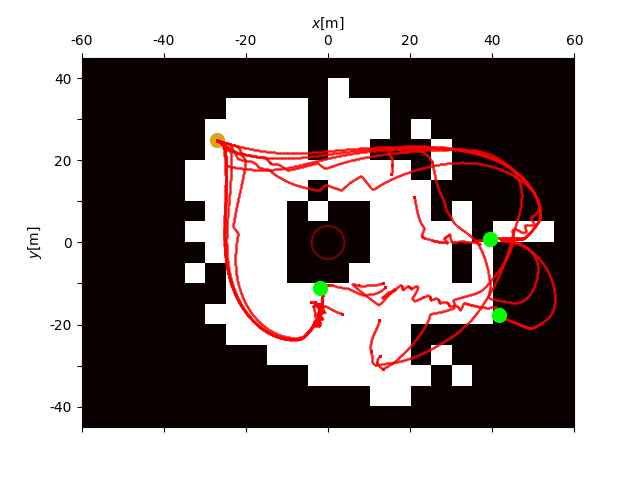}
                \caption{Player 1 - AA}
            \end{subfigure}

            \medskip
            \begin{subfigure}{0.49\textwidth}
                \includegraphics[width=\linewidth]{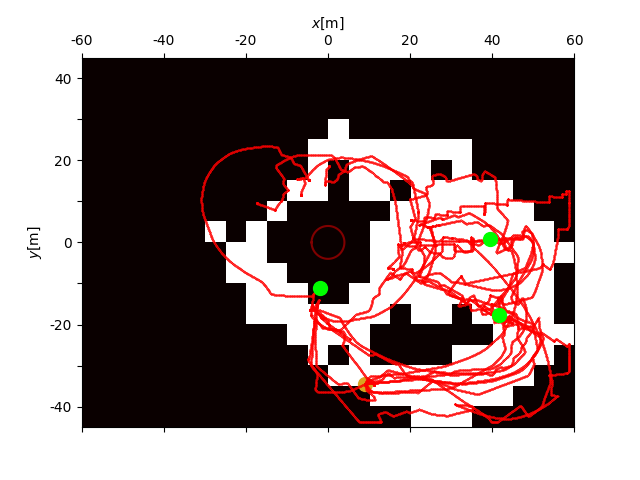}
                \caption{Player 2 - Human}
            \end{subfigure}
            \hfill
            \begin{subfigure}{0.49\textwidth}
                \includegraphics[width=\linewidth]{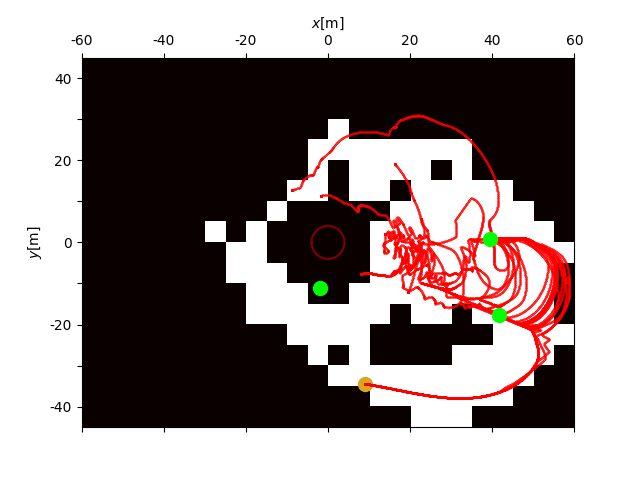}
                \caption{Player 2 - AA}
            \end{subfigure}
        \end{figure}
        \newpage

    \subsection*{Trial \#20}
        \begin{figure}[H]
            \begin{subfigure}{0.49\textwidth}
                \includegraphics[width=\linewidth]{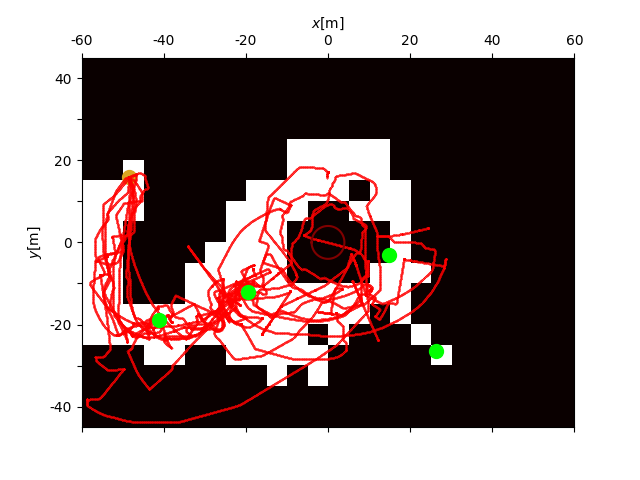}
                \caption{Player 1 - Human}
            \end{subfigure}
            \hfill
            \begin{subfigure}{0.49\textwidth}
                \includegraphics[width=\linewidth]{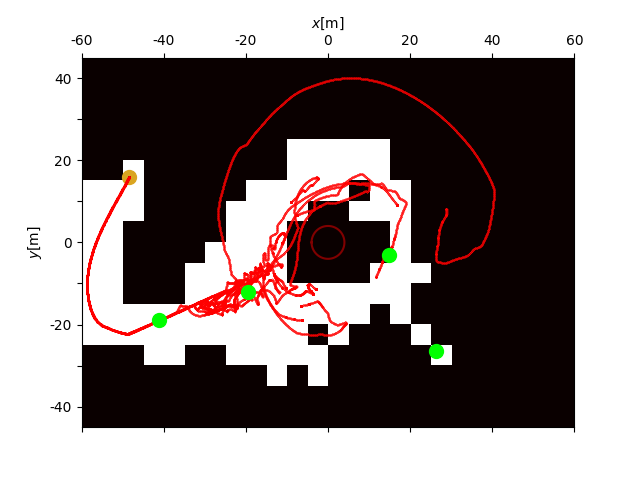}
                \caption{Player 1 - AA}
            \end{subfigure}

            \medskip
            \begin{subfigure}{0.49\textwidth}
                \includegraphics[width=\linewidth]{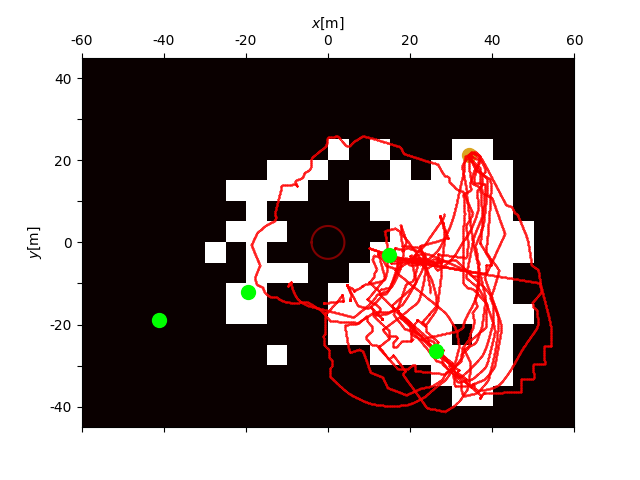}
                \caption{Player 2 - Human}
            \end{subfigure}
            \hfill
            \begin{subfigure}{0.49\textwidth}
                \includegraphics[width=\linewidth]{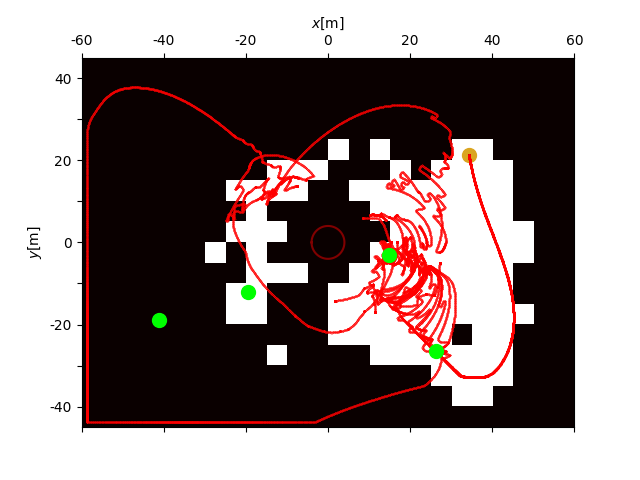}
                \caption{Player 2 - AA}
            \end{subfigure}
        \end{figure}
        \newpage

    \subsection*{Trial \#21}
        \begin{figure}[H]
            \begin{subfigure}{0.49\textwidth}
                \includegraphics[width=\linewidth]{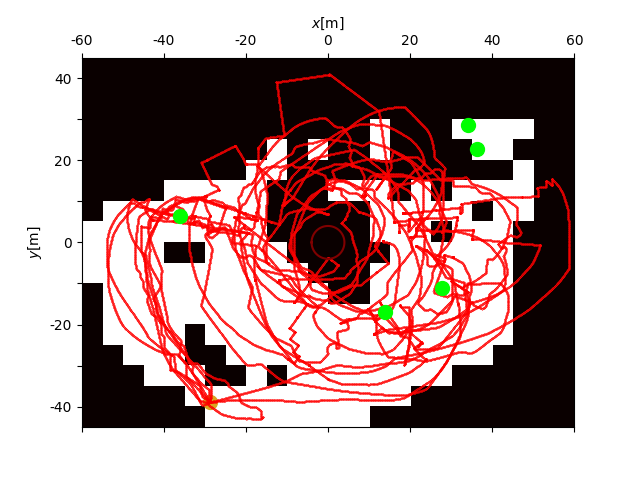}
                \caption{Player 1 - Human}
            \end{subfigure}
            \hfill
            \begin{subfigure}{0.49\textwidth}
                \includegraphics[width=\linewidth]{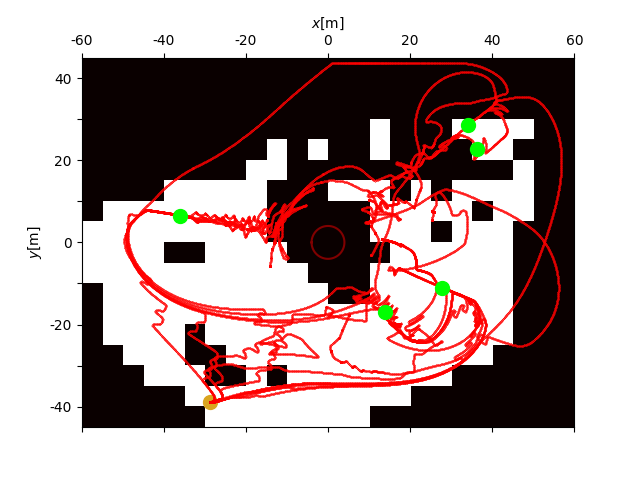}
                \caption{Player 1 - AA}
            \end{subfigure}

            \medskip
            \begin{subfigure}{0.49\textwidth}
                \includegraphics[width=\linewidth]{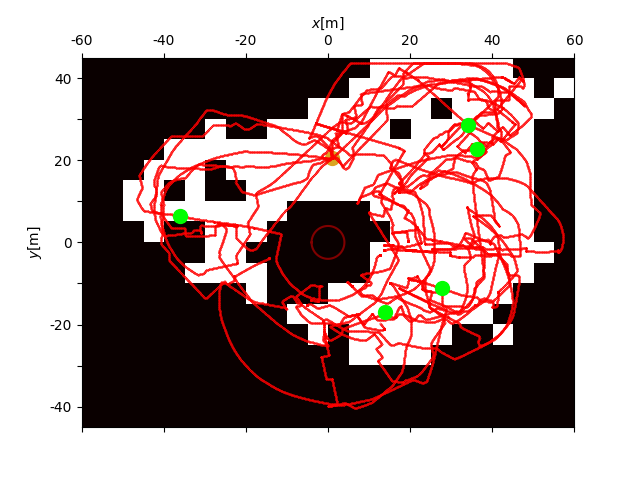}
                \caption{Player 2 - Human}
            \end{subfigure}
            \hfill
            \begin{subfigure}{0.49\textwidth}
                \includegraphics[width=\linewidth]{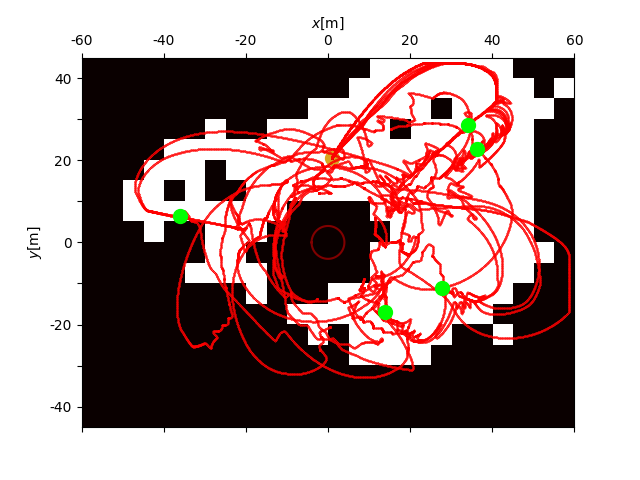}
                \caption{Player 2 - AA}
            \end{subfigure}
        \end{figure}
        \newpage

    \subsection*{Trial \#22}
        \begin{figure}[H]
            \begin{subfigure}{0.49\textwidth}
                \includegraphics[width=\linewidth]{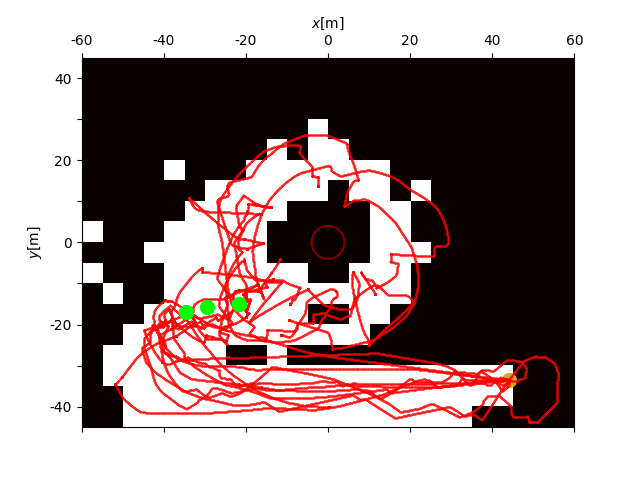}
                \caption{Player 1 - Human}
            \end{subfigure}
            \hfill
            \begin{subfigure}{0.49\textwidth}
                \includegraphics[width=\linewidth]{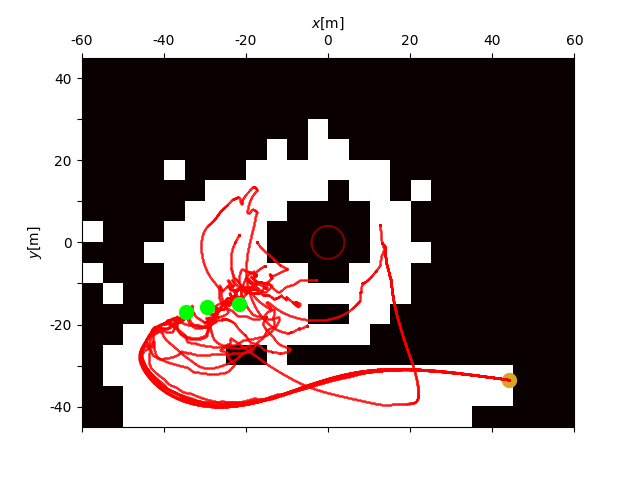}
                \caption{Player 1 - AA}
            \end{subfigure}

            \medskip
            \begin{subfigure}{0.49\textwidth}
                \includegraphics[width=\linewidth]{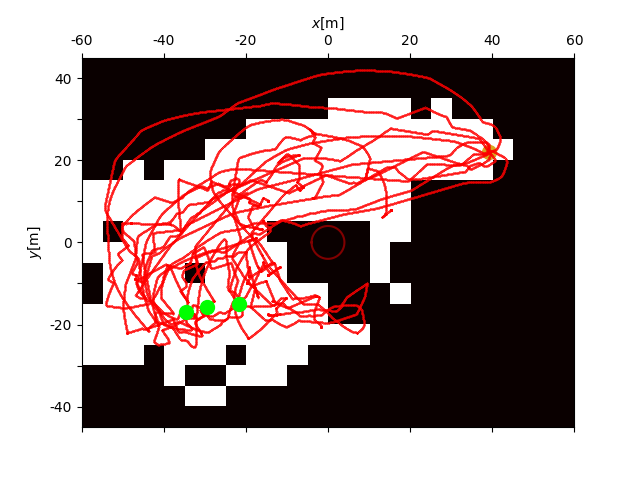}
                \caption{Player 2 - Human}
            \end{subfigure}
            \hfill
            \begin{subfigure}{0.49\textwidth}
                \includegraphics[width=\linewidth]{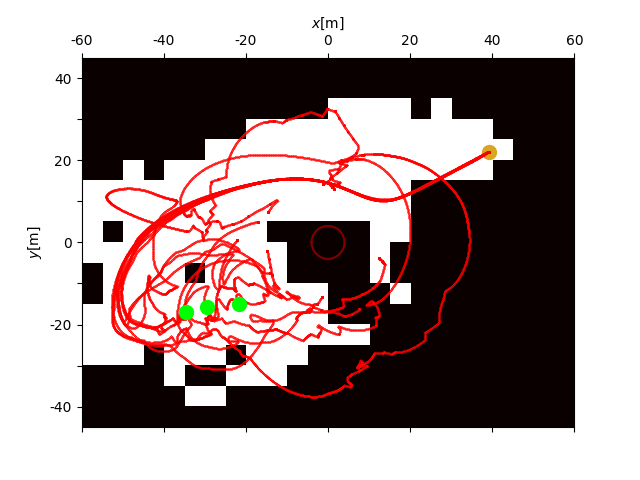}
                \caption{Player 2 - AA}
            \end{subfigure}
        \end{figure}
        \newpage

    \subsection*{Trial \#23}
        \begin{figure}[H]
            \begin{subfigure}{0.49\textwidth}
                \includegraphics[width=\linewidth]{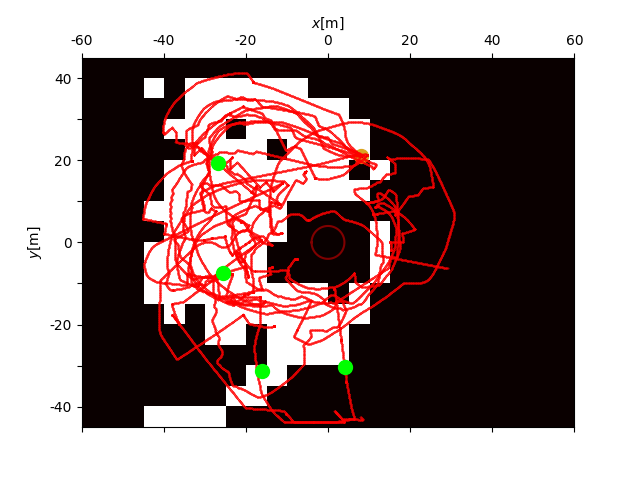}
                \caption{Player 1 - Human}
            \end{subfigure}
            \hfill
            \begin{subfigure}{0.49\textwidth}
                \includegraphics[width=\linewidth]{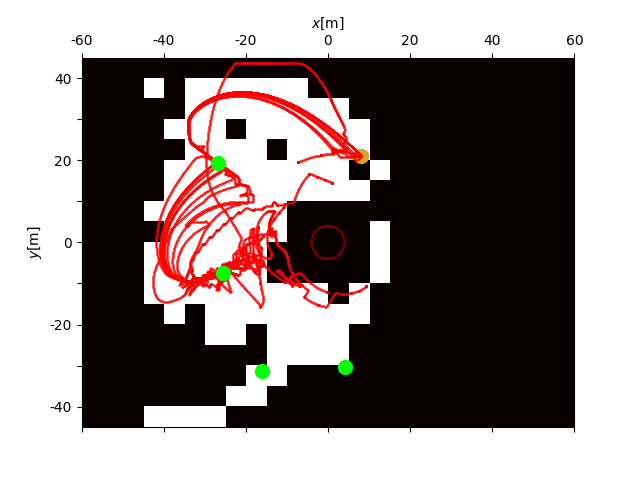}
                \caption{Player 1 - AA}
            \end{subfigure}

            \medskip
            \begin{subfigure}{0.49\textwidth}
                \includegraphics[width=\linewidth]{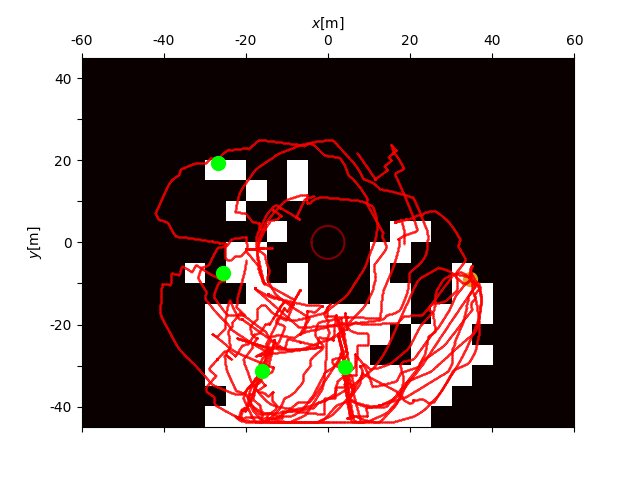}
                \caption{Player 2 - Human}
            \end{subfigure}
            \hfill
            \begin{subfigure}{0.49\textwidth}
                \includegraphics[width=\linewidth]{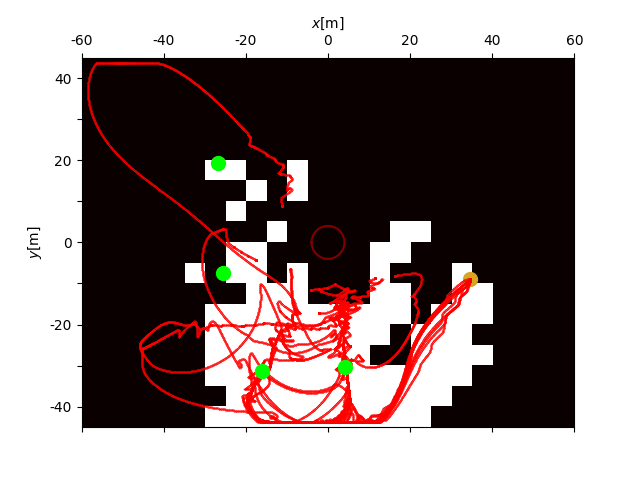}
                \caption{Player 2 - AA}
            \end{subfigure}
        \end{figure}
        \newpage

    \subsection*{Trial \#24}
        \begin{figure}[H]
            \begin{subfigure}{0.49\textwidth}
                \includegraphics[width=\linewidth]{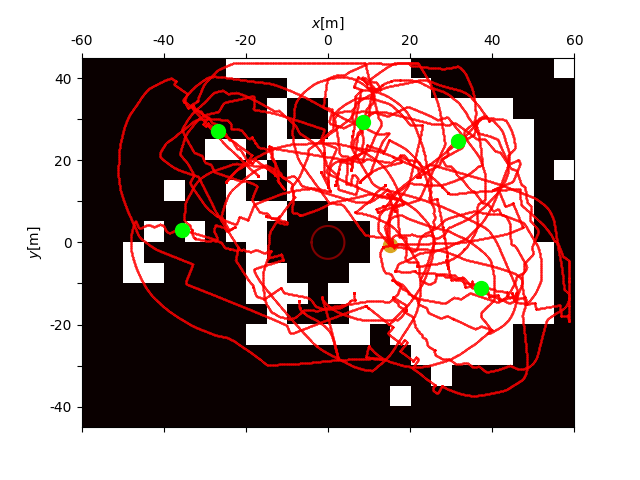}
                \caption{Player 1 - Human}
            \end{subfigure}
            \hfill
            \begin{subfigure}{0.49\textwidth}
                \includegraphics[width=\linewidth]{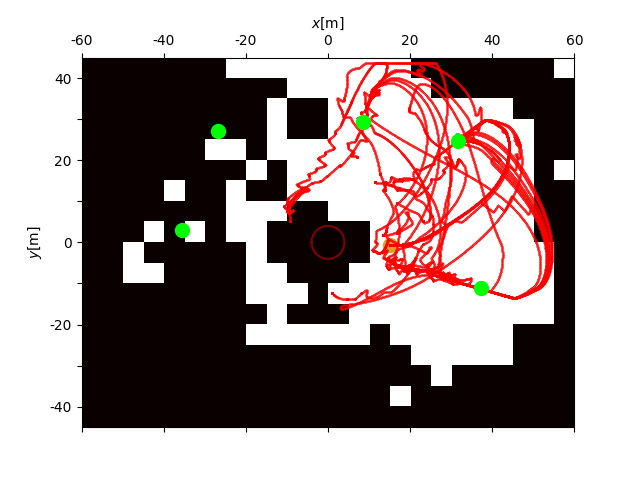}
                \caption{Player 1 - AA}
            \end{subfigure}

            \medskip
            \begin{subfigure}{0.49\textwidth}
                \includegraphics[width=\linewidth]{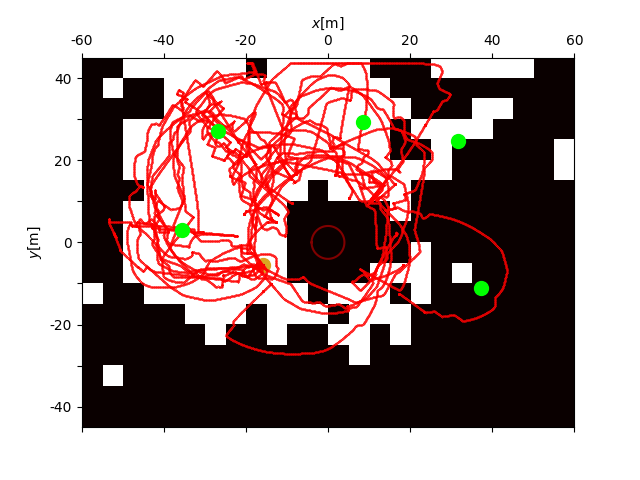}
                \caption{Player 2 - Human}
            \end{subfigure}
            \hfill
            \begin{subfigure}{0.49\textwidth}
                \includegraphics[width=\linewidth]{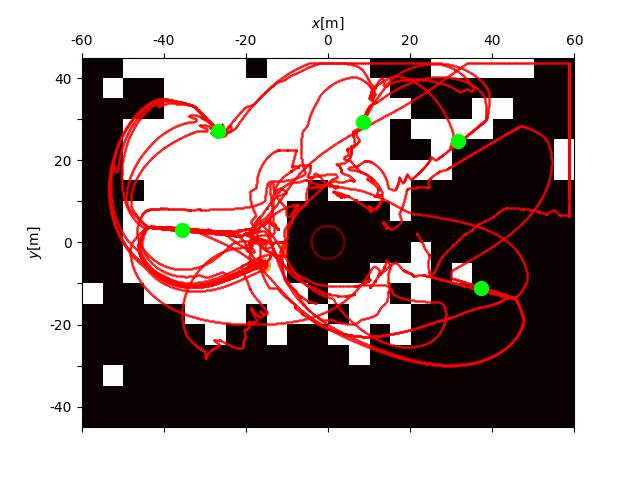}
                \caption{Player 2 - AA}
            \end{subfigure}
        \end{figure}
        \newpage

\section*{Self-Play Team}

    \subsection*{Trial \#7}
        \begin{figure}[H]
            \begin{subfigure}{0.49\textwidth}
                \includegraphics[width=\linewidth]{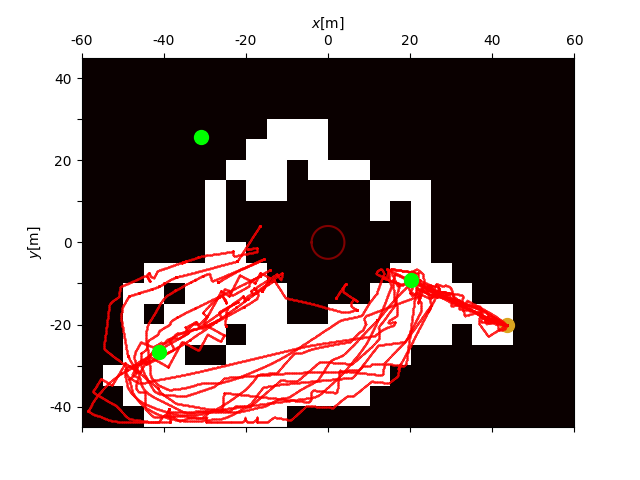}
                \caption{Player 1 - Human}
            \end{subfigure}
            \hfill
            \begin{subfigure}{0.49\textwidth}
                \includegraphics[width=\linewidth]{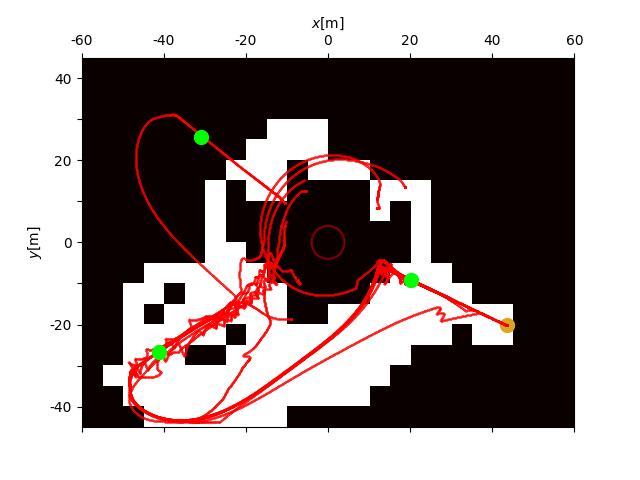}
                \caption{Player 1 - AA}
            \end{subfigure}

            \medskip
            \begin{subfigure}{0.49\textwidth}
                \includegraphics[width=\linewidth]{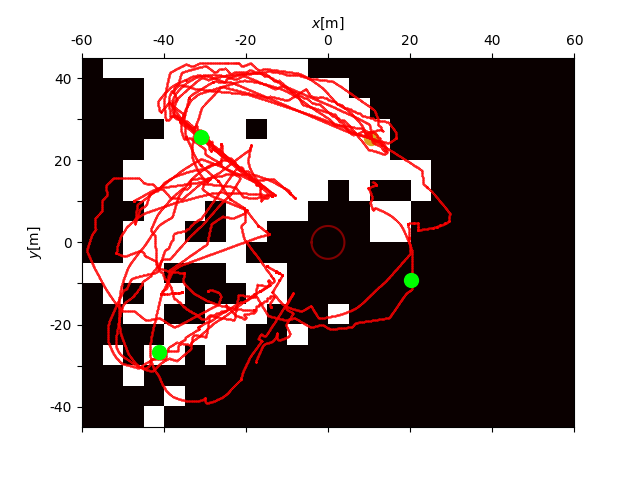}
                \caption{Player 2 - Human}
            \end{subfigure}
            \hfill
            \begin{subfigure}{0.49\textwidth}
                \includegraphics[width=\linewidth]{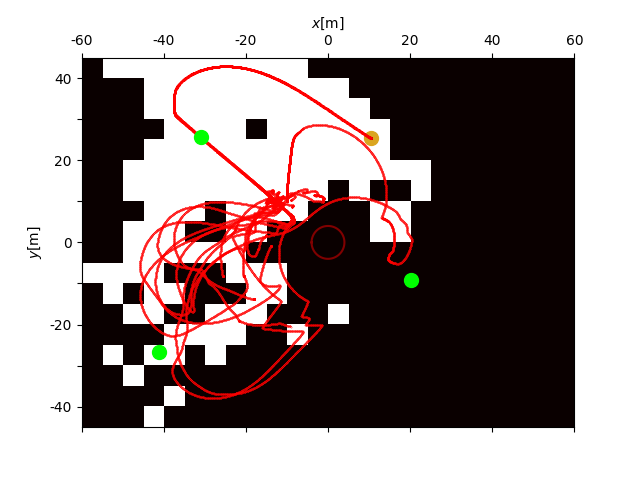}
                \caption{Player 2 - AA}
            \end{subfigure}
        \end{figure}
        \newpage

    \subsection*{Trial \#8}
        \begin{figure}[H]
            \begin{subfigure}{0.49\textwidth}
                \includegraphics[width=\linewidth]{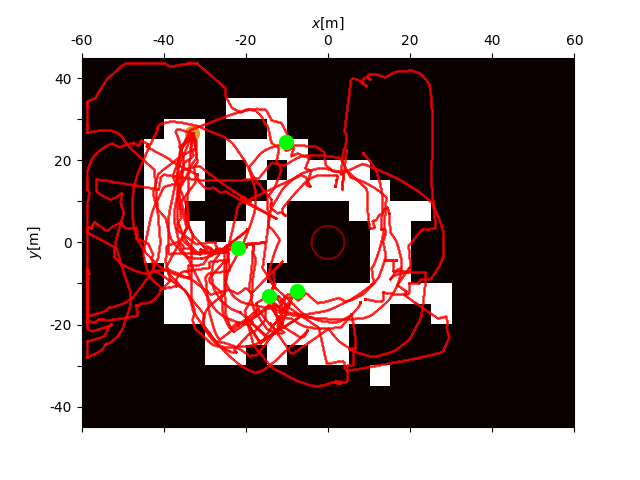}
                \caption{Player 1 - Human}
            \end{subfigure}
            \hfill
            \begin{subfigure}{0.49\textwidth}
                \includegraphics[width=\linewidth]{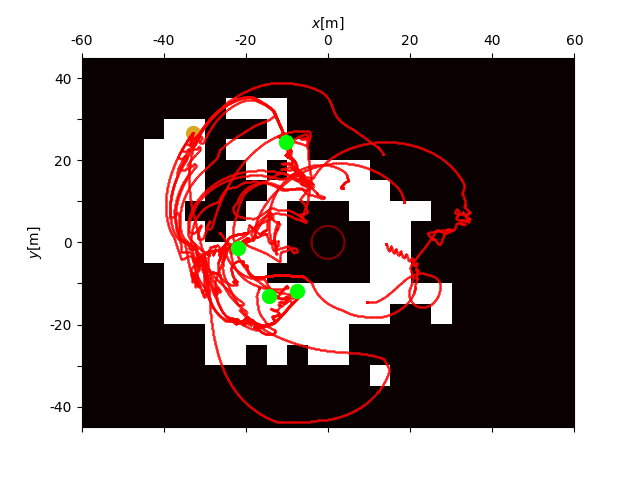}
                \caption{Player 1 - AA}
            \end{subfigure}

            \medskip
            \begin{subfigure}{0.49\textwidth}
                \includegraphics[width=\linewidth]{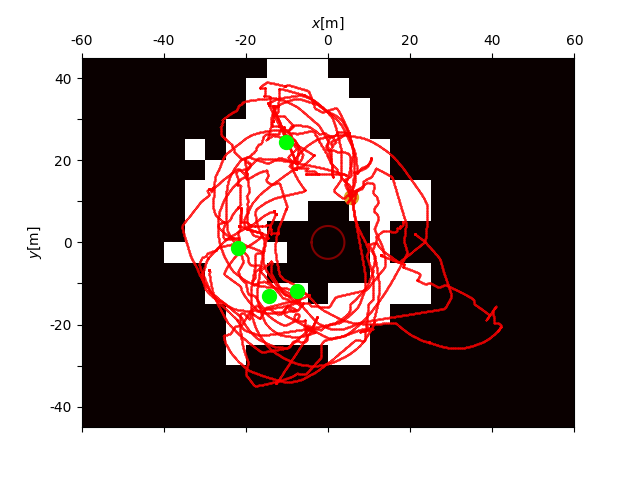}
                \caption{Player 2 - Human}
            \end{subfigure}
            \hfill
            \begin{subfigure}{0.49\textwidth}
                \includegraphics[width=\linewidth]{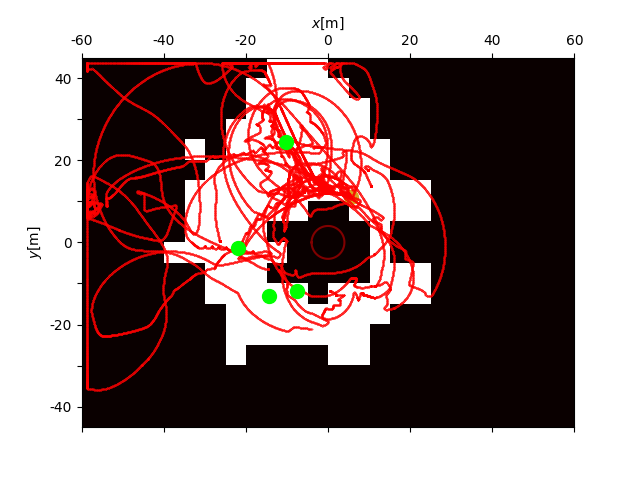}
                \caption{Player 2 - AA}
            \end{subfigure}
        \end{figure}
        \newpage

    \subsection*{Trial \#9}
        \begin{figure}[H]
            \begin{subfigure}{0.49\textwidth}
                \includegraphics[width=\linewidth]{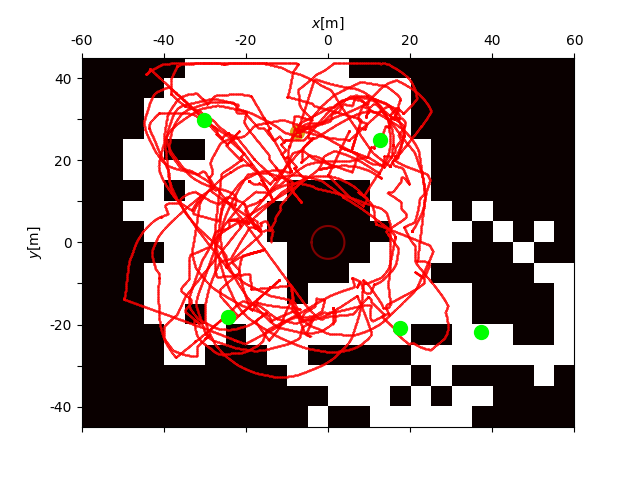}
                \caption{Player 1 - Human}
            \end{subfigure}
            \hfill
            \begin{subfigure}{0.49\textwidth}
                \includegraphics[width=\linewidth]{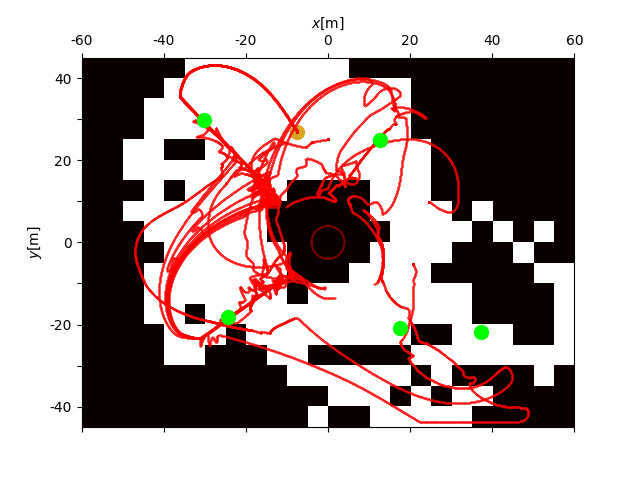}
                \caption{Player 1 - AA}
            \end{subfigure}

            \medskip
            \begin{subfigure}{0.49\textwidth}
                \includegraphics[width=\linewidth]{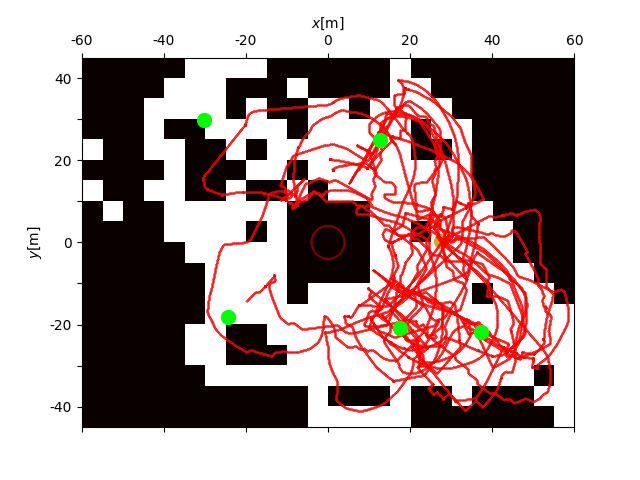}
                \caption{Player 2 - Human}
            \end{subfigure}
            \hfill
            \begin{subfigure}{0.49\textwidth}
                \includegraphics[width=\linewidth]{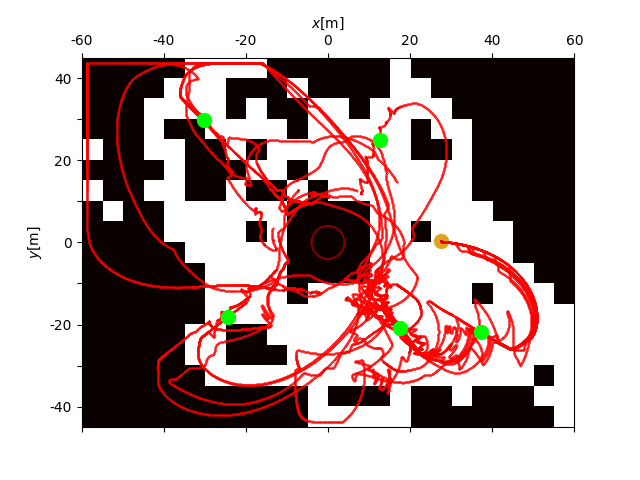}
                \caption{Player 2 - AA}
            \end{subfigure}
        \end{figure}
        \newpage

    \subsection*{Trial \#10}
        \begin{figure}[H]
            \begin{subfigure}{0.49\textwidth}
                \includegraphics[width=\linewidth]{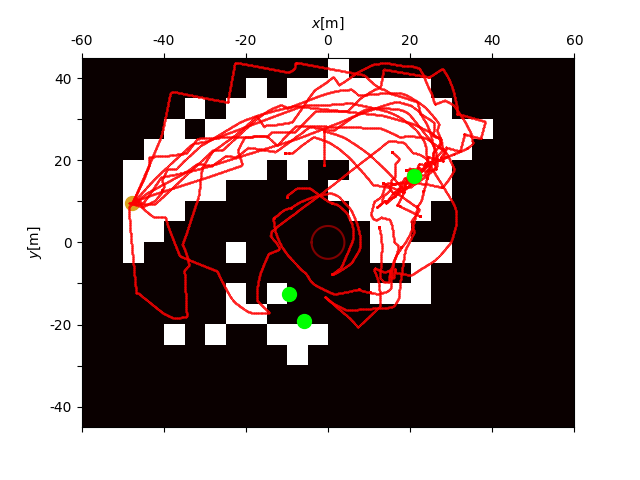}
                \caption{Player 1 - Human}
            \end{subfigure}
            \hfill
            \begin{subfigure}{0.49\textwidth}
                \includegraphics[width=\linewidth]{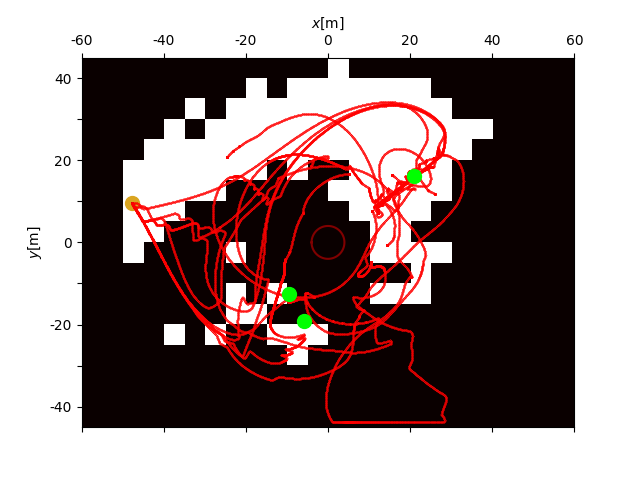}
                \caption{Player 1 - AA}
            \end{subfigure}

            \medskip
            \begin{subfigure}{0.49\textwidth}
                \includegraphics[width=\linewidth]{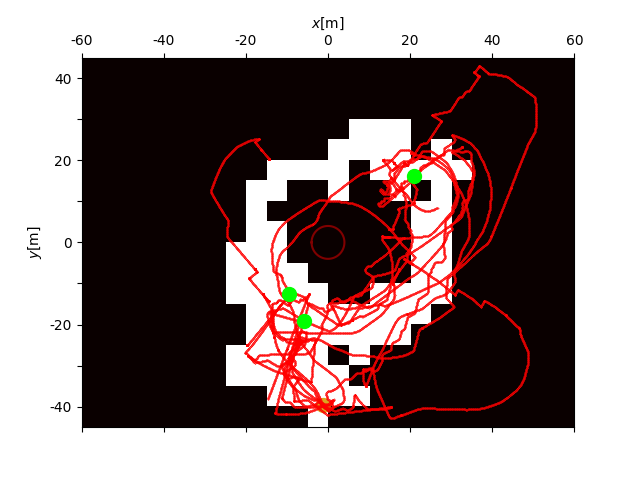}
                \caption{Player 2 - Human}
            \end{subfigure}
            \hfill
            \begin{subfigure}{0.49\textwidth}
                \includegraphics[width=\linewidth]{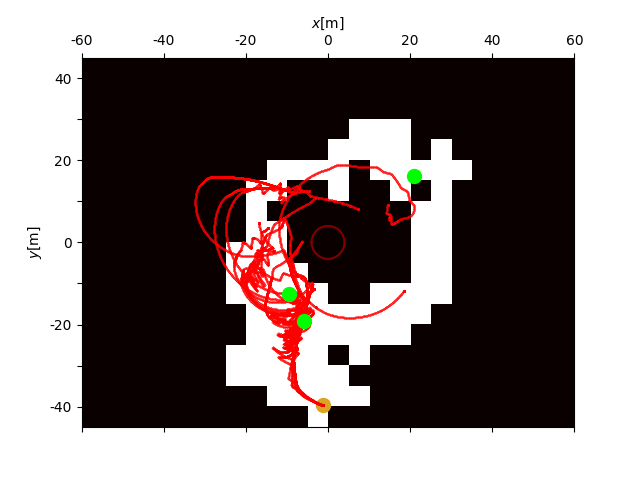}
                \caption{Player 2 - AA}
            \end{subfigure}
        \end{figure}
        \newpage

    \subsection*{Trial \#11}
        \begin{figure}[H]
            \begin{subfigure}{0.49\textwidth}
                \includegraphics[width=\linewidth]{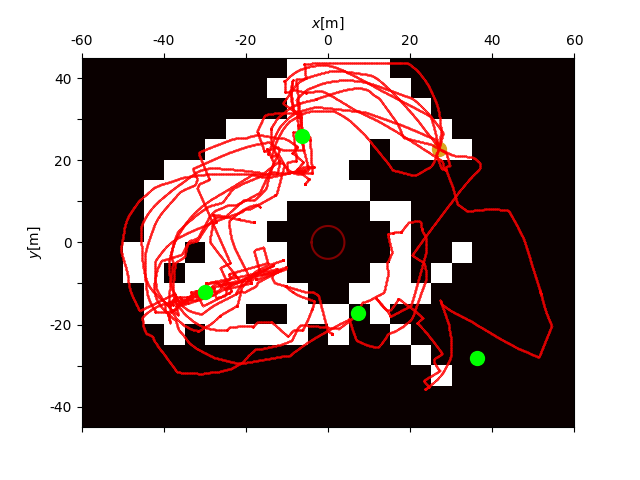}
                \caption{Player 1 - Human}
            \end{subfigure}
            \hfill
            \begin{subfigure}{0.49\textwidth}
                \includegraphics[width=\linewidth]{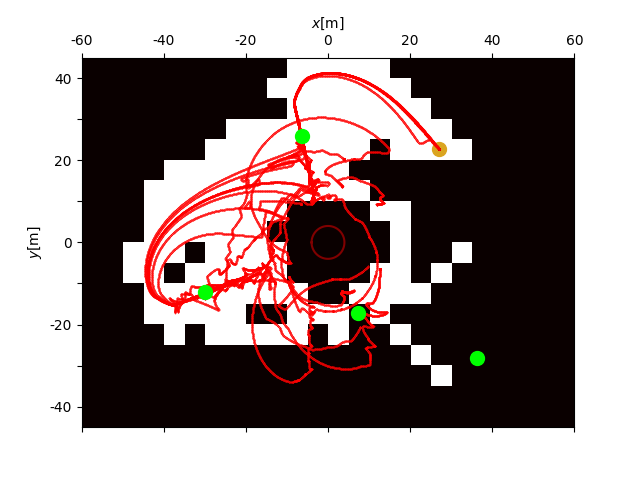}
                \caption{Player 1 - AA}
            \end{subfigure}

            \medskip
            \begin{subfigure}{0.49\textwidth}
                \includegraphics[width=\linewidth]{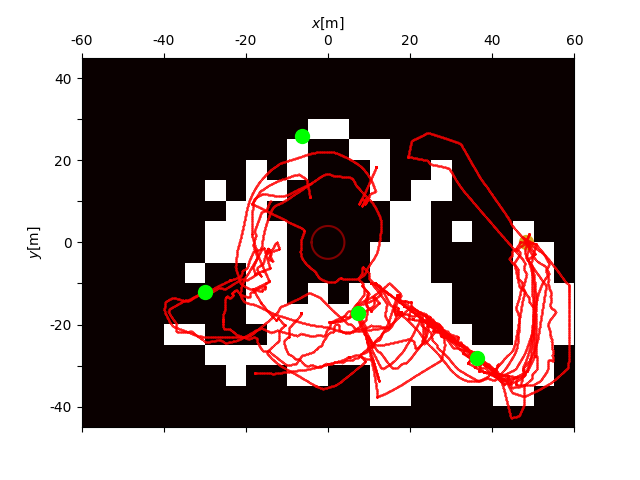}
                \caption{Player 2 - Human}
            \end{subfigure}
            \hfill
            \begin{subfigure}{0.49\textwidth}
                \includegraphics[width=\linewidth]{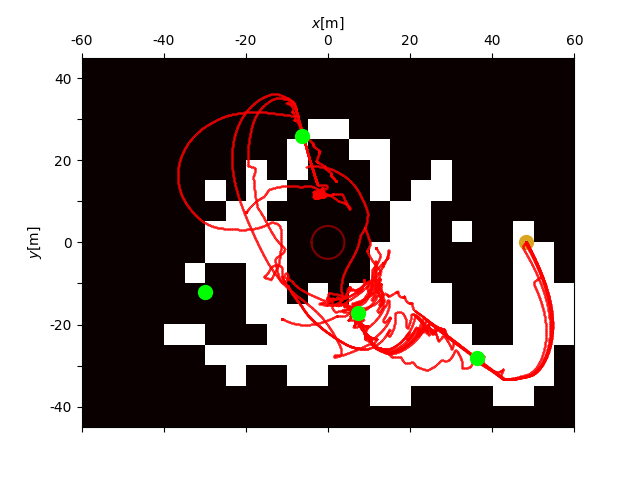}
                \caption{Player 2 - AA}
            \end{subfigure}
        \end{figure}
        \newpage

    \subsection*{Trial \#12}
        \begin{figure}[H]
            \begin{subfigure}{0.49\textwidth}
                \includegraphics[width=\linewidth]{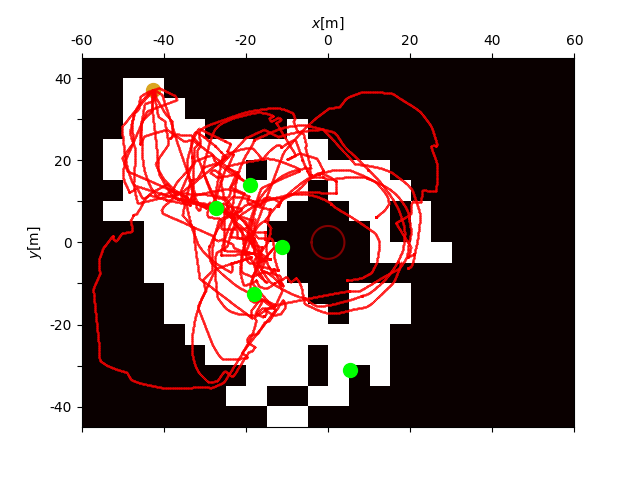}
                \caption{Player 1 - Human}
            \end{subfigure}
            \hfill
            \begin{subfigure}{0.49\textwidth}
                \includegraphics[width=\linewidth]{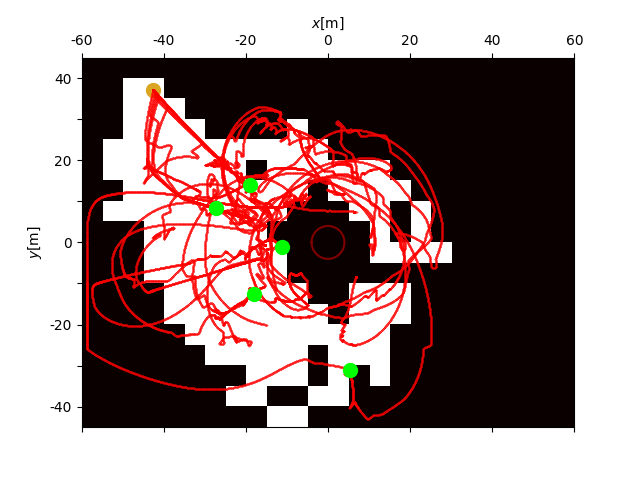}
                \caption{Player 1 - AA}
            \end{subfigure}

            \medskip
            \begin{subfigure}{0.49\textwidth}
                \includegraphics[width=\linewidth]{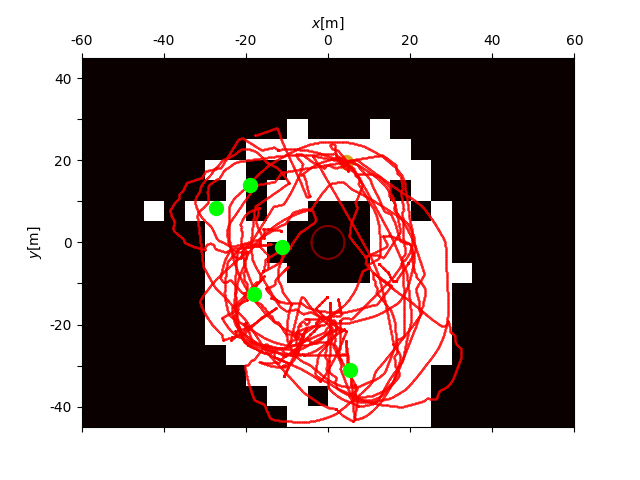}
                \caption{Player 2 - Human}
            \end{subfigure}
            \hfill
            \begin{subfigure}{0.49\textwidth}
                \includegraphics[width=\linewidth]{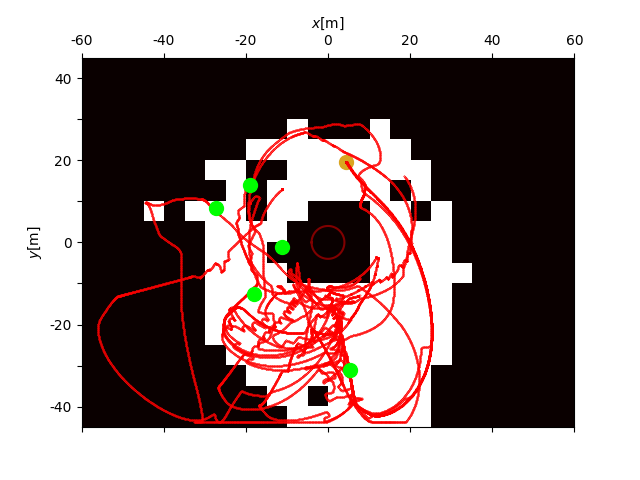}
                \caption{Player 2 - AA}
            \end{subfigure}
        \end{figure}
        \newpage

    \subsection*{Trial \#13}
        \begin{figure}[H]
            \begin{subfigure}{0.49\textwidth}
                \includegraphics[width=\linewidth]{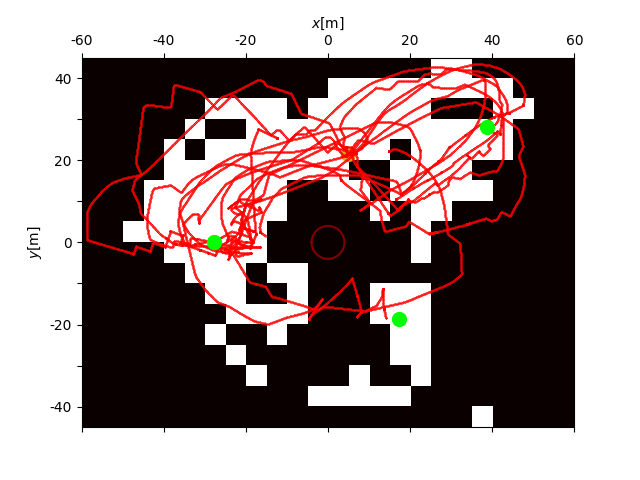}
                \caption{Player 1 - Human}
            \end{subfigure}
            \hfill
            \begin{subfigure}{0.49\textwidth}
                \includegraphics[width=\linewidth]{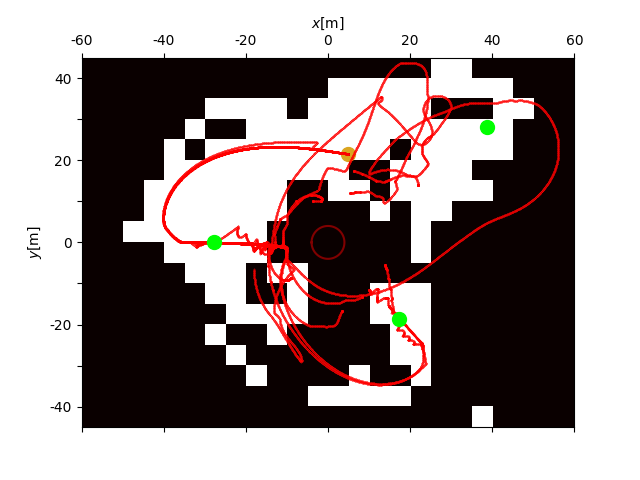}
                \caption{Player 1 - AA}
            \end{subfigure}

            \medskip
            \begin{subfigure}{0.49\textwidth}
                \includegraphics[width=\linewidth]{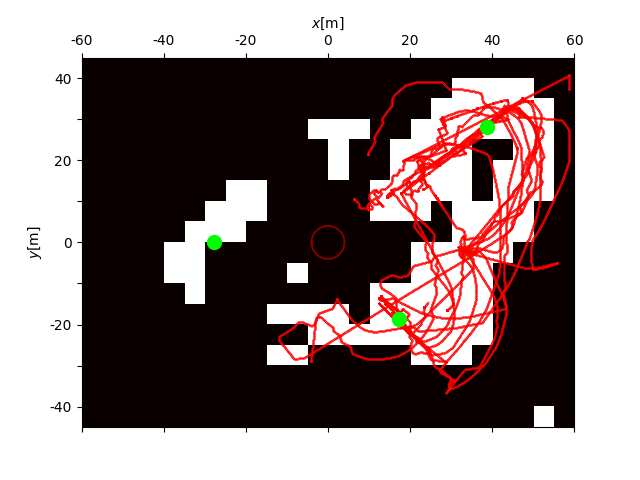}
                \caption{Player 2 - Human}
            \end{subfigure}
            \hfill
            \begin{subfigure}{0.49\textwidth}
                \includegraphics[width=\linewidth]{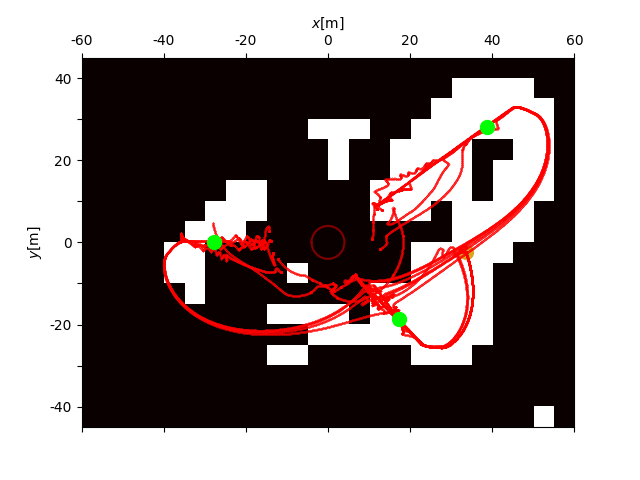}
                \caption{Player 2 - AA}
            \end{subfigure}
        \end{figure}
        \newpage

    \subsection*{Trial \#14}
        \begin{figure}[H]
            \begin{subfigure}{0.49\textwidth}
                \includegraphics[width=\linewidth]{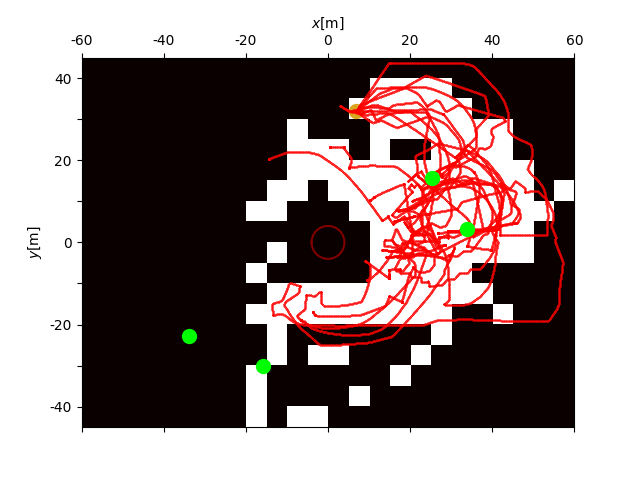}
                \caption{Player 1 - Human}
            \end{subfigure}
            \hfill
            \begin{subfigure}{0.49\textwidth}
                \includegraphics[width=\linewidth]{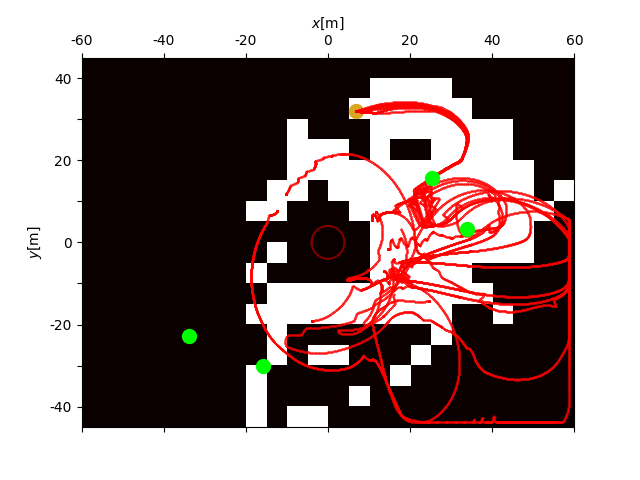}
                \caption{Player 1 - AA}
            \end{subfigure}

            \medskip
            \begin{subfigure}{0.49\textwidth}
                \includegraphics[width=\linewidth]{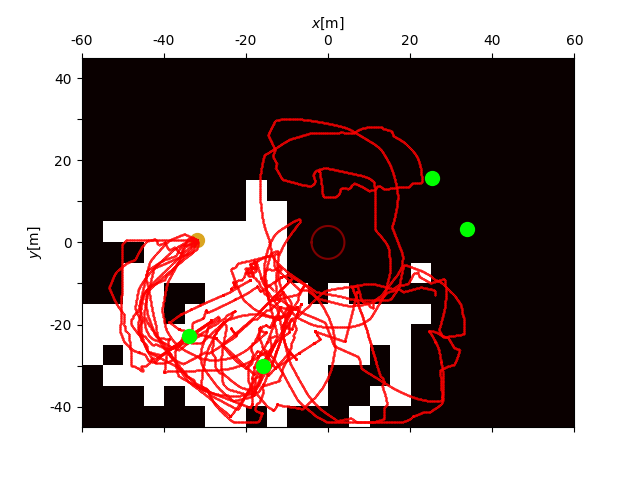}
                \caption{Player 2 - Human}
            \end{subfigure}
            \hfill
            \begin{subfigure}{0.49\textwidth}
                \includegraphics[width=\linewidth]{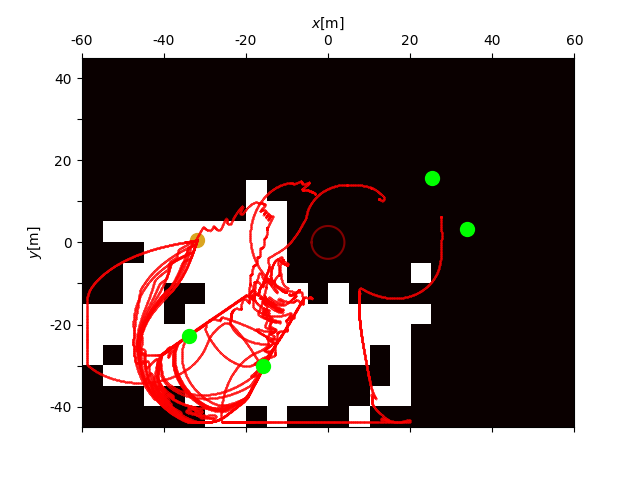}
                \caption{Player 2 - AA}
            \end{subfigure}
        \end{figure}
        \newpage

    \subsection*{Trial \#15}
        \begin{figure}[H]
            \begin{subfigure}{0.49\textwidth}
                \includegraphics[width=\linewidth]{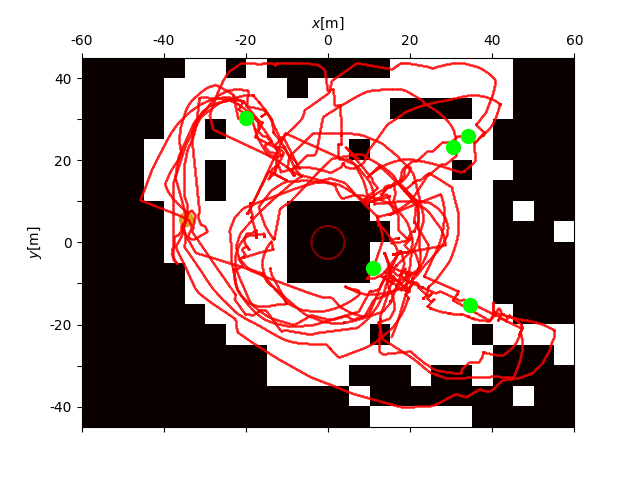}
                \caption{Player 1 - Human}
            \end{subfigure}
            \hfill
            \begin{subfigure}{0.49\textwidth}
                \includegraphics[width=\linewidth]{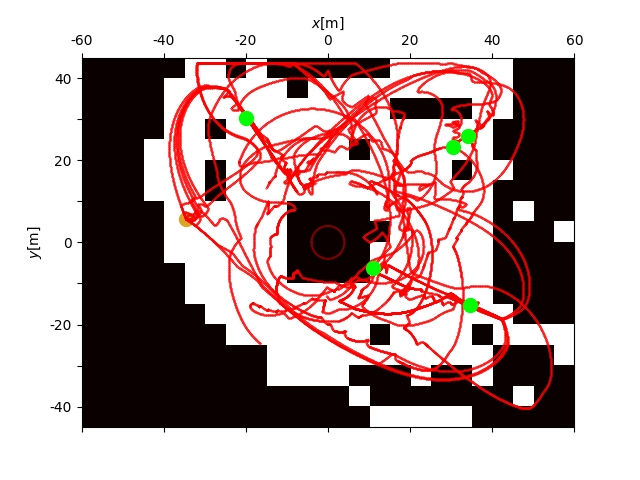}
                \caption{Player 1 - AA}
            \end{subfigure}

            \medskip
            \begin{subfigure}{0.49\textwidth}
                \includegraphics[width=\linewidth]{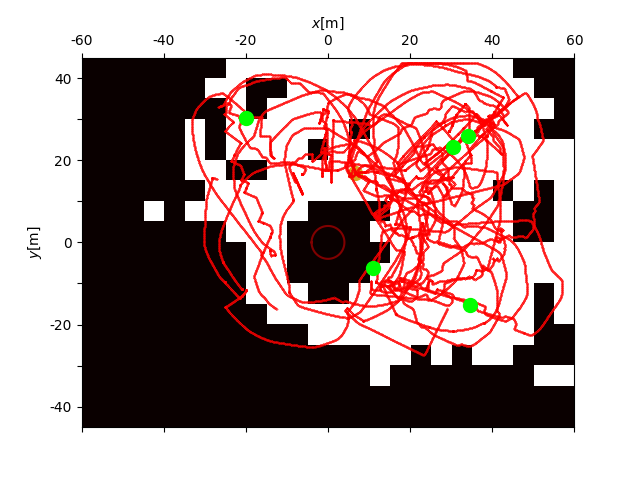}
                \caption{Player 2 - Human}
            \end{subfigure}
            \hfill
            \begin{subfigure}{0.49\textwidth}
                \includegraphics[width=\linewidth]{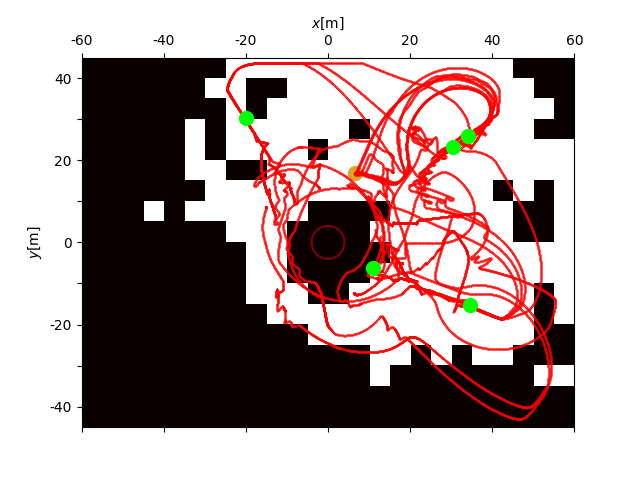}
                \caption{Player 2 - AA}
            \end{subfigure}
        \end{figure}
        \newpage

    \subsection*{Trial \#16}
        \begin{figure}[H]
            \begin{subfigure}{0.49\textwidth}
                \includegraphics[width=\linewidth]{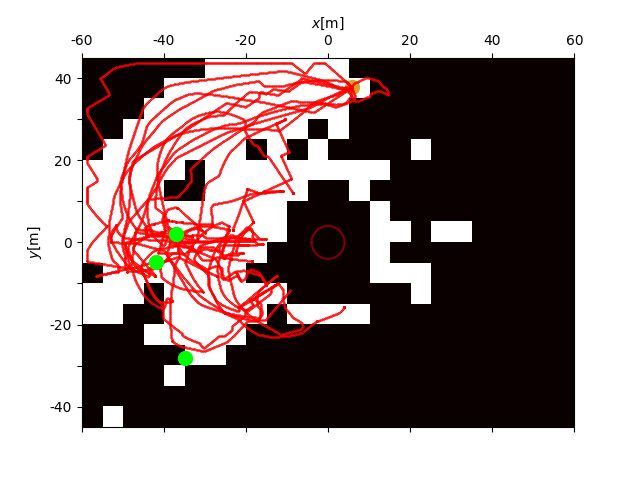}
                \caption{Player 1 - Human}
            \end{subfigure}
            \hfill
            \begin{subfigure}{0.49\textwidth}
                \includegraphics[width=\linewidth]{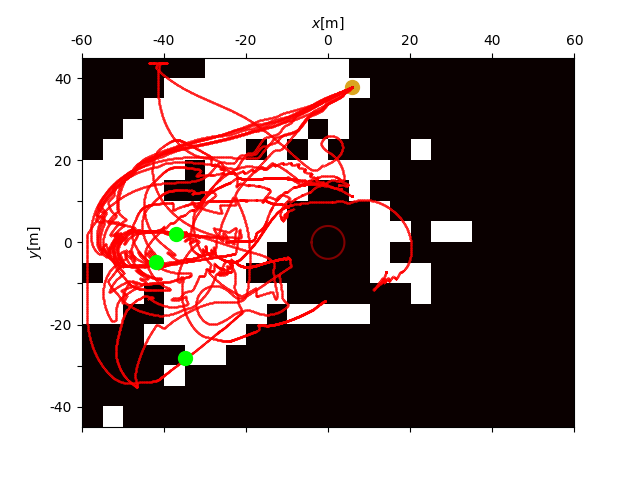}
                \caption{Player 1 - AA}
            \end{subfigure}

            \medskip
            \begin{subfigure}{0.49\textwidth}
                \includegraphics[width=\linewidth]{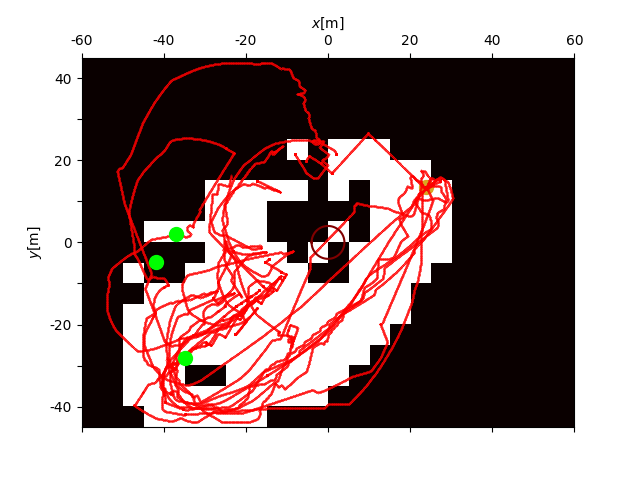}
                \caption{Player 2 - Human}
            \end{subfigure}
            \hfill
            \begin{subfigure}{0.49\textwidth}
                \includegraphics[width=\linewidth]{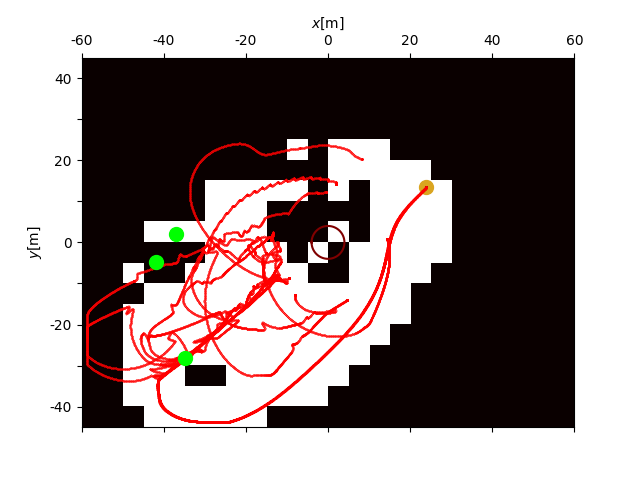}
                \caption{Player 2 - AA}
            \end{subfigure}
        \end{figure}
        \newpage

    \subsection*{Trial \#17}
        \begin{figure}[H]
            \begin{subfigure}{0.49\textwidth}
                \includegraphics[width=\linewidth]{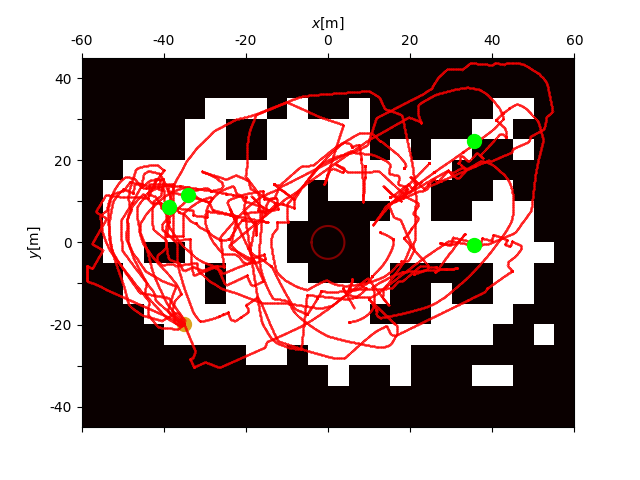}
                \caption{Player 1 - Human}
            \end{subfigure}
            \hfill
            \begin{subfigure}{0.49\textwidth}
                \includegraphics[width=\linewidth]{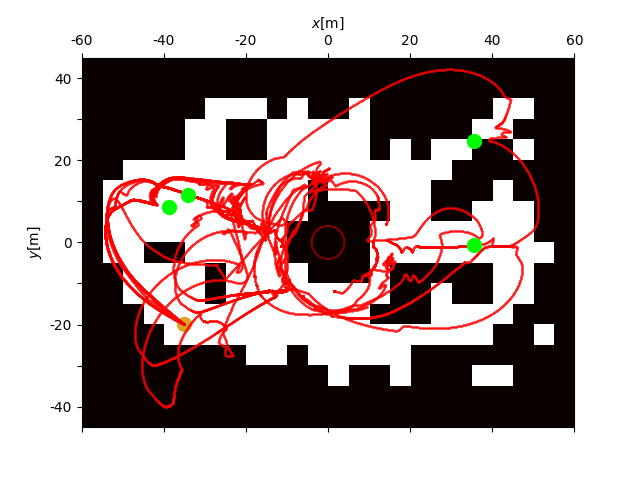}
                \caption{Player 1 - AA}
            \end{subfigure}

            \medskip
            \begin{subfigure}{0.49\textwidth}
                \includegraphics[width=\linewidth]{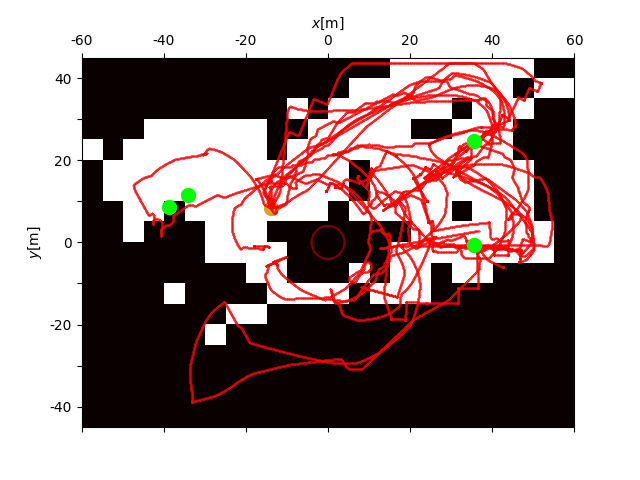}
                \caption{Player 2 - Human}
            \end{subfigure}
            \hfill
            \begin{subfigure}{0.49\textwidth}
                \includegraphics[width=\linewidth]{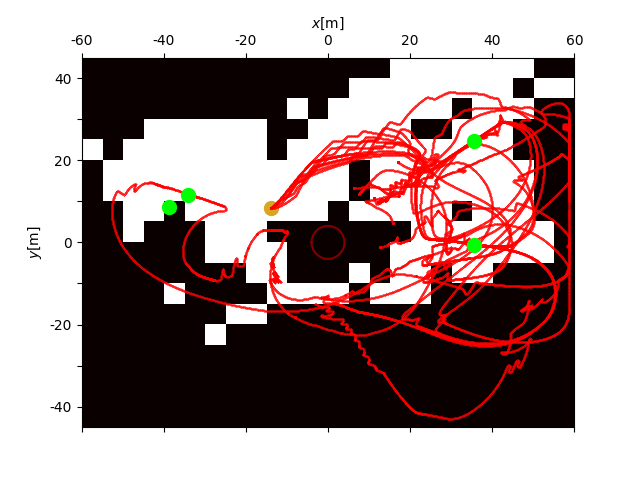}
                \caption{Player 2 - AA}
            \end{subfigure}
        \end{figure}
        \newpage

    \subsection*{Trial \#18}
        \begin{figure}[H]
            \begin{subfigure}{0.49\textwidth}
                \includegraphics[width=\linewidth]{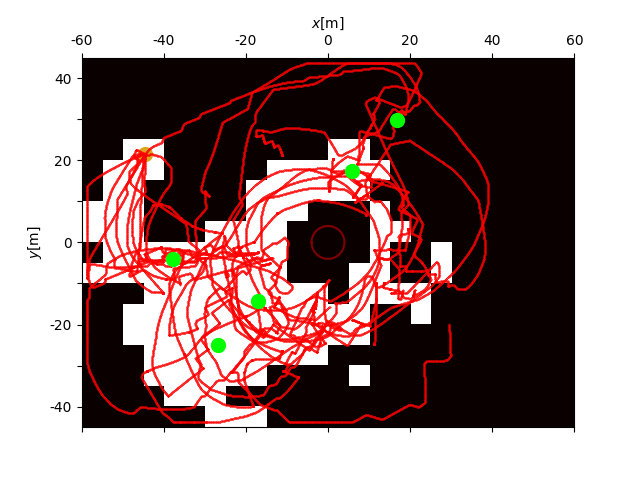}
                \caption{Player 1 - Human}
            \end{subfigure}
            \hfill
            \begin{subfigure}{0.49\textwidth}
                \includegraphics[width=\linewidth]{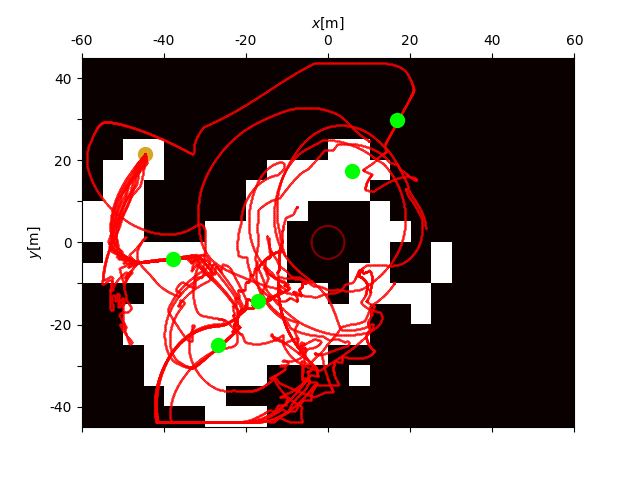}
                \caption{Player 1 - AA}
            \end{subfigure}

            \medskip
            \begin{subfigure}{0.49\textwidth}
                \includegraphics[width=\linewidth]{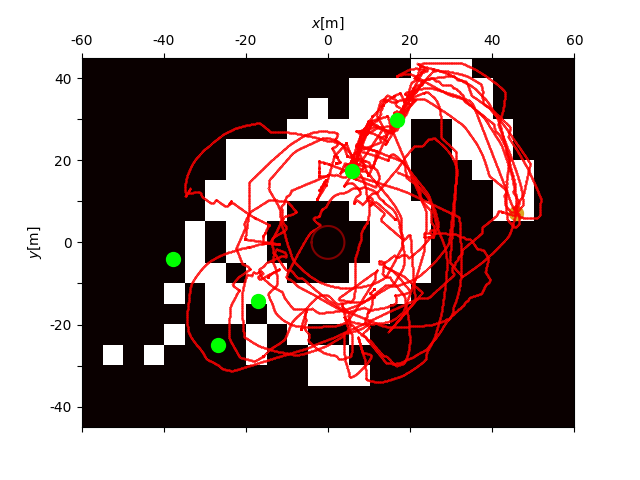}
                \caption{Player 2 - Human}
            \end{subfigure}
            \hfill
            \begin{subfigure}{0.49\textwidth}
                \includegraphics[width=\linewidth]{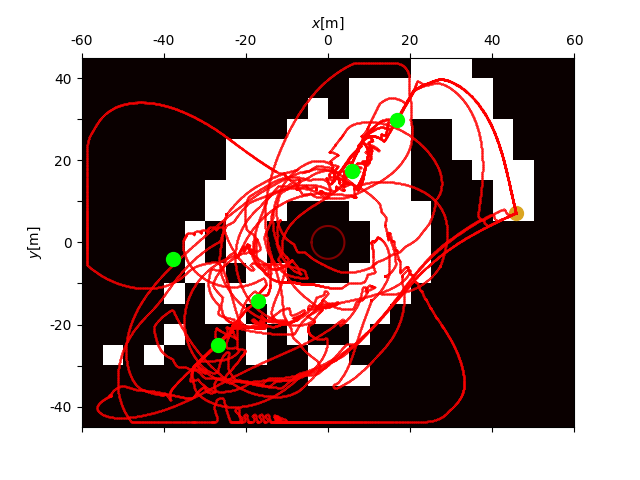}
                \caption{Player 2 - AA}
            \end{subfigure}
        \end{figure}
        \newpage

    \subsection*{Trial \#19}
        \begin{figure}[H]
            \begin{subfigure}{0.49\textwidth}
                \includegraphics[width=\linewidth]{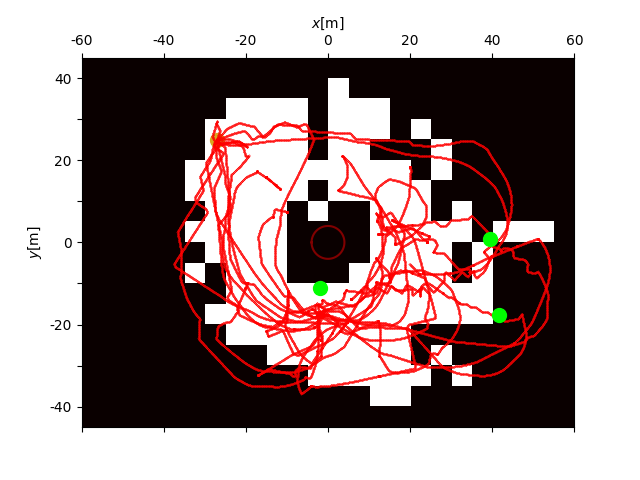}
                \caption{Player 1 - Human}
            \end{subfigure}
            \hfill
            \begin{subfigure}{0.49\textwidth}
                \includegraphics[width=\linewidth]{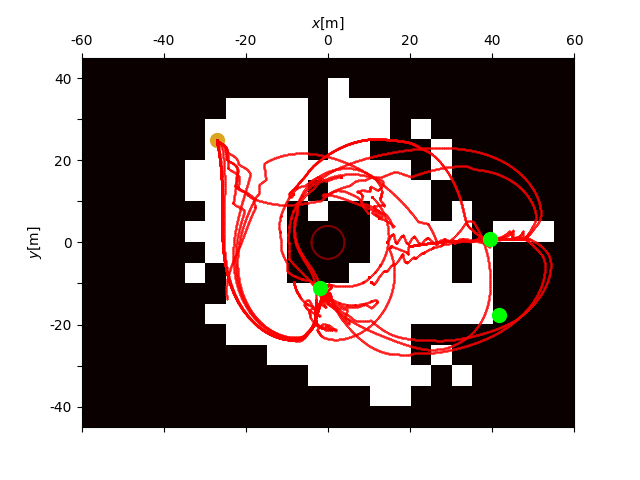}
                \caption{Player 1 - AA}
            \end{subfigure}

            \medskip
            \begin{subfigure}{0.49\textwidth}
                \includegraphics[width=\linewidth]{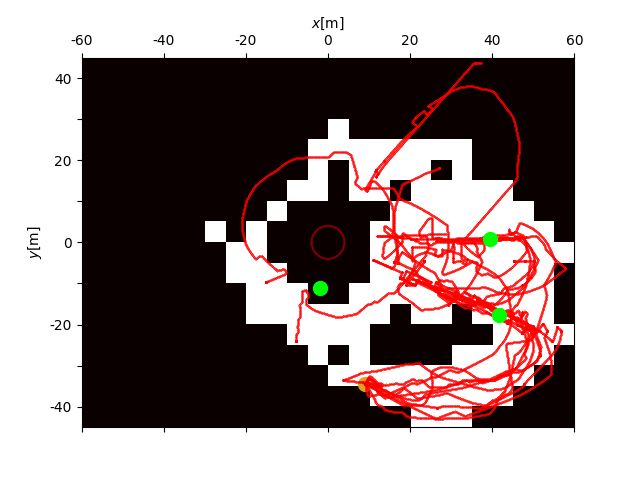}
                \caption{Player 2 - Human}
            \end{subfigure}
            \hfill
            \begin{subfigure}{0.49\textwidth}
                \includegraphics[width=\linewidth]{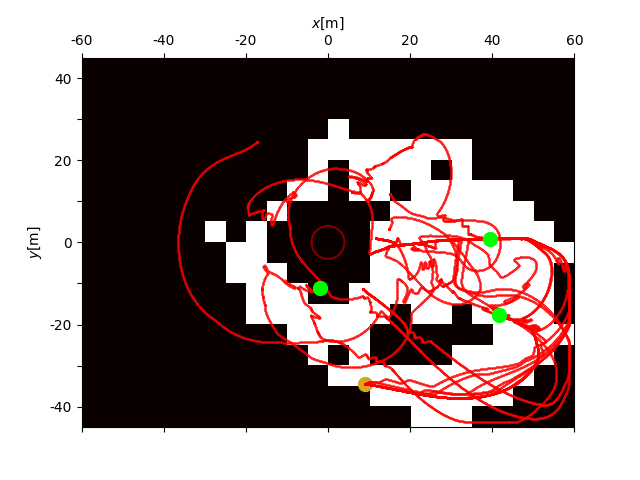}
                \caption{Player 2 - AA}
            \end{subfigure}
        \end{figure}
        \newpage

    \subsection*{Trial \#20}
        \begin{figure}[H]
            \begin{subfigure}{0.49\textwidth}
                \includegraphics[width=\linewidth]{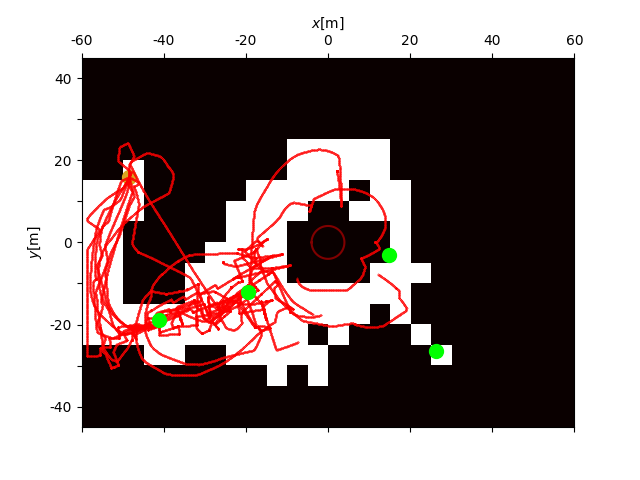}
                \caption{Player 1 - Human}
            \end{subfigure}
            \hfill
            \begin{subfigure}{0.49\textwidth}
                \includegraphics[width=\linewidth]{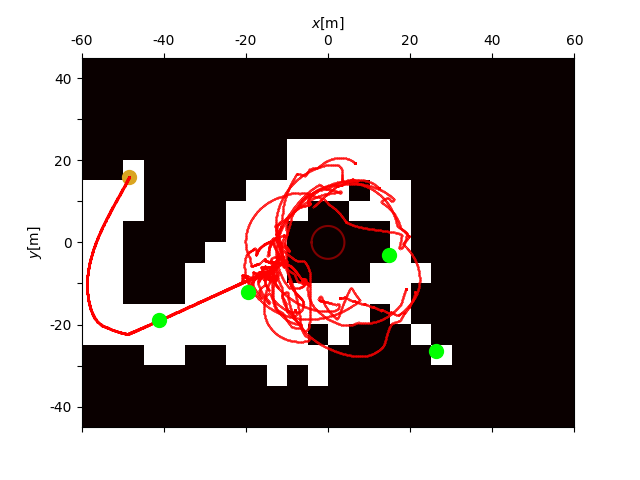}
                \caption{Player 1 - AA}
            \end{subfigure}

            \medskip
            \begin{subfigure}{0.49\textwidth}
                \includegraphics[width=\linewidth]{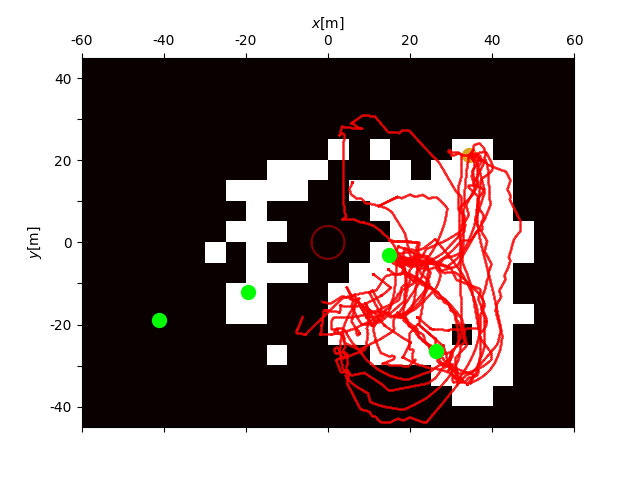}
                \caption{Player 2 - Human}
            \end{subfigure}
            \hfill
            \begin{subfigure}{0.49\textwidth}
                \includegraphics[width=\linewidth]{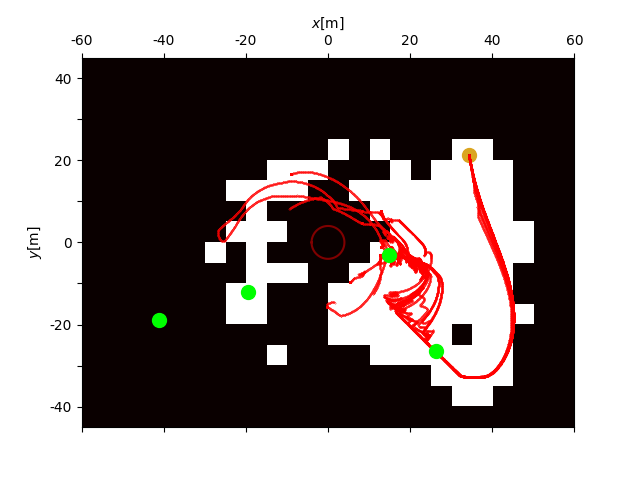}
                \caption{Player 2 - AA}
            \end{subfigure}
        \end{figure}
        \newpage

    \subsection*{Trial \#21}
        \begin{figure}[H]
            \begin{subfigure}{0.49\textwidth}
                \includegraphics[width=\linewidth]{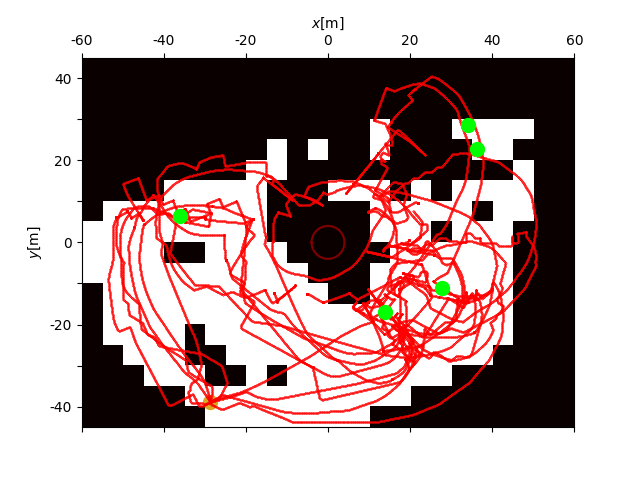}
                \caption{Player 1 - Human}
            \end{subfigure}
            \hfill
            \begin{subfigure}{0.49\textwidth}
                \includegraphics[width=\linewidth]{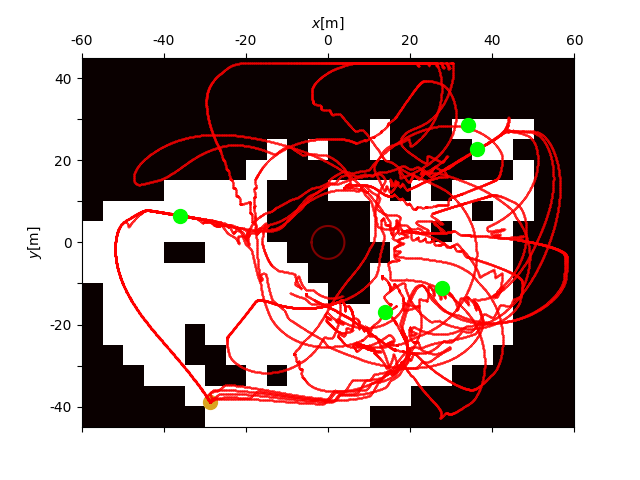}
                \caption{Player 1 - AA}
            \end{subfigure}

            \medskip
            \begin{subfigure}{0.49\textwidth}
                \includegraphics[width=\linewidth]{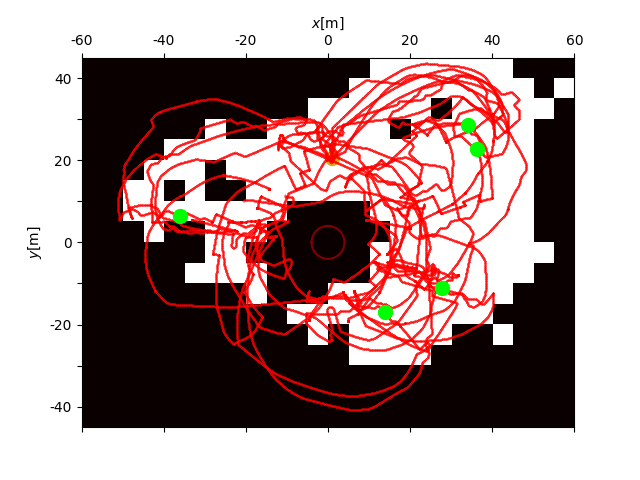}
                \caption{Player 2 - Human}
            \end{subfigure}
            \hfill
            \begin{subfigure}{0.49\textwidth}
                \includegraphics[width=\linewidth]{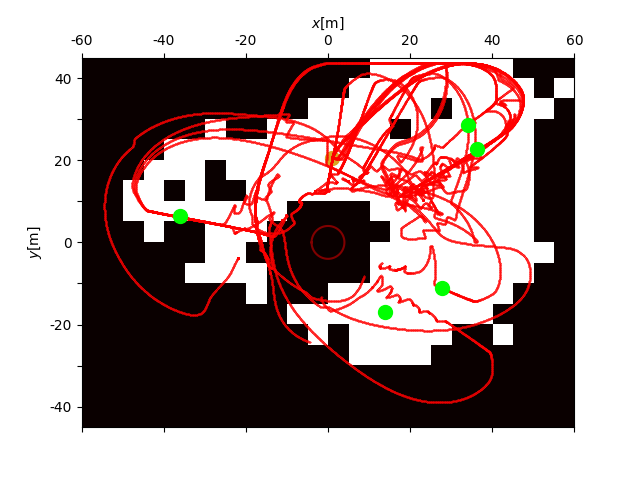}
                \caption{Player 2 - AA}
            \end{subfigure}
        \end{figure}
        \newpage

    \subsection*{Trial \#22}
        \begin{figure}[H]
            \begin{subfigure}{0.49\textwidth}
                \includegraphics[width=\linewidth]{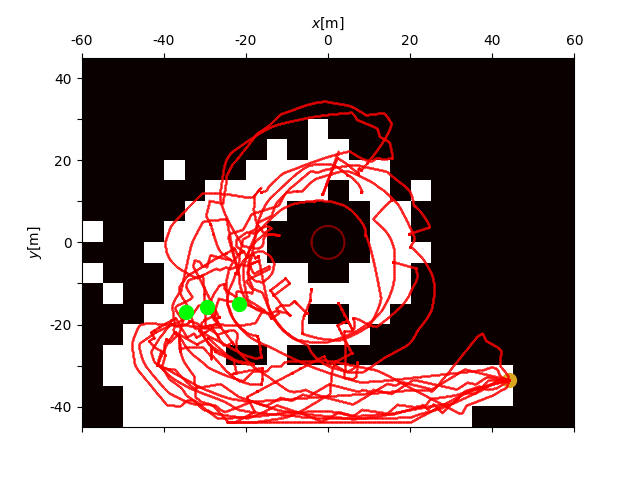}
                \caption{Player 1 - Human}
            \end{subfigure}
            \hfill
            \begin{subfigure}{0.49\textwidth}
                \includegraphics[width=\linewidth]{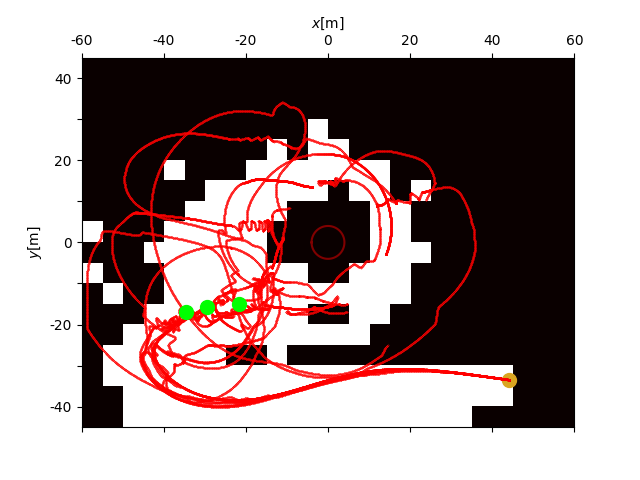}
                \caption{Player 1 - AA}
            \end{subfigure}

            \medskip
            \begin{subfigure}{0.49\textwidth}
                \includegraphics[width=\linewidth]{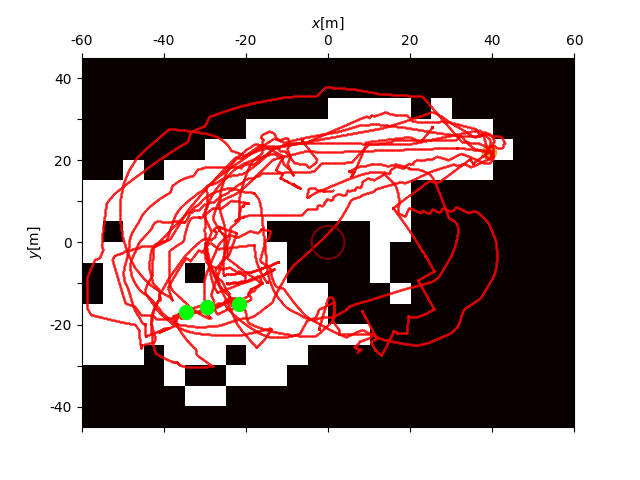}
                \caption{Player 2 - Human}
            \end{subfigure}
            \hfill
            \begin{subfigure}{0.49\textwidth}
                \includegraphics[width=\linewidth]{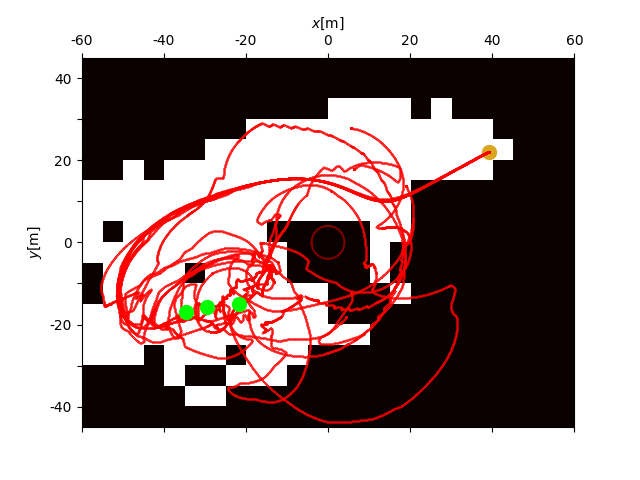}
                \caption{Player 2 - AA}
            \end{subfigure}
        \end{figure}
        \newpage

    \subsection*{Trial \#23}
        \begin{figure}[H]
            \begin{subfigure}{0.49\textwidth}
                \includegraphics[width=\linewidth]{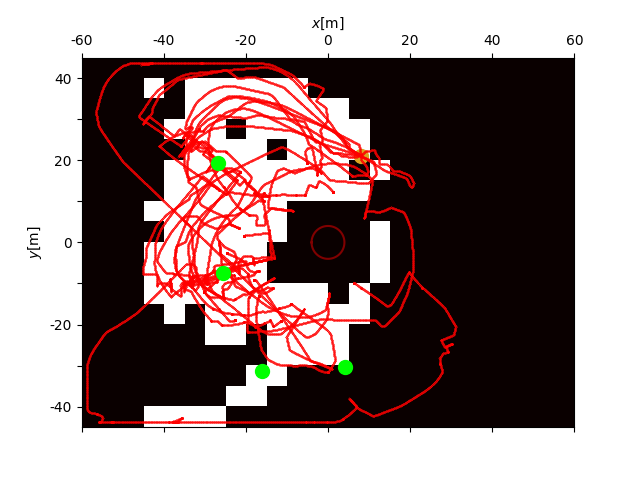}
                \caption{Player 1 - Human}
            \end{subfigure}
            \hfill
            \begin{subfigure}{0.49\textwidth}
                \includegraphics[width=\linewidth]{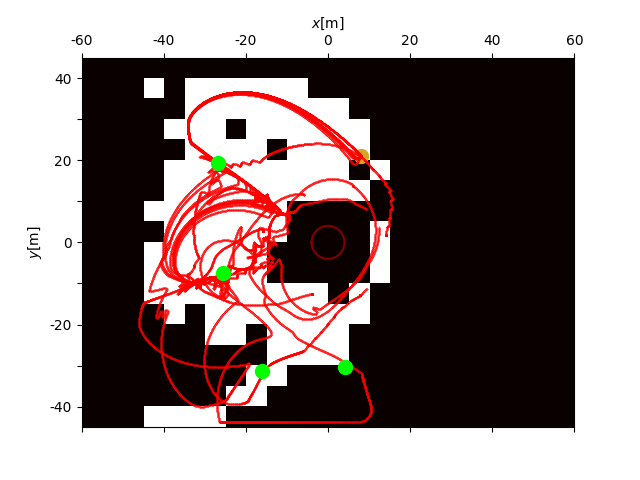}
                \caption{Player 1 - AA}
            \end{subfigure}

            \medskip
            \begin{subfigure}{0.49\textwidth}
                \includegraphics[width=\linewidth]{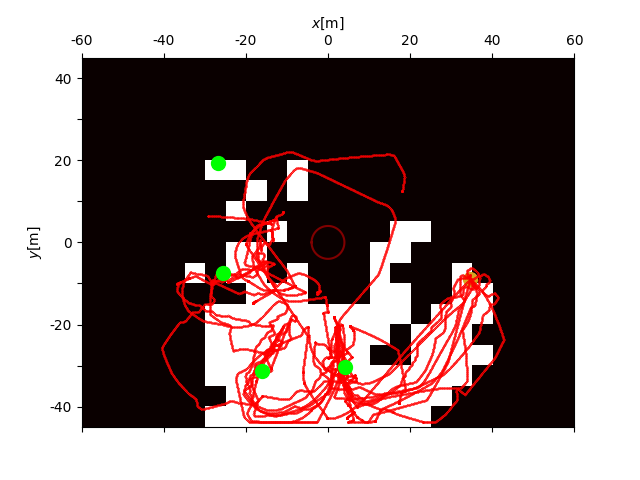}
                \caption{Player 2 - Human}
            \end{subfigure}
            \hfill
            \begin{subfigure}{0.49\textwidth}
                \includegraphics[width=\linewidth]{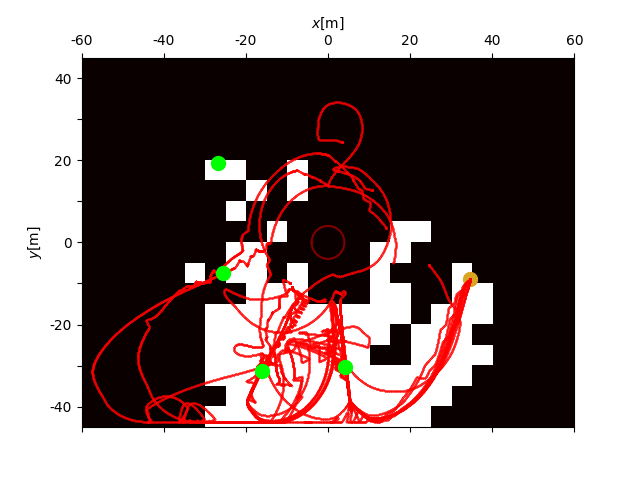}
                \caption{Player 2 - AA}
            \end{subfigure}
        \end{figure}
        \newpage

    \subsection*{Trial \#24}
        \begin{figure}[H]
            \begin{subfigure}{0.49\textwidth}
                \includegraphics[width=\linewidth]{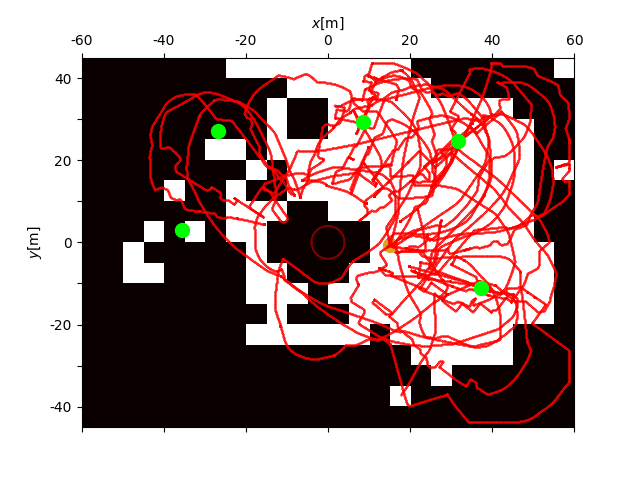}
                \caption{Player 1 - Human}
            \end{subfigure}
            \hfill
            \begin{subfigure}{0.49\textwidth}
                \includegraphics[width=\linewidth]{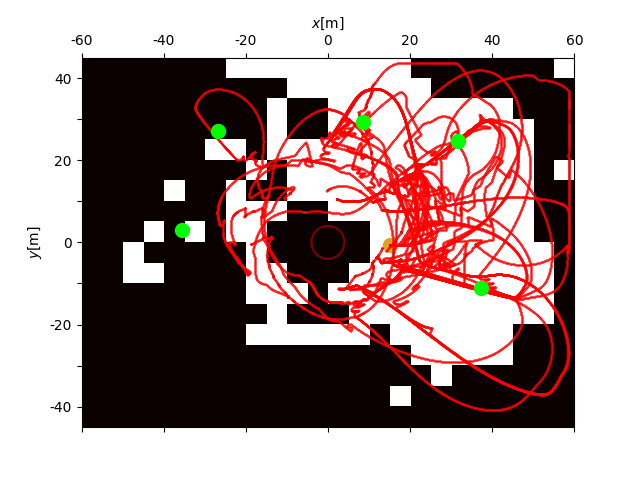}
                \caption{Player 1 - AA}
            \end{subfigure}

            \medskip
            \begin{subfigure}{0.49\textwidth}
                \includegraphics[width=\linewidth]{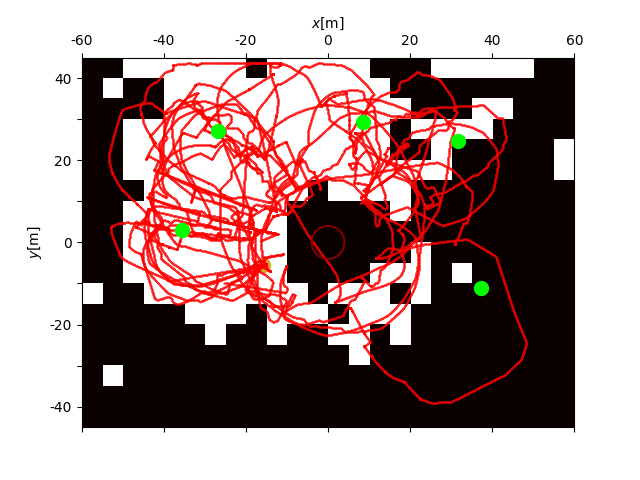}
                \caption{Player 2 - Human}
            \end{subfigure}
            \hfill
            \begin{subfigure}{0.49\textwidth}
                \includegraphics[width=\linewidth]{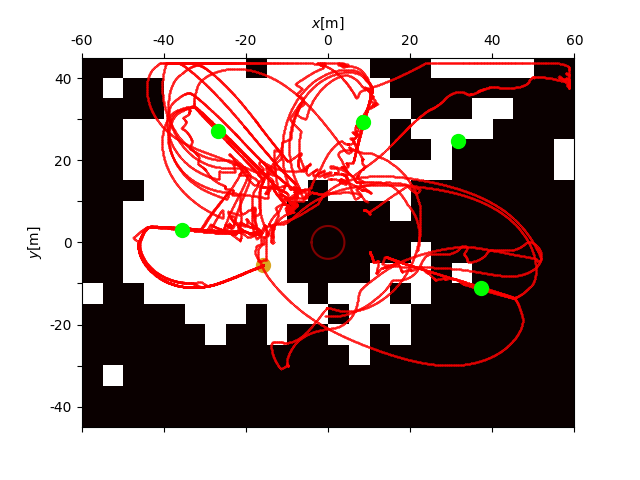}
                \caption{Player 2 - AA}
            \end{subfigure}
        \end{figure}
        \newpage